\documentclass[aps, superscriptaddress,  twocolumn]{revtex4}
\usepackage{CJK}
\usepackage{mathtools}
\usepackage[normalem]{ulem}
\usepackage{graphicx}
\usepackage{dcolumn}
\usepackage[mathlines]{lineno}
\usepackage{hyperref}
\usepackage{bm}
\usepackage{amssymb}
\usepackage{amsmath}
\usepackage{braket}
\usepackage{color}
\usepackage{xcolor}
\usepackage{soul} 
\usepackage{graphics}
\usepackage{bbm}

\newcommand{\bea}{\begin{eqnarray}}
\newcommand{\eea}{\end{eqnarray}}

\begin{document}
\title{The topological spectrum of high dimensional quantum states}

\author{Robert de Mello Koch}
\email{robert.demellokoch@gmail.com}
\affiliation{School of Science, Huzhou University, Huzhou 313000, China}
\affiliation{Mandelstam Institute for Theoretical Physics, School of Physics, University of the Witwatersrand, Private Bag 3, Wits 2050, South Africa}

\author{Pedro Ornelas}
\affiliation{School of Physics, University of the Witwatersrand, Private Bag 3, Wits 2050, South Africa}

\author{Neelan Gounden}
\affiliation{School of Physics, University of the Witwatersrand, Private Bag 3, Wits 2050, South Africa}

\author{Bo-Qiang Lu}
\affiliation{School of Science, Huzhou University, Huzhou 313000, China}


\author{Isaac Nape}
\affiliation{School of Physics, University of the Witwatersrand, Private Bag 3, Wits 2050, South Africa}

\author{Andrew Forbes}
\email{andrew.forbes@wits.ac.za}
\affiliation{School of Physics, University of the Witwatersrand, Private Bag 3, Wits 2050, South Africa}

\begin{abstract}
\noindent \textbf{Topology has emerged as a fundamental property of many systems, manifesting in cosmology, condensed matter, high-energy physics and waves. Despite the rich textures, the topology has largely been limited to low dimensional systems that can be characterised by a single topological number, e.g., a Chern number in matter or a Skyrme number in waves.  Here, using photonic quantum states as an example, we harness the synthetic dimensions of orbital angular momentum (OAM) to discover a rich tapestry of topological maps in high dimensional spaces. Moving beyond spin textured fields, we demonstrate topologies using only one degree of freedom, the OAM of light. By interpreting the density matrix as a non-Abelian Higgs potential, we are able to predict topologies that exist as high dimensional manifolds which remarkably can be deconstructed into a multitude of simpler maps from disks to disks and spheres to spheres, giving rise to the notion of a topological spectrum rather than a topological number. We confirm this experimentally using quantum wave functions with an underlying topology of 48 dimensions and a topological spectrum spanning over 17000 maps, an encoding alphabet with enormous potential.  We show that the topological spectrum allows the simultaneous ability to be robust to and probe for perturbation, the latter made possible by observing emergent signatures in the non-topological (trivial) spaces of the spectrum. Our experimental approach benefits from easy implementation, while our theoretical framework is cast in a manner that can be extrapolated to any particle type, dimension and degree of freedom.  Our work opens exciting future possibilities for quantum sensing and communication with topology.}
\end{abstract}

\maketitle

\noindent Topology has proven a powerful tool by which to characterise and understand complex systems.  Although often associated with physical objects \cite{ashbridge2022knotting}, such as the equivalence of coffee mugs and donuts, it can manifest in many diverse systems, including cosmology \cite{cruz2007cosmic}, condensed matter \cite{ozawa2019topological}, high-energy physics \cite{eto2024tying,faddeev1997knots}, quantum states \cite{hall2016tying}, acoustic  \cite{ge2021observation,xue2022topological} and water \cite{wang2025topological} waves. A recent development has been the emergence of optical topologies \cite{shen2024optical}, observed as Skyrmions in evanescent waves \cite{tsesses2018optical}, spin-textured vectorial light \cite{gutierrez2021optical}, non-paraxial light \cite{du2019deep}, and quantum entangled photons \cite{ornelas2024non}. The topology of all such systems defines a map from one space to another, captured by a topological number, e.g., a Chern number in condensed matter or a Skyrme number in waves, that remains invariant to smooth deformations of the map. This notion of topological resilience has proven very powerful across diverse fields, including information robustness \cite{wang2024topological}, storage and transfer \cite{yu2017room}, entanglement protection \cite{blanco2018topological}, light steering \cite{zhao2019non} as well as finding new unintended applications such as the development of neuromorphic computing architectures by mimicking biological synapses \cite{song2020skyrmion}.


All these experiments have been restricted to low dimensional systems characterised by a single topological number, yet altering order and dimensionality in topological systems has many benefits \cite{gobel2021beyond,benalcazar2017quantized,lustig2021topological}.  Advances in this direction include three-dimensional knots in crystals \cite{tai2019three}, two-dimensional plasmonic quasicrystals governed by four-dimensional vectors \cite{tsesses2025four}, using three-dimensional space and a free parameter to map to a four-dimensional hypersphere \cite{sugic2021particle,shen2023topological,ehrmanntraut2023optical}. Though it has been suggested theoretically that combining space and time could facilitate a true four-dimensional topology \cite{marco2022optical,lin2024space}, this is yet to be realised. 

Here we report the discovery of the rich topological structure in high dimensional spaces using photonic quantum states as an example.  We show that topology can be found in the same degree of freedom of two entangled particles, allowing us to demonstrate the first Skyrmion topology using only the orbital angular momentum (OAM) of light.  Next, we harness OAM as a synthetic dimension to realise high dimensional topologies embedded in high dimensional ($d$) quantum states.  We discover intricate structures whose description requires a topological spectrum of invariants, revealing a rich tapestry of maps. We outline a new theory of high dimensional topology where, inspired by the description of 't Hooft-Polyakov monopoles in non-Abelian Yang-Mills theory \cite{Harvey:1996ur,Irwin:1997ew}, we show that Lie algebra structures naturally organize and classify topological maps across dimensions. We are able to predict topologies in $d^2 - 1$ dimensional manifolds with a spectrum of $\binom{d^2-1}{3}$ possible mappings, each associated with a wrapping number,  $\frac{1}{4}d(d-1)(d-2)(d+3)$ of which are independent topological invariants. We confirm the theory experimentally, reaching topological manifolds in 48 dimensions, with signatures of beyond 17000 topological numbers, the highest dimensional topological signatures ever observed.  We find that the trivial signatures (those without topology) within the spectrum show emergent topology in the presence of perturbation while the non-trivial signatures remain largely unchanged, opening an exciting path to simultaneous resilience to, and probing of, complex channels. Our work is not restricted to OAM or photons and can easily be extended to other degrees of freedom and particles.  The new notion of a topological spectrum has the potential to unlock an extraordinarily large topological alphabet for information encoding, while opening new possibilities in sensing and metrology with topological states.

\section{Results}
\begin{figure*}[t!]
\includegraphics[width=0.8\linewidth]{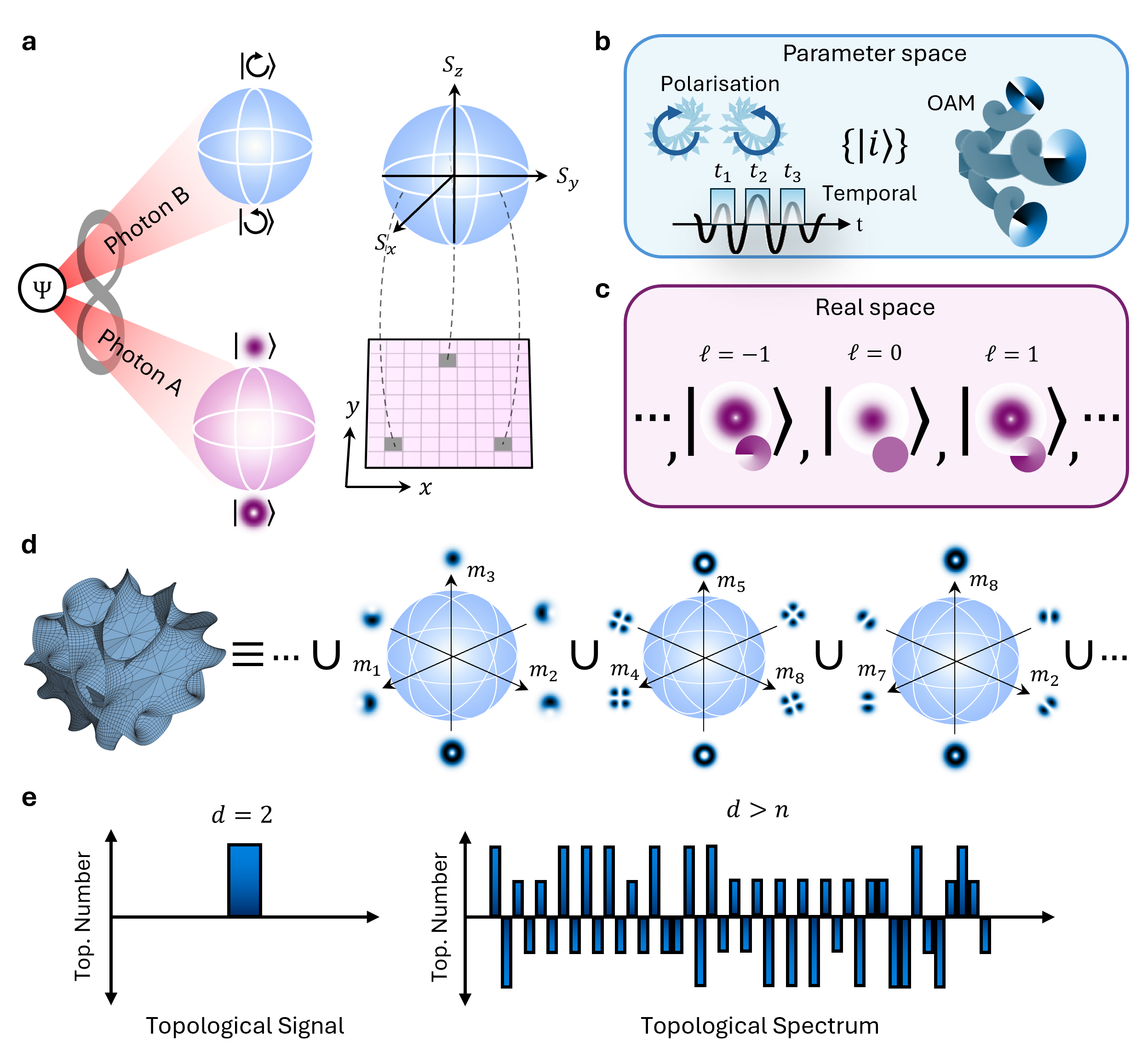}
\caption{\textbf{Topology in high dimensions.} \textbf{a}, A real space, $R^2$, to parameter space, $S^2$, mapping derived from photons entangled in space (photon A) and polarization (photon B). \textbf{b}, Beyond polarization, the parameter space may be labeled by any arbitrary DoF such as time and orbital angular momentum (OAM) with \textbf{c}, real-space labeled by a spatial DoF, an example with cylindrical symmetry being OAM. Using high dimensional degrees of freedom, such as OAM, then allows for the engineering of high dimensional quantum states. \textbf{d}, High-dimensional spaces (shown conceptually as a complex 3D manifold) are notoriously difficult to conceptualize and parametrize with measurable DoFs, thus mappings derived from high-dimensional states which seek to map high-dimensional manifolds to other high-dimensional manifolds is difficult to achieve. An adaptation is made which represents the high-dimensional manifold as a collection of 2-spheres, simplifying a single high-dimensional map into several simpler mappings embedded in the high-dimensional state. \textbf{e}, This generalizes the characterization of 2D quantum states by a singular topological signal to the characterization of high-dimensional quantum states by their topological spectrum made of several independent topological signals, each characterizing the number of wrappings around a different parameter space embedded within the high-dimensional space.}
\label{fig:ConceptFig}
\end{figure*}


\noindent \textbf{Concept and theory.} We use in our study quantum entangled states as a means to access high dimensional spaces, motivated by the interpretation of the quantum wave function as a mapping function that defines topology, and the ease with which such states can be accessed experimentally.  The treatment that follows can be extended to other mechanisms to access high dimensional spaces. To reveal the topology of high dimensional quantum states requires several novel steps, each an advance in their own right, which we illustrate graphically in Figure~\ref{fig:ConceptFig}.  First, we abandon the limiting notion of a spin textured field using polarisation as one of the degrees of freedom \cite{lei2021photonic}, replacing it with a degree of freedom that has unlimited dimensionality.  We select orbital angular momentum (OAM) as our example \cite{forbes2024orbital}, stressing that this can be replaced with another. Next, we use only OAM for both entangled photons, a shift away from the implicit assumption in optical topologies that at least two different degrees of freedom are needed \cite{yao2024multi}.  Now, with both photons expressed in a synthetic dimension with unlimited capacity, we are able to explore high dimensional topologies.  

Higher-dimensional spaces host intricate topological structures that ``live'' in high-dimensional manifolds. To systematically explore this topology, we provide a full theoretical treatment that establishes the Lie algebra of $\text{SU}(d)$ as a powerful framework for classifying such mappings. This allows us to extend the familiar Pauli projections in Stokes measurements for qubits to their higher-dimensional analogs, Gell-Mann projections, for $d=3$ and, more generally, projections onto the basis of $\text{su}(d)$. This allows us to reveal topological manifolds of $d^2 - 1$ dimensions for such systems. To probe this topology, we show that it is natural to construct maps from a reference two-sphere to embedded two-spheres within the higher-dimensional topological space. Each map is specified by a choice of three Lie algebra basis projections (see Supplementary Information), yielding $\binom{d^2-1}{3}$ possible mappings, each associated with a wrapping number counting how many times the reference two-sphere wraps the embedded one. This procedure sometimes yields a disk-to-disk map, which we then extend to a sphere-to-sphere map. The optimal choice of triples is guided by their association with an $\text{SU}(2)$ subgroup embedded within $\text{SU}(d)$, mirroring the classification of monopole charges in non-Abelian Yang-Mills theory, where different $\text{SU}(2)$ subgroups correspond to distinct topological charges. These invariants constitute a topological spectrum, encoding key features of the space’s topology. For example, in the qutrit case ($d=3$), our wave function is topological in 8-dimensions (our high-dimensional manifold) with non-trivial, independent topological invariants from 9 of the 56 possible maps that make up the spectrum, already unveiling a rich topological landscape.  
\begin{figure*}[ht!]
\includegraphics[width=0.8\linewidth]{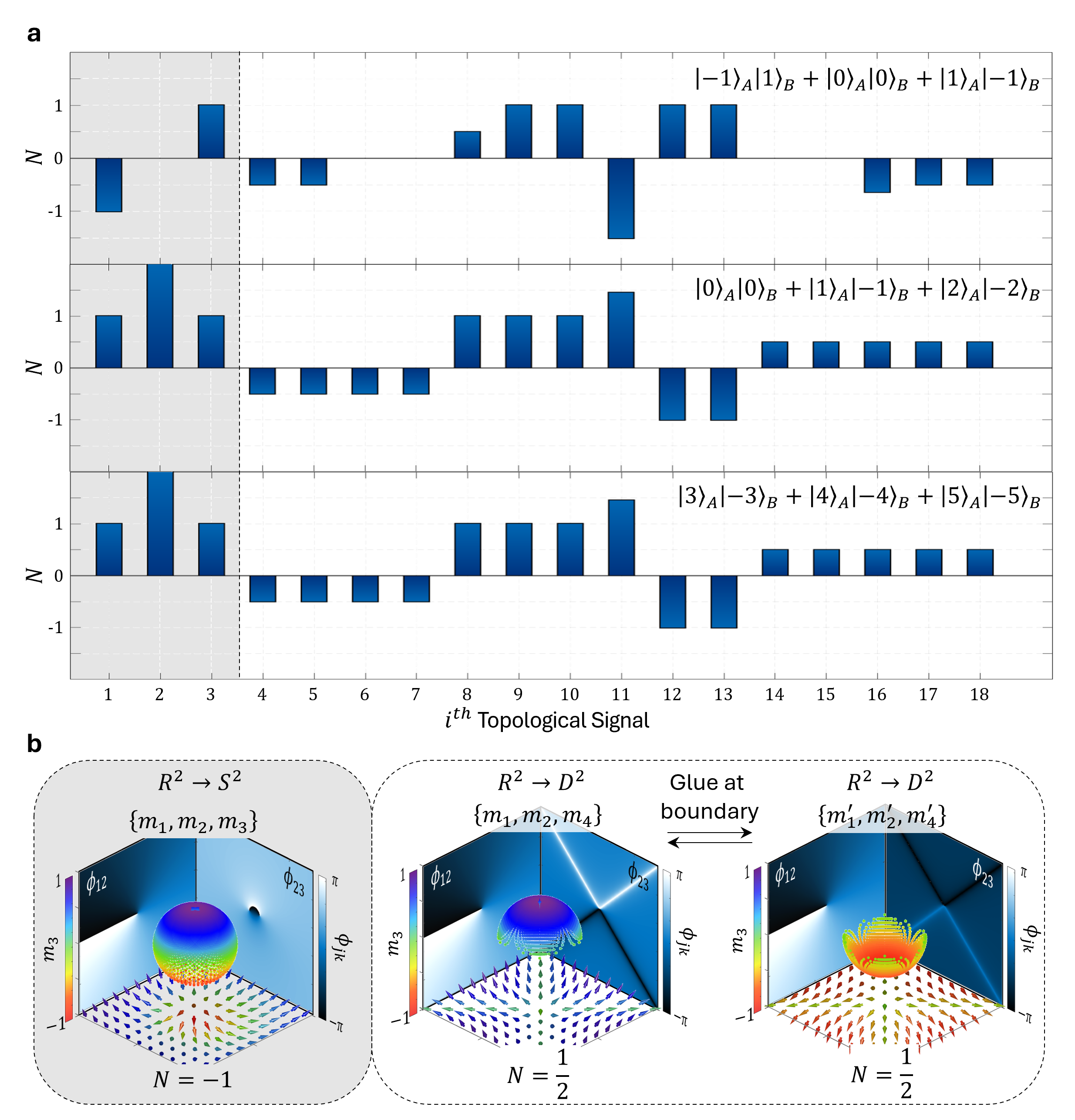}
\caption{\textbf{Topology of high dimensional states.} The topological information of high-dimensional states compiled in the form of a \textbf{a}, topological spectrum. As was the case for 2D states, topology can be used to characterize states of arbitrary dimension, where states with different topological spectra (such as the spectra shown in the first and second panels) fall into separate topological classes and unique states with identical topological spectra (such as the spectra shown in the second and third panels) are characterized under the same topological class. \textbf{b}, A typical $R^2 \to S^2$ mapping obtained from analyzing the triplet $(m_1,m_2,m_3)$ (shaded block) and new $R^2 \to D^2$ mapping obtained from analyzing the triplet $(m_1,m_2,m_4)$ both derived from high-dimensional entangled states, characterized by integer and half-integer wrapping numbers, respectively. By duplicating an $R^2 \to D^2$ mapping and then gluing them along their shared boundaries they can be extended to $R^2 \to S^2$ mappings characterized by integer wrapping numbers.}
\label{fig:Maptypes}
\end{figure*}

We unpack these ideas using our two-photon entangled state as an explicit example. The state, $\ket{\Psi}$, describing the entanglement between a pair of photons (A and B), forms a mapping between the space of one photon to the space of the other.  In photonic systems this is usually the spatial degree of freedom and polarisation, as shown in Figure~\ref{fig:ConceptFig}\textbf{a}. 
However, by fully embracing the mathematical origin of these maps it is possible to extend this idea to arbitrary degrees of freedom (see SI). To illustrate this, we consider entangled states written in the convenient form 
\begin{equation}
    |\psi (\vec{r})\rangle_{AB} = \sum_{i=0}^{d-1}  F_i(\vec{r}_A)|i\rangle_B,
    \label{Eq: SimpSpatialDoFEntState}
\end{equation}
where photon A lives in a spatial mode basis, $\ket{\psi(\vec{r}_A)}_A$, which admits the position change of basis, $\langle\vec{r}|\psi(\vec{r}_A)\rangle_A = F_i(\vec{r}_A)$ with $F_i(\vec{r}_A)$ being a complex scalar field, photon B is written in the computational basis and $d$ refers to the dimension of the state. 
As depicted in Figure~\ref{fig:ConceptFig}\textbf{b}, Photon B's physical DoF may be polarization, discrete temporal states formed by time bins or the orbital angular momentum (OAM) of the photon to name but a few. In this work we will keep both photons in the OAM DoF, including that of Photon A, as shown in Figure~\ref{fig:ConceptFig}\textbf{c}. The form of equation~\ref{Eq: SimpSpatialDoFEntState} intuitively reveals the correspondence between a chosen position for photon A and particular state for photon B. The map between the spaces of the photons, $\vec{m}(x,y)$, can then be constructed by computing the spatially-dependent, generalized Lie algebra (GLA) observables (the definition of the Lie algebra basis used is given in the supplementary information) of the state using $m_i(\vec{r}) = \text{Tr}(T_i\rho)$ where $i\in\{1,...,d^2-1\}$, 
and $\rho=\rho(\vec{r})=|\psi (\vec{r})\rangle_{AB}{}_{AB}\langle\psi(\vec{r})|$ is the spatially varying density matrix of the state. For a 2-dimensional state the observables in question are determined by the Pauli-spin matrices which are a basis for $su(2)$ and also the components of a state vector pointing along the surface of the Bloch sphere. What remains is to analyze the topology of the state. 
For each point $(x,y)$, the GLAs define a corresponding point in the higher-dimensional space. The domain of this mapping, $\mathbb{R}^2$, can be compactified into a two-sphere, $S^2$. Maps between two spheres exhibit non-trivial topology, characterized by a wrapping number, counting how many times one sphere wraps around the other. We exploit this by mapping our reference $S^2$ (the compactified $\mathbb{R}^2$) to $S^2$ subspaces embedded within the Lie algebra of $\text{SU}(d)$. As illustrated in Figure~\ref{fig:ConceptFig}\textbf{d}, this mapping is defined by three components of the spatially varying GLA vector, denoted $m_i$, $m_j$, and $m_k$. The result is a spectrum of topological invariants whose number increases with the quantum state dimension $d$ for $d > 2$, as shown in Figure~\ref{fig:ConceptFig}\textbf{e}.
 The number of possible topological invariants is bounded by the number of distinct two-spheres that can be defined, determined by the independent choices of triples $(m_i, m_j, m_k)$. With $d^2 - 1$ available $m_{i,j,k}$, this leads to a total of $\binom{d^2-1}{3}$ possible mappings. However, the subset of invariants that are non-trivial and independent depends on the specific wave function. For states of the form (\ref{Eq: SimpSpatialDoFEntState}), we show in the supplementary information that there are ${1\over 4}d(d-1)(d-2)(d+3)$ independent invariants that are non-trivial, i.e., with non-zero topoligical numbers. In the qutrit case ($d=3$), we explicitly construct all nine such invariants. For qubits, the single invariant determines whether the states $|0\rangle_B$ and $|1\rangle_B$ are entangled. For qutrits, certain nonzero invariants indicate that $|0\rangle_B$ is entangled with both $|1\rangle_B$ and $|2\rangle_B$, revealing sensitivity beyond simple bipartite entanglement.

Our work utilizes a spectrum of topological invariants rather than a single invariant, a distinction that arises naturally. In the Supplementary Information, we argue that the maps defined by the density matrix can be identified with the Higgs potential in the vacuum sector of a Yang-Mills-Higgs theory. Beyond perturbative excitations, this model hosts 't Hooft-Polyakov monopoles, whose magnetic charge is determined by a topological invariant. Our identification relies on the fact that the triplet $\vec{S}$ defines a Higgs field residing in the vacuum sector and that its wrapping number precisely matches the magnetic charge of the corresponding 't Hooft-Polyakov monopole. In Yang-Mills-Higgs theory, monopoles carry a spectrum of charges, each associated with distinct embeddings of $SU(2)$ subgroups within the larger $SU(d)$ gauge group. Under our identification, these different embeddings correspond exactly to the various choices available for our usual $S^2 \to S^2$ mappings.

We illustrate this concept theoretically in Figure~\ref{fig:Maptypes}, where the topological spectra for three qutrit states given in the spectrum panels and various mapping forms are shown. We predict the emergence of a spectrum comprised of trivial (zero) topologies, non-trivial (integer) Skyrmion topologies that map spheres to spheres (shaded region), and half-integer maps, with the spectra and example maps shown in Figure~\ref{fig:Maptypes}\textbf{a} and \textbf{b}, respectively. To illustrate the power of the idea, note that the two states shown in the top and middle panels are not orthogonal yet belong to distinct topological families by virtue of unique topological spectra. The two states shown in the middle and bottom panels are distinct (orthogonal) yet belong to the same topological family by virtue of their shared topological spectrum. This suggests that complex high dimensional quantum states can be classified by their topological spectrum. The prediction of the existence of half-integer maps is also worth investigating as it arises purely from the study of states with $d>2$. The half-integer maps arises because, for these cases, the usual one-point compactification of $\mathbb{R}^2$ fails and the map is from a disk ($D^2$) to another disk. These mappings may be adjusted to be spheres to spheres by a glueing process that is topologically allowed (see Supplementary Information), where the true wrapping number is twice the initial half-integer value, restoring integer topology. A representative example, for $d=3$, is the triple $(m_1(x,y),m_2(x,y),m_4(x,y))$, whose map is shown in Figure~\ref{fig:Maptypes}\textbf{b}. In the Supplementary Information, we show that this choice leads to the map ($x=r\cos(\phi)$, $y=r\sin(\phi)$, $l_{ab}=l_a-l_b$)
\begin{eqnarray}
&&\vec{S}(r,\phi)=N\Big(\cos(l_{01}\phi),-\sin(l_{01}\phi),r^{|l_2|-|l_1|}\cos(l_{02}\phi) \Big)
\cr
&&N^{-1}=\sqrt{1+r^{2|l_2|-2|l_1|}\cos\Big(l_{02}\phi\Big)^2}\,,
\end{eqnarray}
that it is a disk-to-disk map and the associated topological invariant. After extending to a sphere-to-sphere map, as shown through a gluing process illustrated in Figure~\ref{fig:Maptypes}\textbf{b}, the topological invariant becomes $\hat N_{124}=l_{01}{\rm sgn}(|l_2|-|l_1|)$. 
Henceforth we perform this glueing process and report only integer topological numbers.  

\vspace{0.5cm}

\noindent \textbf{OAM topology in two dimensions.} To verify this concept we generated a high-dimensional OAM-OAM entangled state from a spontaneous parametric downconversion (SPDC) source, with full experimental details provided in the Supplementary Information. The state produced can be described by $\ket{\Psi}_{AB} = \sum c_{\ell}\ket{\ell}_A\ket{-\ell}_B $ where $c_{\ell}$ are complex scalar coefficients and $\ket{\ell}$ is the photonic state with OAM $\ell \hbar$ per photon. From this initial high-dimensional state it is possible to select states of various dimension and topology. 

\begin{figure*}[th!]
\includegraphics[width=0.8\linewidth]{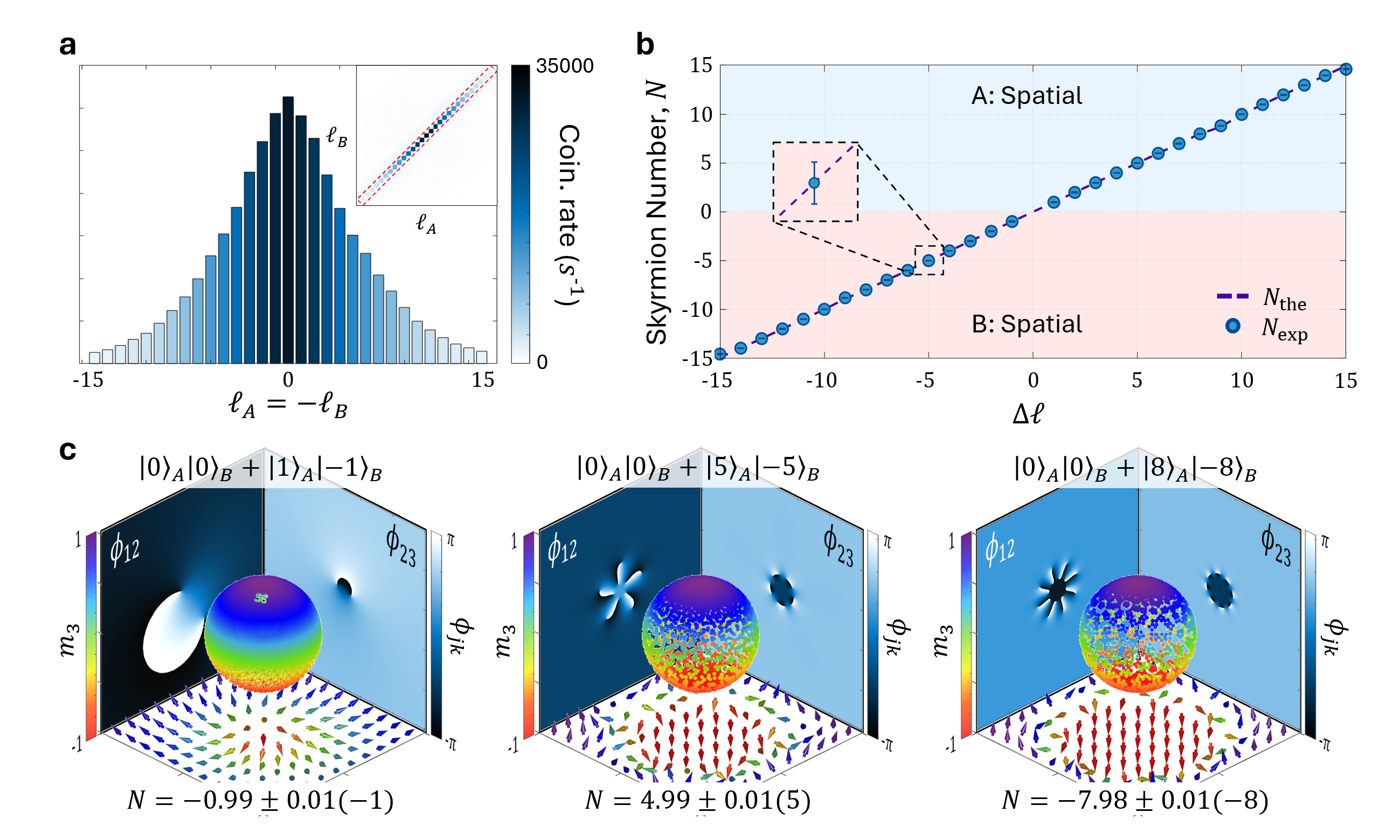}
\caption{\textbf{Experimental results for 2D entangled states.} 
\textbf{a}, Spiral bandwidth from the SPDC process shown in a false colourmap revealing the coincidence rate when performing a particular OAM measurement on photon A given a conjugate OAM measurement made on photon B. This taken from the anti-diagonal of the extended full spiral bandwidth measurement shown as an inset. \textbf{b}, Experimental Skyrmion number for 15 distinct 2D states of the form $\ket{\Psi} = \frac{1}{\sqrt{2}}\left( \ket{\ell_{0}}_A\ket{-\ell_{0}}_B + \ket{\ell_{1}}_A\ket{-\ell_{1}}_B\right)$ against the difference in OAM, $\Delta \ell = \ell_0-\ell_1$, for photon A (blue region) and, $\Delta \ell = \ell_1-\ell_0$, for photon B (red region). \textbf{c}, Full representation of the reconstructed spatially varying Pauli observables for states with Skyrmion numbers $N = -1,5,-8$.}
\label{fig:2DResults}
\end{figure*}

We first demonstrate that Skyrmionic topology can be found in photonic states with the same DoF, using entangled qubit states of the (unnormalised) form $\ket{\ell_0}_A \ket{-\ell_0}_B + \ket{\ell_1}_A \ket{-\ell_1}_B $, with the results shown in Figure~\ref{fig:2DResults}. The high dimensional $d \approx 17$ OAM-OAM space of the entangled pair, shown graphically as the OAM spectrum taken from the strong anti-diagonal (inset) in Figure~\ref{fig:2DResults}\textbf{a} with $91\%$ Fidelity, can be sub-divided into multiple qubit states.  Our experimental range admits topological numbers up to $|N| \approx 15 $, each related to the difference in OAM values in the state following $\Delta \ell = \ell_1 - \ell_0$ with $|\ell_1| \neq |\ell_0|$.  We construct all 30 such states, inferring their topological number through a quantum state tomography for qubits, finding excellent agreement between theory ($N_\text{the}$) and experiment ($N_\text{exp}$), as shown in Figure~\ref{fig:2DResults}\textbf{b}. The switch from positive to negative Skyrmion number is achieved by altering which photon represents real space (Spatial) and which the parameter space ($S^2$).  Some example topologies are shown in Figure~\ref{fig:2DResults}\textbf{c} for $N=-1,5$ and $-8$, together with the experimentally reconstructed profiles and topological numbers.  The left panel reveals a Bloch vector field with a Ne\'el-type texture, with the middle and right panels showing two higher-order vector textures.  The full sphere coverage confirms the mapping, i.e., that every possible OAM state for photon B is in correspondence with photon A's position. Therefore, $\vec{m}(x,y)$ forms a surjective map between real space, $R^2$, and the Bloch sphere, $S^2$, whose poles correspond to the basis states $\ket{0}$ and $\ket{1}$. The two adjacent panels show the Bloch phases $\phi_{jk}=\arctan(\frac{m_k}{m_j})$ representing the orientation of the Bloch vector about the axis $m_i$ where $i\neq j,k$. Below the these plots is the traditional viewpoint of a spin-textured field. All are in good agreement with theory, confirming the topology within the same DoF. In all these examples, the sphere we map to has co-ordinates $\{m_1,m_2,m_3\}$ defined by the OAM-OAM projections. The order of these three co-ordinates can be considered a matter of convention, with no new topological invariants found by permutation.  

\vspace{0.5cm}

\begin{figure*}[t]
\includegraphics[width=0.8\linewidth]{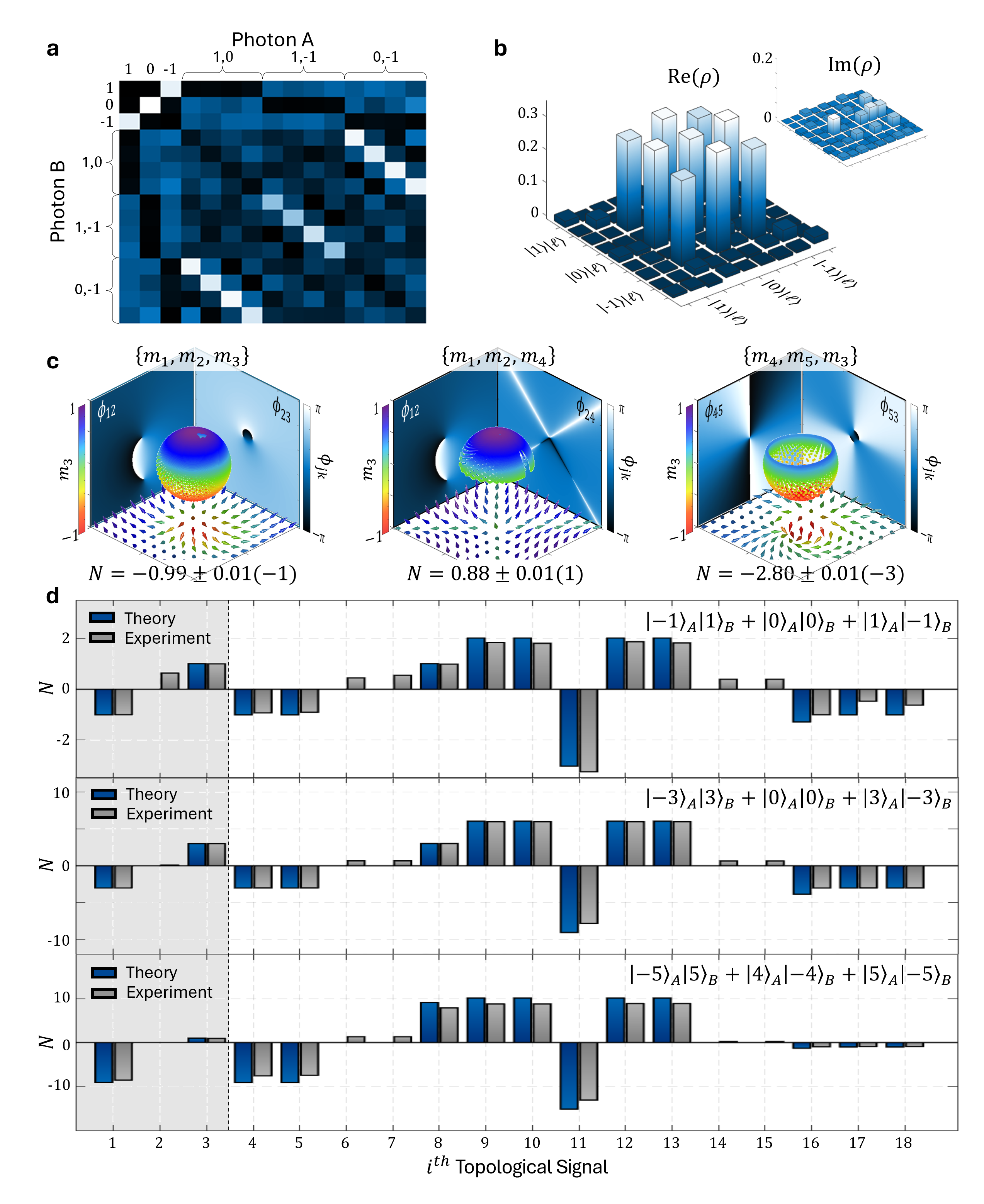}
\caption{\textbf{Experimental results for 3D entangled states.} \textbf{a}, A QST was performed on the selected entangled state, $\ket{\Psi} = \frac{1}{\sqrt{3}}\left(\ket{-1}_A\ket{1}_B + \ket{0}_A\ket{0}_B + \ket{1}_A\ket{-1}_B\right)$, by OAM projection measurements on photon A (columns) in the subspace $\{\ell_0=1,\ell_1=0,\ell_2=-1\}$ and photon B (rows) in the subspace $\{\ell_0=-1,\ell_1=0,\ell_2=1\}$, with coincidences collected for all the outcomes shown in colour from low (black) to high (white). \textbf{b}, Real and imaginary (shown as inset) parts of the density matrix of the reconstructed state, indicative of the desired state. \textbf{c}, Full representation of selected spatially varying Gell-Mann observable triplet combinations, namely $\{m_1,m_2,m_3\}, \{m_1,m_2,m_4\}$ and $\{m_4,m_5,m_1\}$. The plotted points on a sphere show coverage over different OAM bloch spheres and vectors (below spheres) in the plane showing a Nee\'l and higher-order type textures. The Gell-Mann phases $\phi_{ij}=\arctan(\frac{m_j}{m_i})$ represent the orientation of the bloch vector about the axis $m_k$ where $k\neq i,j$ and contain singularities where $m_k = \pm1$. \textbf{d}, Reduced topological spectra for states possessing the OAM values $\{-1,0,1\}, \{-3,0,3\}$ and $\{-5,4,5\}$ showing only the topological signals which may be non-trivial depending on the OAM values present within the state.}
\label{fig:3DMainResults}
\end{figure*}

\noindent \textbf{OAM topology in three dimensions.} We now use the SPDC possibilities of Figure~\ref{fig:2DResults}\textbf{a} to sub-divide the high dimensional space into qutrit spaces of the (unnormalised) form $ \ket{\ell_0}_A \ket{-\ell_0}_B + \ket{\ell_1}_A \ket{-\ell_1}_B +  \ket{\ell_2}_A \ket{-\ell_2}_B$, with comprehensive results shown in Figure~\ref{fig:3DMainResults}.  Since quantum state tomographies of high dimensional states are non-trivial \cite{agnew2011tomography}, we first show the raw projective data in Figure~\ref{fig:3DMainResults}\textbf{a} together with the reconstructed density matrix in Figure~\ref{fig:3DMainResults}\textbf{b} for the exemplary case of $\ell_{0,1,2}=-1,0,1$, returning a fidelity of $F = 0.95$ when compared to a expected maximally entangled state.  

We are now in a position to reveal the first high dimensional topological structures. We have an 8 dimensional topological space spanned by vectors $m_1$ through $m_8$.  We have shown theoretically that this high dimensional manifold gives rise to 9 distinct and independent maps with non-trivial topology, some maps from disks to disks and some from spheres to spheres. We confirm this in Figure~\ref{fig:3DMainResults}\textbf{c} with three illustrative examples.  Each shows Bloch vector fields derived from the single qutrit entangled state. The first map shown from sphere to sphere with the associated projection of $\{m_1,m_2,m_3\}$ giving rise to a topological number $-0.99$ in good agreement with the theoretical value of $-1$. This value is corroborated by the experimental reconstruction of the coverage, phase projections and vector field for visualisation. The second and third examples show disk to disk maps, a new topology not previously observed experimentally.  They represent  $R^2 \to D^2$ maps, from the (uncompactified) plane to a disk embedded within the higher dimensional space, shown for the triplet components $\{m_1, m_2, m_4\}$ and $\{m_4, m_5, m_3\}$ giving rise to Bloch vector fields which cover the upper and lower halves of the sphere, respectively. This is further confirmed for both maps by the fact that as $r\to\infty$, photon A's position states map to the equator of photon B's state space. We have shown the data without the glueing process, but report the topological numbers in glued form (integer), with the experimental values ($N \approx 0.88$ and $N \approx -2.8$) in good agreement with the predicted values of $N = 1$ and $N = -3$, respectively. It should be noted that these topologies reside in a single qutrit state rather than produced by multiple qubit states.

We perform the analysis of Figure~\ref{fig:3DMainResults}\textbf{c} for every map in the full topological space to reconstruct the topological spectra shown in Figure~\ref{fig:3DMainResults}\textbf{d}, the first observation of this in a physical system.  We select three example states for illustrative purposes, with the chosen state given in each panel, with more exhaustive examples in the Supplementary Information.  Each independent map is plotted on the horizontal axis together with its associated topological number.  The shaded regions depict the maps from spheres to spheres (left) and those of disks to disks (right) after the glueing process. The prediction from our theory is shown aside experimental data, in all instances showing very good agreement.  We can quantify this by similarity measures, returning $>90\%$ (when considering the cosine similarity score)for all states (see Table S2 in the Supplementary Information). 

We can observe that the form of the first two spectra are predicted to be identical except for a scaling factor in the topological invariant associated with the individual maps, with individual signals as high as N=10, as expected from the relative OAM values present in the states.  We confirm both the form and value of the spectra experimentally. The bottom panel has a similar OAM difference in state construction as the top panel, but has vastly different signature values as seen from the scale of the topological numbers.  These examples highlight the importance of a new perspective on topology when it manifests in high dimensions.

\vspace{0.5cm}

\begin{figure*}[th!]
\includegraphics[width=0.75\linewidth]{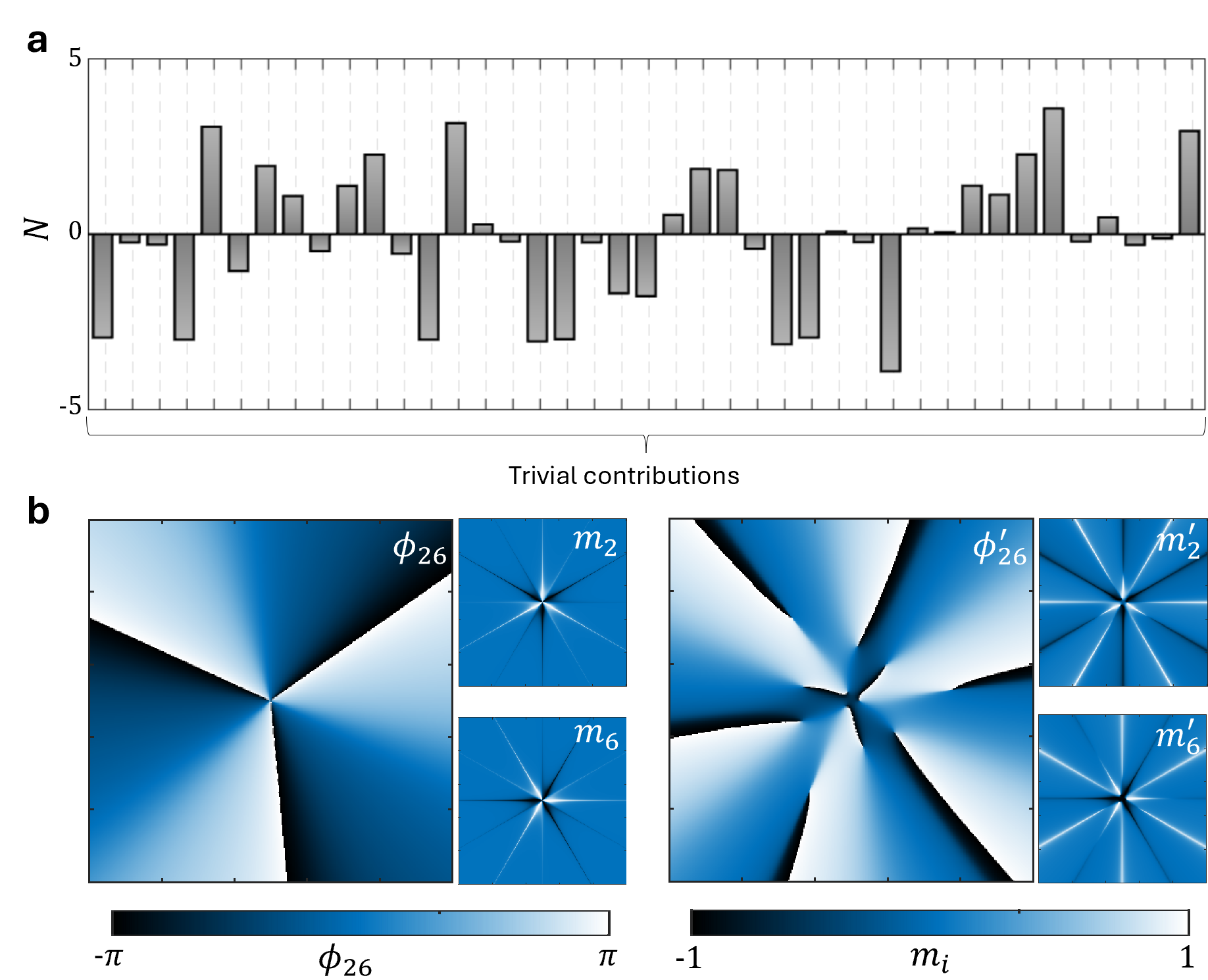}
\caption{\textbf{Emergence of topology.} \textbf{a}, Extended topological spectrum for the state $\ket{\Psi} = \frac{1}{\sqrt{3}}\left(\ket{-3}_A\ket{3}_B + \ket{0}_A\ket{0}_B + \ket{3}_A\ket{-3}_B\right)$ showing the emergence of topology due to the presence of entanglement within different subspaces of the full biphoton hilbert space. \textbf{b}, Gell-Mann phases and vector components for a chosen trivial topological signal derived from the state $\ket{\Psi}$ and the partially separable state $\ket{\Psi'}$.}
\label{fig:EmergeTopology}
\end{figure*}

\noindent \textbf{Emergence of topology.} We note in Figure \ref{fig:3DMainResults}\textbf{d} small signatures in some spaces that should have no topology, a feature that we explore thoroughly in the SI with an illustrative case given in Figure~\ref{fig:EmergeTopology}. 
 We find that when a perturbation alters the amplitudes and phases of the original states, the non-trivial part of the spectrum remains unchanged.  But when new entanglement subspaces are introduced into the larger Hilbert space, the previous trivial part of the spectrum can reveal emergent topology as a result of the emergent entanglement.  In Figure~\ref{fig:EmergeTopology}\textbf{a} we focus only on the previously trivial part of the spectrum for the state $\ket{\Psi} = \frac{1}{\sqrt{3}}\left(\ket{-3}_A\ket{3}_B + \ket{0}_A\ket{0}_B + \ket{3}_A\ket{-3}_B\right)$, where we have deliberately introduced new subspaces within the qutrit Hilbert space, e.g., $\ket{0}_A\ket{-3}_B,\ket{0}_A\ket{3}_B,\ket{-3}_A\ket{0}_B$, and $\ket{3}_A\ket{0}_B$.  In a real-world scenario this could be due to passing through a complex channel that introduces modal noise, or due to an eaversdropper introducing errors into the original state. This is a significant advantage over low dimensional systems: observing the entire spectrum opens the possibility to remain robust to perturbations yet also probe for them.  We can unravel the origin of these emergent topological features by studying the Gell-Mann triplets from which they are derived. In Figure~\ref{fig:EmergeTopology}\textbf{b} the Gell-Mann phases, $\phi_{26}$, and Gell-Mann components, $m_2$ and $m_6$, taken from the triplet $\{m_2, m_5, m_6\}$, are shown for an ideal state (left) as well as a deliberately perturbed state (right).  Here it is evident that the perturbation of the state results in the emergence of new singular topological features that are not originally present in the ideal state, while the original topological signatures remain intact (see SI).\\ 

\begin{figure*}[t]
\includegraphics[width=1\linewidth]{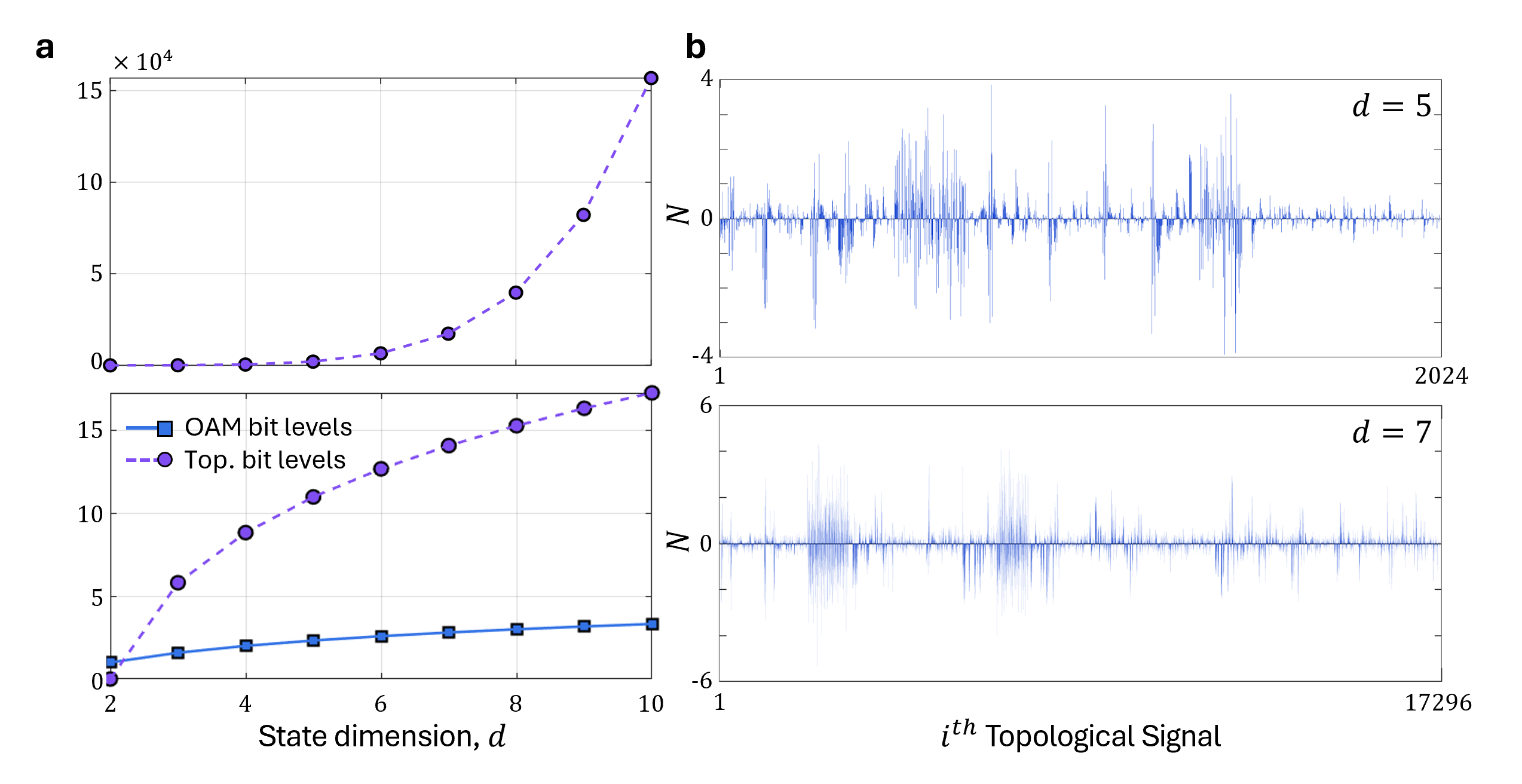}
\caption{\textbf{Topological information encoding potential for high-dimensional states} \textbf{a}, The number of topological signals (top panel) and OAM bit-encoding levels and topological bit-encoding levels (bottom panel) are plotted against the state dimension clearly showing the increased information encoding potential of topology Vs the number of independent OAM states of the system. b, Full experimental topological spectrum shown for a 5D (top panel) and 7D (top panel) state of the form $\ket{\Psi} = \frac{1}{\sqrt{d}} \sum_{\ell=-d}^{d}c_{\ell}\ket{\ell}_A\ket{-\ell}_B$. The full spectrum consists of all the topological signals calculated from all possible Gell-Mann triplet combinations (there are $2024$ and $17296$ different combinations for a 5D and 7D state, respectively) for a given state.}
\label{fig:HighDimCap}
\end{figure*}

\noindent \textbf{Towards high dimensions.} Having used $d=2$ to highlight the possibility of topology with one DoF, and $d=3$ to introduce the notion of a topological spectrum, we now show that indeed this approach can be pushed to high dimensions with a potentially enormous encoding alphabet.  The space of density matrices corresponds to the Lie algebra $\text{SU}(d)$, with basis elements $T_a$. Each topological invariant is associated with a choice of three basis elements, yielding ${d^2-1}\choose{3}$ possible mappings, which scales rapidly as $d$ increases. As shown in Figure~\ref{fig:HighDimCap}\textbf{a}, this growth is dramatic: for $d=5$, there are 2024 candidate invariants, rising to 17,296 for $d=7$. While the specific non-zero and independent invariants depend on the quantum state, this scaling highlights the rapidly increasing complexity of the topological spectrum.  Consequently, one can achieve an enormous information capacity by encoding information into the topological spectrum. To illustrate this, in the bottom panel of Figure~\ref{fig:HighDimCap}\textbf{a} we have plotted the information stored (using a simple binary bit encoding scheme) in the topological spectrum and that stored in the number of basis states, against the dimension of the state. These results are compelling and reveal a prospective application of these states to high capacity information encoding.  Simply put: while OAM is an exciting pathway to high information capacity per photon, it seems that the underlying topology of such systems makes for an even more appealing prospect.  In Figure~\ref{fig:HighDimCap}\textbf{b} we show experimentally measured spectra corresponding to $d=5$ and $d=7$, for topological manifolds of dimension 24 and 48, respectively. In the latter example, we have deciphered over 17000 signatures, a clear indication that this approach is highly scalable. These results represent the highest dimensions reported in any topological system.

\section{Discussion and conclusion}
In the quantum context, both dimensionality and topology have attractive benefits.  The harnessing of topology in quantum systems is mooted to overcome the fragility of such systems to external perturbation, facilitating the robust sharing and storing of quantum information \cite{zhou2020solids,psaroudaki2021skyrmion}.   Higher dimensions in quantum states too has benefits, such as noise resilience, higher bounds for eavesdropping and security against cloning \cite{nape2023quantum}.  The mutual benefits are very attractive and remain to be explored, particularly in cases where information must be distributed by entanglement as a resource, such as in teleportation, quantum networks, and quantum computers.

Intriguingly, OAM entangled photons to access high dimensional spaces has been extensively explored \cite{erhard2018twisted}, but the topological features seem to have been hidden from view.  Our new perspective on how to access topology in synthetic dimensions of photons has revealed that the underlying tapestry is very rich, with potentially many thousands of signatures to be found. It puts into question the information capacity that is thought possible using the spatial degree of freedom of light. We illustrate this in Figure~\ref{fig:HighDimCap}\textbf{a}, comparing an OAM alphabet to a topological alphabet for the same quantum state dimension, the latter offers an exciting future prospect.

 It is possible that this work has scope to expand even further, for instance, the existence of other invariants not yet discovered, which could distinguish states beyond those we have identified. Less speculative is the ability to shape the topological spectrum by controlling the SPDC source, e.g., by pump shaping \cite{bornman2021optimal}, for new exotic textures beyond what we have observed. 


In conclusion, our report represents the first realisation of skyrmionic states with just one degree of freedom (OAM in our example).  Our generalisation to high dimensions has shown the first high dimensional topological states of light, necessitating a new theoretical and experimental framework, while introducing the notion of a topological spectrum rather than a topological number. The implication is a rich topological signature for identifying high dimensional quantum states, potentially opening new paths to communication and sensing with topology.

\section*{Acknowledgements}
This work was supported by the South African National Research Foundation, National Institute for Theoretical and Computational Science and CSIR Rental Pool Programme.  The authors thank Cade Peters for useful discussions.

\section*{Author contributions}
The experiment was performed by P.O. and N.G., R.M.K. contributed the theory with inputs from B.Q.L and P.O.  All authors contributed to the writing of the manuscript and analysis of data. A.F. conceived of the idea and supervised the project.

\section*{Competing Interests}
The authors declare no competing interests.

\section*{Data availability}
The data are is available from the corresponding author on request.


\clearpage
\appendix

\setcounter{section}{0}
\setcounter{figure}{0}
\setcounter{table}{0}
\setcounter{equation}{0}
\setcounter{footnote}{0}
\renewcommand{\thesection}{S\arabic{section}}
\renewcommand{\thefigure}{S\arabic{figure}}
\renewcommand{\thetable}{S\arabic{table}}
\renewcommand{\theequation}{S\arabic{equation}}

\begin{widetext}

\section*{Supplementary: Defining the skyrmion topology for arbitrary biphoton states}
\subsection{Defining arbitrary high dimensional states}

\noindent We study states consisting of two photons (A and B) where the OAM degree of freedom (DoF) of photon A is entangled with an arbitrary d-dimensional DoF for photon B. Such a pure state is described by
\begin{equation}
    |\Psi\rangle = \sum_{i=0}^{d-1} c_{i} |\ell_i\rangle_A|i\rangle_B,
    \label{Eq: OAMDoFEntState}
\end{equation}
%
where $c_{i}$ are complex coefficients that characterize the state, $\ell_i$ denotes an OAM of $\ell_i\hbar$ per photon and the computational basis state $|i\rangle$ denotes a state for an arbitrary DoF of photon B. Whilst compact, the expression given in Eq.~(\ref{Eq: OAMDoFEntState}) is not convenient for the analysis that follows. A better description is achieved by performing a change of basis in the Hilbert space of photon A from OAM to position, $\vec{r} = (r,\phi,z)$. The change of basis is easily performed using the overlap $\langle \vec{r}|\ell_i\rangle = \text{LG}_{\ell_i}(r,\phi,z)$ where $\text{LG}_{\ell_i}(r,\phi,z)$ is the Laguerre Gaussian function with radial index $p=0$, described by 
%
\begin{align*}
\text{LG}_{\ell}(r, \phi, z) = \, C_{\ell} \frac{\omega_0}{\omega(z)} \tilde{f}_{\ell}(r,z) e^{i \ell\phi} e^{-ik\frac{r^2}{2R(z)}} e^{-i\psi_{\ell}(z)} e^{ikz},
\end{align*}
where $C_{\ell}$ is a normalization constant, $\omega_0$ is the beam waist at $z=0$, $ \omega(z) = \omega_0 \sqrt{1 + \left( \frac{z}{z_R} \right)^2}$ is the beam waist after propagation, $z_R = \frac{\pi \omega_0^2}{\lambda}$ is the Rayleigh range, $k=\frac{2\pi}{\lambda}$ is the usual wavenumber with $\lambda$ being the wavelength of the photon, $R(z)$ is the radius of curvature, $\psi_{\ell}(z)$ is the mode order dependent Gouy phase picked up by the state and $\tilde{f}_{\ell}(r,z) = \left( \frac{\sqrt{2}r}{w(z)} \right)^{|\ell|} \exp\left(-\frac{r^2}{w^2(z)}\right)$. After the change of basis, Eq.~(\ref{Eq: OAMDoFEntState}) becomes 
\begin{equation}
  {}_A \langle \vec{r}|\Psi\rangle = |\psi (\vec{r}_A)\rangle = \sum_{i=0}^{d-1} c_i \text{LG}_{\ell_i}(r_A,\phi_A,z_A)|i\rangle_B.
    \label{Eq: SpatialDoFEntState}
\end{equation}
We are interested in topological properties of these states. Since this topology is invariant to relative phase changes between constituent spatial modes, induced by propagation \cite{ornelas2024non}, we can focus on these states at $z=0$ without loss of generality. Furthermore, the topology of the state for $d=2$ is extracted from a state normalized so that $\text{Tr}\left({}_A \langle \vec{r}|\Psi\rangle_{AB}{}_{AB}\langle\Psi|\vec{r}\rangle_A\right) = 1$ where $\text{Tr}(\cdot)$ defines the trace operation. Consequently, we require that Eq.~(\ref{Eq: SpatialDoFEntState}) is normalized so that $\sum_{i=0}^{d-1} |c_{i}|^2|\text{LG}_{\ell_i}(r,\phi,z)|^2=1$ for all positions in space. To extract the topology for higher $d>2$ dimensional states, this normalization is not useful. Implementing these simplifications, Eq.~(\ref{Eq: SpatialDoFEntState}) becomes
\begin{equation}
    |\psi (\vec{r})\rangle = \sum_{i=0}^{d-1}  f_i(r_A) e^{i\ell_i\phi_A}|i\rangle_B,
    \label{Eq: SpatialDoFSimpEntState}
\end{equation}
\noindent where $f_{i}(r_A)=r_A^{|\ell_i|}e^{-r_A^2}$ with $r_A$ the radial coordinate of photon A, normalized to its beam waist. This description treats the biphoton state as a single qudit whose state is determined by the spatially varying amplitude and phase information of its correlated twin.\\ 

\noindent In the theoretical analysis that follows we study the $d=2$ and $d=3$ cases in detail. These results then provide the foundation for the analysis we perform in higher dimensions for which $d>3$. 

\section*{Supplementary: Topological landscape for higher-dimensional states}

In this Section topological invariants that can be extracted from the state Eq.~(\ref{Eq: SpatialDoFSimpEntState}) are discussed. We begin with a brief and informal discussion explaining how we think about higher dimensional topology and, importantly, why this involves multiple topological invariants as opposed to just a single invariant. We then turn to the problem of characterizing the topology of our quantum state. First, there are natural generalizations of the Skyrmion number invariant that is relevant for $d=2$. An important property of the $d=2$ case is that the Skyrmion number density evaluates to a total derivative with respect to the radial coordinate $r_A$. The first class of invariants we describe share the property that the topological number density is a total derivative. These invariants again describe the wrapping number of maps between spheres and the main novelty of these maps is that many independent spheres can be embedded into the space arising from the $d$-dimensional density matrix of photon B. We refer to these as the ``usual sphere to sphere mappings'' although they already represent a non-trivial generalization of the $d=2$ invariants. The second class of invariants we describe again have a topological number density that is a total derivative, but now the invariant takes values in the half-integers. These mapping are naturally thought of as mappings between two discs $D^2\to D^2$ and this naturally explains why these invariants are half integer valued. We conjecture that these are the only non-trivial topological invariants that can be defined and we give arguments that explain why this conjecture is plausible. We also provide a counting for the number of topological invariants as a function of $d$.


\subsection{Framing Topology: Intuition and Foundations}

A central idea in our study, is that the wave function does not give rise to a single topological invariant, but rather to a spectrum of such invariants. This may seem counterintuitive: much of our intuition is built from studying two-dimensional oriented surfaces, which are classified by a single integer, the genus \cite{bredon}. This classification is often illustrated with the by now famous equivalence of a coffee mug and a donut. It is for good reason that this serves as our everyday notion of topology—such structures are easy to visualize.

However, our focus lies in the topology of high-dimensional quantum states, where our low-dimensional intuition fails. In higher dimensions, not only do we lose the ability to visualize topology, but classifying spaces up to topological equivalence generally requires an infinite number of invariants. Higher-dimensional spaces possess richer and more intricate structures, and the number and nature of required invariants depend on the specific equivalence notion under consideration. For example\cite{lee2010introduction}:

\begin{itemize}
    \item \textbf{Homotopy equivalence}: Spaces are classified by a sequence of invariants describing how spheres of different dimensions can be mapped into the space.
    
    \item \textbf{Homology and Cohomology}: These provide algebraic invariants that characterize cycles and holes at various dimensions.
    
    \item \textbf{Characteristic Classes}: These include Pontryagin, Stiefel-Whitney, and Chern classes, which help distinguish smooth and vector bundle structures.
\end{itemize}

In three dimensions, Perelman's proof of the Poincar\'e and Geometrization conjectures \cite{tao2006perelman} provides a classification via Thurston's eight geometries\cite{thurston1997three}. In four dimensions, Donaldson \cite{donaldson1983application} and Seiberg-Witten invariants \cite{witten1994monopoles} reveal the existence of exotic smooth structures. For higher-dimensional spaces, no finite set of invariants generally suffices; instead, an infinite hierarchy of algebraic and geometric invariants is required for complete classification \cite{vonk2005mini}.

In our study, we focus on investigating how two spheres can be mapped into a higher-dimensional space. As suggested by the discussion above, we find that the topology of the high-dimensional wave function is specified by a spectrum of topological invariants, realized through the wrapping numbers of maps and their extensions, which we define in the sections that follow. To build intuition we describe the topology of the torus in the next section, in a way that illustrates our approach to the topology of high dimensional quantum states.

\subsubsection{Torus topology from low dimensional maps}

A torus exhibits non-trivial topology, which can be effectively characterized by analyzing the structure of maps between tori. This fundamental idea forms the basis of this section.

A torus can be mapped to the plane. To visualize the map, imagine the plane as a piece of paper, and curl it into a tube -- the end result is a cylinder. Take the resulting cylinder and bend it to join top edge to bottom edge. The result is a torus, as shown in Figure~\ref{fig:torusexample}(a). The torus has no boundary, so that from the above construction, its clear that we should identify the left edge of the piece of paper with the right edge, as well as bottom edge with top edge. Thus, we can think of the torus as a finite rectangular region with opposite edges identified. We can give the rectangular region coordinates $(x_1,x_2)$ limiting $0\le x_1\le L_1$, $0\le x_2\le L_2$ and identifying
\bea
(x_1,0)\sim (x_1,L_2)\qquad\qquad (0,x_2)\sim (L_1,x_2)
\label{identification}
\eea
It is also possible to allow the coordinates $x_1,x_2$ to range freely as long as we identify coordinates as
$x_1\sim x_1+L_1$ and $x_2\sim x_2+L_2$. One consequence of this identification is that $(x_1,x_2)$ is actually the same point as $(x_1+nL_1,x_2+mL_2)$ for any integers $n$, $m$. This second description will be useful for the maps we describe below. Notice that a horizontal or vertical line drawn across the rectangular region, between two edge points that are identified, are closed paths (circles) drawn on the torus, as illustrated in Figure~\ref{fig:torusexample}(b).

Now consider maps between two tori. Consider a map that takes some point $(x_1,x_2)$ into a new point $(f_1(x_1),f_2(x_2))$. The non-trivial constraint on this function is that it must respect the identification spelled out in Equation~(\ref{identification}). A nice map to consider is obtained by taking
\bea
f_1(x_1)=p x_1\qquad\qquad f_2(x_2)= q x_2\label{torusmap}
\eea
with $p$ and $q$ integers. To see that we respect (\ref{identification}) we must show that points identified before the mapping are still identified after the mapping. This is indeed the case for the mapping (\ref{torusmap}). Indeed, under the map we have 
\bea
(f_1(x_1),f_2(0))=(px_1,0)\qquad\qquad
(f_1(x_1),f_2(L_2))=(px_1,q L_2)
\eea
We know that $(px_1,0)$ and $(px_1,q L_2)$ are the same point on the torus, which demonstrates that our map is consistent with the first identification of (\ref{identification}). The second of (\ref{identification}) is verified with a completely parallel argument.

\begin{figure*}[t]
\includegraphics[width=0.7\linewidth]{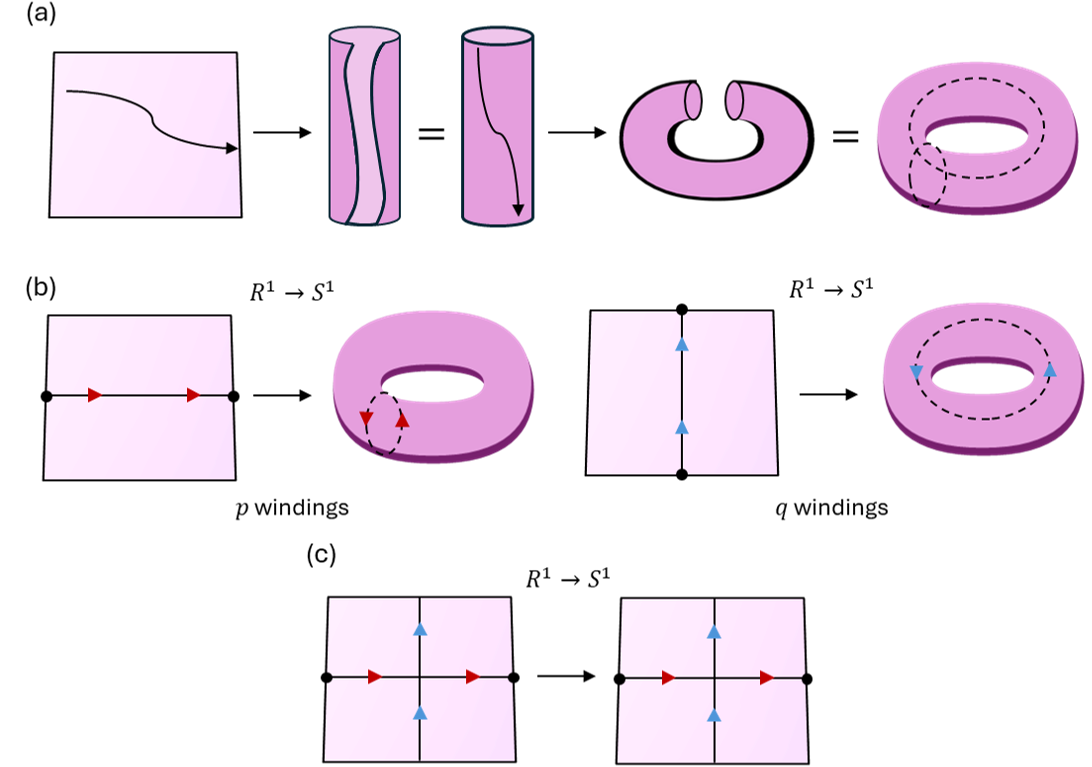}
\caption{\textbf{Topology of the torus.} (a) The torus can be constructed from a piece of paper, by first identifying the left and right edges to produce a tube and then identifying top and bottom edges. (b) A horizontal or vertical line on the plane, starting from one edge and moving to the corresponding point on the oppoosite edge is a closed path on the torus. (c) Maps  between tori can be visualized as maps between two rectangular regions.}
\label{fig:torusexample}
\end{figure*}

The complete set of maps between two tori can be classified by a pair of integers. These are $p$ and $q$ for the map we gave above. Two maps with the same $p$ and $q$ values can be smoothly deformed into each other. The first torus wraps the second torus a total of $pq$ times. However, this integer wrapping number does not characterize the topology. Indeed, the maps with $(p,q)=(1,6)$ and $(p,q)=(2,3)$ both have a wrapping number of 6, but these maps can not be deformed into each other. The wrapping number does not characterize the topology of the map -- there are inequivalent ways to do the wrapping and to distinguish these inequivalent wrappings one needs the pair $(p,q)$. In the same way, the topology of maps between two $d$ dimensional tori can not be specified by a single topological invariant, but rather they are specified by $d$ independent topological invariants.

As dimension increases it becomes increasingly difficult to build up a consistent visualization of a map wrapping a torus an integer number of times around a second torus. In the lowest dimensional case of a circle wrapping a circle we can think of a rubber band wrapped many times around the handle of a broom. Its harder to form the picture of a donut wrapping a donut, and mentally dangerous to consider the same exercise in higher dimensions. 

For the case of the torus, we can simplify the discussion by considering maps that only act on one coordinate, i.e. maps that take $x_1$ into $g(x_1)$ and we do nothing to $x_2$. This is best thought of as mapping a circle $S^1$ (with coordinate $x_1$) into another circle $S^1$ that is embedded on the torus. We know that $S^1\to S^1$ maps have an integer winding number. By considering suitable $S^1\to S^1$ maps we can recover the integers $p,q$ that characterize the topology of maps between tori. 

There are a number of important lessons from this example.
\begin{itemize}
    \item It is not possible to summarize the topology of the maps we described above by a single integer. There must be two integers and they encode important information about the topology of the torus - there is one integer for each non-contractible cycle of the torus. For a $d$-dimensional torus we would need $d$ integers. In this way the topology of the torus is captured by multiple topological invariants.
    \item These invariants are constructed by looking at lower dimensional maps -- maps between two circles, embedded in the torus. This is equally applicable for the $d$ dimensional torus for any $d$ -- in this case we could again study maps between circles. We did not dissect the space into smaller pieces, but rather, these lower dimensional maps are defined using circles that are embedded in the torus. 
    \item Although we considered one dimensional maps, they detect an important property of the $d$ dimensional torus -- that there are $d$ distinct closed paths that can be drawn, that are not the boundary of anything. These paths are one dimensional, so they are naturally detected by maps between circles, but they determine properties of the topology of the $d$ dimensional torus.
\end{itemize}

Our study of the topology of higher dimensional wave functions will also produce multiple topological invariants. The challenge in exploring the topology becomes the challenge of finding useful lower dimensional maps that exhibit a non-trivial topology. In our example these invariants will not be diagnosed using maps between two circles $S^1$ as in the torus example above. Rather, we will use maps between two two-spheres $S^2$. These $S^2$s are also embedded in the higher dimensional space in complete analogy to the above discussion.

As a final remark, proving that two maps are topologically inequivalent—meaning they cannot be smoothly deformed into one another—requires only a single topological invariant that evaluates to different values for the two maps. This is why computing just one invariant is often sufficient to distinguish between different topologies, a principle frequently applied in physics. However, a deeper question asks what set of topological invariants is necessary to ensure that two maps yielding the same values for all invariants are truly topologically equivalent. Addressing this question requires multiple invariants. Since each invariant encodes independent information about the state, the problem of encoding information into topology is fundamentally connected to understanding the full set of invariants needed for classification.

\subsubsection{Embedded spheres}

To reveal the topology encoded in our higher-dimensional quantum state, we study maps from an $S^2$ defined by the one-point compactification of the 
$(r,\phi)$ position-space plane of photon A, into an $S^2$ embedded in the higher-dimensional space associated with photon B. The presence of a topological spectrum of invariants arises from the multiple distinct ways in which the embedded $S^2$ can be defined. These invariants correspond to the independent wrapping numbers of the various maps.

The embedded sphere is described as the locus of points traced by a vector $\vec{S}$, which we conveniently normalize to have unit length. This normalization simplifies equations and ensures that, when visualizing the embedded manifold, it naturally appears as a sphere.

\begin{figure*}[t]
\includegraphics[width=0.35\linewidth]{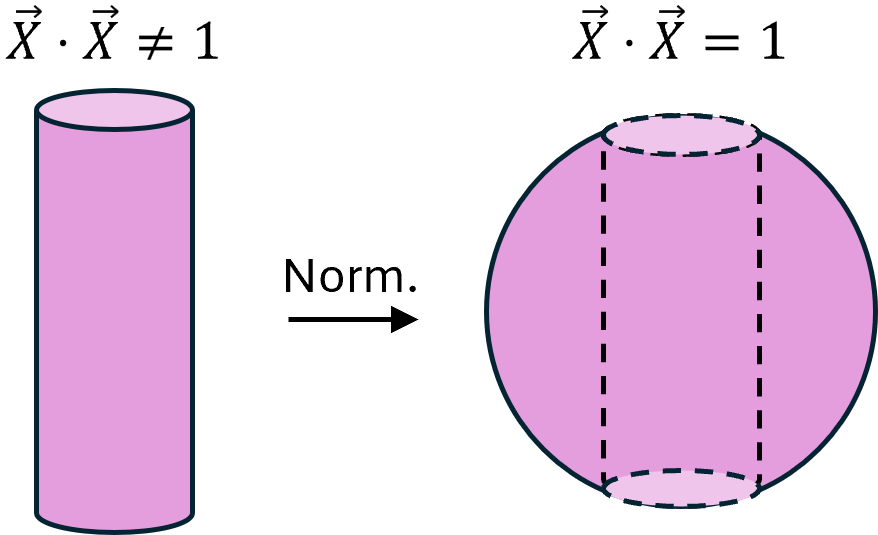}
\caption{This figure depicts a cylinder. In the left side figure, position vectors originating from the cylinder’s central point to points on its surface vary across different points. In the right panel, these vectors have been rescaled to unit length. All points on the cylinder will now also lie on the surface of the unit sphere. The topology remains unchanged -- the space is still a cylinder.}
\label{fig:spherenorm}
\end{figure*}

At this point, a potential source of confusion must be addressed. Normalizing $\vec{S}$ to unit length does not imply that the locus of its tip forms a sphere -- this step is merely a convenience and plays no essential role in the argument. To illustrate the key idea, consider Figure \ref{fig:spherenorm}, which depicts a cylinder. In the left panel, the lengths of position vectors originating from the cylinder’s central point vary across different points on its surface. In the right panel, these vectors have been rescaled to unit length, transforming the cylinder so it now wraps a sphere with regions around the poles not wrapped. Despite this transformation, the topology remains unchanged—the space is still a cylinder.

The embedded space we are mapping into has an intrinsic topology that is independent of how we normalize $\vec{S}$. Our analysis will rigorously confirm that it is indeed a sphere by explicitly constructing a map from a reference $S^2$ onto the embedded space and demonstrating that this map possesses an integer wrapping number. As always, it is the topological invariant that ultimately verifies the presence of a sphere.

\subsection{Topological invariants for Qubit ($d=2$) states}

\noindent This section reviews the computation of the topological invariant derived from entangled states with $d=2$. The discussion outlines the general structure of the argument so that it is immediately relevant for the derivation of topological invariants derived from higher dimensional $d>2$ scenarios. We explicitly outline what aspects of the $d=2$ treatment apply verbatim, for the higher dimensional problem.\\

\noindent Our starting point is the state given in Eq.~(\ref{Eq: SpatialDoFSimpEntState}) with $d=2$
\bea
|\psi(\vec{r})\rangle&=&f_0(r_A)e^{il_0\phi_A}|0\rangle_B+f_1(r_A)e^{il_1\phi_A}|1\rangle_B\,\,=\,\,\left[\begin{array}{c} f_0(r_A)e^{il_0\phi_A}\\ f_1(r_A)e^{il_1\phi_A}\end{array}\right]
\eea
To streamline the discussion, we drop the subscript $A$ on coordinates in what follows. This state defines three naive Bloch components, as follows
\bea
\tilde{m}_x&=&\langle\psi|\sigma_x|\psi\rangle\qquad \tilde{m}_y\,\,=\,\,\langle\psi|\sigma_y|\psi\rangle\qquad \tilde{m}_z\,\,=\,\,\langle\psi|\sigma_z|\psi\rangle
\eea
The adjective naive indicates that in general $\tilde{m}_x^2+\tilde{m}_y^2+\tilde{m}_z^2\ne 1$. The topological invariant is defined using Bloch vectors that obey 
\bea 
\vec{m}\cdot\vec{m}=1\label{gen1}
\eea
This condition is necessary since the target of the map is $S^2$. A standard computation gives the naive Bloch components \footnote{Introducing the naive components $\tilde{m}_a$ might seem unnecessary: we can normalize the wave function to automatically produce normalized Bloch components. This is special to $d=2$. For $d>2$ there is no normalization of the wave function that produces correctly normalized Bloch components.}, which can then be normalized to obtain
%
%
%
\bea
m_x&=&{2f_0(r)f_1(r)\cos\left(\phi (l_0-l_1)\right)\over f_0(r)^2+f_1(r)^2}\,\,=\,\,{2v\cos\left(\phi (l_0-l_1)\right)\over 1+v^2}\cr\cr 
m_y&=&-{2 f_0(r)f_1(r) \sin (\phi(l_0-l_1))\over f_0(r)^2+f_1(r)^2}\,\,=\,\,-{2v\sin\left(\phi (l_0-l_1)\right)\over 1+v^2}\cr\cr 
m_z&=&{f_0(r)^2-f_1(r)^2\over f_0(r)^2+f_1(r)^2}\,\,=\,\,{1-v^2\over 1+v^2}\label{NES}
\eea
where $v={f_1(r)\over f_0(r)}$. Differentiating the normalization condition Equation~(\ref{gen1}) with respect to $r$ gives
\bea
m_x{\partial m_x\over\partial r}+m_y{\partial m_y\over\partial r}+m_z{\partial m_z\over\partial r}&=&0\label{gen2}
\eea
Equations (\ref{gen1}) and (\ref{gen2}) hold for any normalized triple of Bloch components. No property of these components beyond normalization, is used. Since our higher dimensional invariants are defined using a unit 3-dimensional vector of parameters, these two relations also hold for the higher dimensional $d>2$ topological invariants. In addition to these general properties, there are three relations that make detailed use of the form of the Bloch components for this problem, given by
\bea
{\partial m_x\over\partial\phi}&=&\Delta l m_y\qquad\qquad {\partial m_y\over\partial\phi}\,\,=\,\,-\Delta l m_x\qquad\qquad {\partial m_z\over\partial\phi}\,\,=\,\,0\label{specrel}
\eea
where $\Delta l=l_0-l_1$. Equation (\ref{specrel}) will not hold for many of the higher dimensional topological invariants we define. We call the maps for which Equation (\ref{specrel}) holds the ``usual sphere to sphere'' mappings.

\subsubsection{Topological number density as a total derivative}

It is worth stressing a key feature of all topological invariants we define: the topological number density is a total derivative with respect to the radius $r$. This is a highly non-trivial property that guided our search for topological invariants. To exhibit this property, note that the topological number density is given by
\bea
\vec{m}\cdot \left({\partial\vec{m}\over\partial r}\times{\partial\vec{m}\over\partial\phi}\right)
=m_x {\partial m_y\over\partial r}{\partial m_z\over\partial\phi}-m_x {\partial m_z\over\partial r}{\partial m_y\over\partial\phi}+m_y {\partial m_z\over\partial r}{\partial m_x\over\partial\phi}-m_y {\partial m_x\over\partial r}{\partial m_z\over\partial\phi}+m_z {\partial m_x\over\partial r}{\partial m_y\over\partial\phi}-m_z {\partial m_y\over\partial r}{\partial m_x\over\partial\phi}\cr\label{topnumbdens}
\eea
Using Equation (\ref{specrel}) all $\phi$ derivatives can be evaluated to find
\bea
\vec{m}\cdot \left({\partial\vec{m}\over\partial r}\times{\partial\vec{m}\over\partial\phi}\right)=0-m_x {\partial m_z\over\partial r}(-\Delta l m_x)+m_y {\partial m_z\over\partial r}(\Delta l m_y)-0+m_z {\partial m_x\over\partial r}(-\Delta l m_x)-m_z {\partial m_y\over\partial r}(\Delta l m_y)\nonumber
\eea
Next, collect terms and add and subtract exactly the same term, shown in bold
\bea
\vec{m}\cdot \left({\partial\vec{m}\over\partial r}\times{\partial\vec{m}\over\partial\phi}\right)=\Delta l \left( m_x^2+m_y^2+{\bf m_z^2}\right){\partial m_z\over\partial r}
-\Delta l m_z\left(m_x{\partial m_x\over\partial r}+m_y{\partial m_y\over\partial r}{\bf+m_z{\partial m_z\over\partial r}}\right)\nonumber
\eea
At this point the universal relations, Equation (\ref{gen1}) and Equation (\ref{gen2}) play an important role. Use (\ref{gen1}) in the first three terms above and note that the last three terms vanish by (\ref{gen2}). Thus, we obtain
\bea
\vec{m}\cdot \left({\partial\vec{m}\over\partial r}\times{\partial\vec{m}\over\partial\phi}\right)=(l_0-l_1) {\partial m_z\over\partial r}= {\partial\over\partial r}\Big((l_0-l_1) m_z\Big)
\eea
This demonstrates, as advertised, that the topological number density is a total derivative with respect to $r$. An immediate consequence of this demonstration is that it is now simple to obtain the invariant: the above integrand is a total derivative with respect to $r$, so the $r$ integration is immediate. Further, the integrand is independent of $\phi$ so that the $\phi$ integration is equally simple. The resulting invariant is
\bea
N&=&{1\over 4\pi}\int_0^{2\pi} d\phi \int_0^\infty dr \,\,\vec{m}\cdot \left({\partial\vec{m}\over\partial r}\times{\partial\vec{m}\over\partial\phi}\right)\cr\cr\cr
&=&{l_0-l_1\over 2}\Big(m_z(r=\infty)-m_z(r=0)\Big)\cr\cr
&=&(l_0-l_1)\left({1\over 1+\left({f_0(0)\over f_1(0)}\right)^2}-{1\over 1+\left({f_0(\infty)\over f_1(\infty)}\right)^2}\right)\label{Eq: SkyNumSupp}
\eea
which is the usual answer \cite{gao2020paraxial}. To proceed we recall that
$f_{a}(r)=r^{|\ell_a|}e^{-r^2}$ which immediately implies $\left({f_0(r)\over f_1(r)}\right)^2={r^{2|\ell_0|}\over r^{2|\ell_1|}}$.
Consequently, as long as $|\ell_0|\ne|\ell_1|$\footnote{The condition $\ell_0\ne\ell_1$ is implicitly assumed since it is needed to ensure the wave function is entangled. The condition $|\ell_0|\ne|\ell_1|$ is stronger. The case that $|\ell_0|=|\ell_1|$ is a case of trivial topology and it gives $N=0$.} we know that $\left({f_0(0)\over f_1(0)}\right)^2$ and $\left({f_0(\infty)\over f_1(\infty)}\right)^2$ take the value $0$ or $\infty$ demonstrating that $N$ is indeed integer. We can write $N$ as
\bea
N=(l_0-l_1)\Big(\theta(|\ell_0|-|\ell_1|)-\theta(|\ell_1|-|\ell_0|)\Big)
\eea
with $\theta(x)$ the Heaviside function equal to 1 if its argument is positive and equal to zero otherwise. This formula is not sensitive to the particular convention we employ for $\theta(x)$ at $x=0$. 

\subsection{Topological invariants for Qutrit ($d=3$) states}

\noindent Consider the construction of topological invariants for $d=3$. The wave functions relevant for $d=3$ take the form
\bea
|\psi(\vec{r})\rangle&=&f_0(r_A)e^{il_0\phi_A}|0\rangle_B+f_1(r_A)e^{il_1\phi_A}|1\rangle_B+f_2(r_A)e^{il_2\phi_A}|2\rangle_B
\eea
Again, for clarity, drop the $A$, $B$ subscripts. The density matrix for qubit B is a $3\times 3$ hermitian matrix with unit trace. Consequently it is specified by 8 independent components. This motivates the definition of the 8 dimensional vector\footnote{The position $\vec{r}$ is 3 dimensional while the vector $\vec{m}$ is 8 dimensional.}
\bea
\vec{m}=\langle\psi(\vec{r})|\vec{\lambda}|\psi(\vec{r})\rangle
\eea
whose components are obtained by expanding the density matrix in the basis provided by the Gell-Mann matrices
\bea
\lambda_1=\left[\begin{array}{ccc}0&1&0\\1&0&0\\0&0&0\end{array}\right]\quad
&&\lambda_2=\left[\begin{array}{ccc}0&-i&0\\i&0&0\\0&0&0\end{array}\right]\qquad
\lambda_3=\left[\begin{array}{ccc}1&0&0\\0&-1&0\\0&0&0\end{array}\right]\qquad
\lambda_4=\left[\begin{array}{ccc}0&0&1\\0&0&0\\1&0&0\end{array}\right]\cr\cr\cr
\lambda_5=\left[\begin{array}{ccc}0&0&-i\\0&0&0\\i&0&0\end{array}\right]\quad
&&\lambda_6=\left[\begin{array}{ccc}0&0&0\\0&0&1\\0&1&0\end{array}\right]\qquad
\lambda_7=\left[\begin{array}{ccc}0&0&0\\0&0&-i\\0&i&0\end{array}\right]\qquad
\lambda_8={\frac {1}{\sqrt {3}}}\left[\begin{array}{ccc}1&0&0\\0&1&0\\0&0&-2\end{array}\right]
\eea
Explicit computation establishes that
\bea
m_1&=&2 f_0(r) f_1(r) \cos \Big(\phi  (l_0-l_1)\Big) \qquad\quad m_2\,\,=\,\,-2 f_0(r) f_1(r) \sin \Big(\phi  (l_0-l_1)\Big)\cr\cr
m_3&=&f_0(r)^2-f_1(r)^2 \qquad\qquad\qquad\qquad m_4\,\,=\,\,2f_0(r)f_2(r)\cos\Big(\phi  (l_0-l_2)\Big)\cr\cr
m_5&=&-2f_0(r)f_2(r)\sin\Big(\phi(l_0-l_2)\Big) \qquad m_6\,\,=\,\,2 f_1(r) f_2(r)\cos\Big(\phi(l_1-l_2)\Big)\cr\cr
m_7&=&-2f_1(r)f_2(r)\sin\Big(\phi(l_1-l_2)\Big) \qquad m_8\,\,=\,\,\frac{f_0(r)^2+f_1(r)^2-2 f_2(r)^2}{\sqrt{3}}\label{explicitms}
\eea

There is a topological invariant given by the wrapping number of a map from the $S^2$ associated to the compactified $(r,\phi)$ plane to an $S^2$ embedded in the above 8 dimensional space. Since there are many ways that an $S^2$ can be embedded in 8 dimensions, there are many candidate $S^2\to S^2$ maps. The data needed to define such a map, is a three dimensional unit vector which is a function of the $(r,\phi)$ coordinates. This unit vector defines points on the embedded $S^2$. Since each point on the compactified plane defines a unit vector, this maps each point on the plane to a point on the embedded $S^2$ i.e. it defines an $S^2\to S^2$ map. Note that this unit vector replaces the Bloch vector of the $d=2$ problem. Any three components $m_i$ of the density matrix in the Gell-Mann matrix basis defines a candidate vector\footnote{Unlike the $d=2$ case, there is no choice of normalization of the wave function that will ensure that the Gell-Mann vector is normalized, regardless of which triple of $m_i$ is chosen.}
\bea
\vec{\tilde S}&=&(\tilde{S}_1,\tilde{S}_2,\tilde{S}_3)\,\,=\,\,(m_i,m_j,m_k)
\eea
where $i\neq j \neq k$. This gives the analog of the naive Bloch components. The unit vector $\vec{S}$ is obtained by normalizing $\vec{\tilde{S}}$. After normalization, $\vec{S}(S_1,S_2,S_3)$ is a candidate map from a two sphere to a two sphere. It is worth stressing that the relations embodied in Equations~(\ref{gen1}) and (\ref{gen2}) clearly hold for these maps.

\subsubsection{Usual $S^2 \to S^2$ maps}\label{usualmaps}

\noindent The first $d=3$ topological invariants we define are associated with $S^2 \to S^2$ maps. These maps enjoy the special property given in Equation~(\ref{specrel}). It is for this reason that we refer to these as the ``usual $S^2\to S^2$ maps''. They are the natural generalization of the invariants defined for the qubit state. Note however, that this is already a non-trivial generalization of the qubit case, since the construction of the relevant $S^2\to S^2$ maps requires selecting a pair of integers from $\ell_0,\ell_1$ and $\ell_2$. There are three ways to select a pair and consequently we find the following three invariants

\begin{itemize}
\item[1.] The first topological invariant is given by selecting $(\tilde{S}_1,\tilde{S}_2,\tilde{S}_3)=(m_1,m_2,m_3)$. Repeating the analysis of the previous section, we obtain the topological invariant
\bea
N_{123}&=&(l_0-l_1)\Big(\theta(|l_0|-|l_1|)-\theta(|l_1|-|l_0|)\Big)
\label{frsts2s2}
\eea

\item[2.] The second topological invariant selects $(\tilde{S}_1,\tilde{S}_2,\tilde{S}_3)=(m_4,m_5,(m_3+\sqrt{3}m_8)/2)$. Notice that we have selected a linear combination of the components $m_i$ to define $\tilde{S}_3$. The topological invariant for this case is
\bea
N_{45*}&=&(l_0-l_2)\Big(\theta(|l_0|-|l_2|)-\theta(|l_2|-|l_0|)\Big)
\label{scnds2s2}
\eea
where $*$ indicates we used the linear combination $(m_3+\sqrt{3}m_8)/2$ for $\tilde{S}_3$.

\item[3.] The final topological invariant in this case is obtained by selecting $(\tilde{S}_1,\tilde{S}_2,\tilde{S}_3) = (m_6,m_7,(-m_3+\sqrt{3}m_8)/2)$. The topological invariant is
\bea
N_{67*}&=&(l_1-l_2)\Big(\theta(|l_1|-|l_2|)-\theta(|l_2|-|l_1|)\Big)\label{thrds2s2}
\eea
\end{itemize}

\subsubsection{15 more exotic invariants}\label{exoticmaps}

The $d=3$ topological invariant we define next can not be regarded as a generalization of the topological invariant defined from the qubit ($d=2$) case. The discussion is illustrated using the mapping obtained by choosing
\bea
\vec{\tilde S}=(\tilde{S}_1,\tilde{S}_2,\tilde{S}_3)=(m_1,m_2,m_4)
\eea
This example is selected because it effectively highlights the key feature of interest. The unit vector defined by this choice is
%
%
%
\bea
S_1(r,\phi)&=&{f_1(r) \cos \Big(\phi  (l_0-l_1)\Big)\over \sqrt{f_1(r)^2+f_2(r)^2\cos\Big(\phi  (l_0-l_2)\Big)^2}}\cr\cr 
S_2(r,\phi)&=&-{f_1(r) \sin \Big(\phi  (l_0-l_1)\Big)\over \sqrt{f_1(r)^2+f_2(r)^2\cos\Big(\phi  (l_0-l_2)\Big)^2}}\cr\cr
S_3(r,\phi)&=&{f_2(r)\cos\Big(\phi  (l_0-l_2)\Big)\over \sqrt{f_1(r)^2+f_2(r)^2\cos\Big(\phi  (l_0-l_2)\Big)^2}}\label{IntMap}
\eea
The denominators above arise from normalization. The complicated $\phi$ dependence of these expressions is a new feature and this dependence makes it clear that Equation~(\ref{specrel}) does not hold. This is a clear departure from the maps used in the previous subsection and those used in the qubit analysis. Despite this, a remarkable and highly non-trivial property of this unit vector, is that it defines a topological number density that is again a total derivative
\bea
\vec{S}\cdot \left({\partial\vec{S}\over\partial r}\times{\partial\vec{S}\over\partial\phi}\right)=(l_0-l_1){\partial S_3\over\partial r}={\partial\over\partial r}\Big((l_0-l_1)S_3\Big)
\label{remres}
\eea
Thus, Equation (\ref{remres}) continues to hold which is compelling evidence that Equation~(\ref{IntMap}) defines an interesting invariant. Equation~(\ref{remres}) is remarkable and deserves explanation. Our previous analysis fails because the last equation in (\ref{specrel}) no longer holds. The terms that were forced to zero because $S_3$ is independent of $\phi$ are
\bea
{\partial S_3\over\partial \phi}\left(S_1{\partial S_2\over\partial r}-S_2{\partial S_1\over\partial r}\right)
\eea
Our formulas for $S_1$ and $S_2$ in Equation~(\ref{IntMap}) have the following schematic form
\bea
S_1(r,\phi)&=&g(r,\phi)\cos (\phi\Delta l)\qquad S_2(r,\phi)\,\,=\,\,-g(r,\phi)\sin (\phi\Delta l)\label{defnicepairs}
\eea
so that derivatives with respect to $r$ act only on $g(r,\phi)$. Consequently
\bea
S_1{\partial S_2\over\partial r}=S_2{\partial S_1\over\partial r}\label{remarkablepairs}
\eea
so the term proportional to ${\partial S_3\over\partial \phi}$ vanishes because its coefficient vanishes. This explanation makes the important point that Equation~(\ref{IntMap}) relies only on properties of the pair $(S_1,S_2)$, i.e. of the pair $(m_1,m_2)$. Two more pairs, $(S_1,S_2)=(m_4,m_5)$ and $(S_1,S_2)=(m_6,m_7)$, share these properties. Consequently, we have 15 new invariants of the form
\bea
N_{12x}\qquad\qquad x&=&4,5,6,7,*\cr\cr
N_{45x}\qquad\qquad x&=&1,2,6,7,*\cr\cr
N_{67x}\qquad\qquad x&=&1,2,4,5,*
\eea
where $*$ is a linear combination of $m_3$ and $m_8$ that we have not spelled out yet. For these pairs the terms proportional to ${\partial S_3\over\partial \phi}$ vanish, and the wrapping number densities have the form
\bea
\vec{S}\cdot \left({\partial\vec{S}\over\partial r}\times{\partial\vec{S}\over\partial\phi}\right)=(l_0-l_1){\partial S_3\over\partial r}\qquad\qquad &&{\rm for}\,\,\,N_{12x}\cr\cr
\vec{S}\cdot \left({\partial\vec{S}\over\partial r}\times{\partial\vec{S}\over\partial\phi}\right)=(l_0-l_2){\partial S_3\over\partial r}\qquad\qquad &&{\rm for}\,\,\,N_{45x}\cr\cr
\vec{S}\cdot \left({\partial\vec{S}\over\partial r}\times{\partial\vec{S}\over\partial\phi}\right)=(l_1-l_2){\partial S_3\over\partial r}\qquad\qquad &&{\rm for}\,\,\,N_{67x}
\eea
Up to this point we have focused on the topological number density. The invariant is obtained by integrating this density over $r$ and $\phi$. Given the above result, the integral over $r$ is simple to evaluate. The result is 
\bea
N_{124}&=&{1\over 4\pi}\int\vec{S}\cdot\left({\partial\vec{S}\over\partial r}\times{\partial\vec{S}\over\partial\phi}\right)dr d\phi\,\,=\,\,{(l_0-l_1)\over 4\pi}\int_0^{2\pi}d\phi\int_0^{\infty}dr{\partial S_3\over\partial r}\cr\cr
&=&{(l_0-l_1)\over 4\pi}\int_0^{2\pi}d\phi\left[S_3(r=\infty,\phi)-S_3(r=0,\phi)\right]\cr\cr
&=&{(l_0-l_1)\over 4\pi}\int_0^{2\pi} d\phi\left[{\cos\Big((l_0-l_2)\phi\Big)\over\sqrt{{f_1(r=\infty)^2\over f_2(r=\infty)^2}+\cos\Big((l_0-l_2)\phi\Big)^2}}-{\cos\Big((l_0-l_2)\phi\Big)\over\sqrt{{f_1(r=0)^2\over f_2(r=0)^2}+\cos\Big((l_0-l_2)\phi\Big)^2}}\right]\cr\cr
&&\label{124ans}
\eea
This result has a number of interesting features, worth commenting on. The $\phi$ dependence of the integrand looks complicated, but this is an illusion. If $|l_1|=|l_2|$ we have ${f_1(r=0)^2\over f_2(r=0)^2}={f_1(r=\infty)^2\over f_2(r=\infty)^2}$ and the integrand vanishes. If $|l_1|\ne|l_2|$ then ${f_1(r=0)^2\over f_2(r=0)^2}$ and $f_1(r=\infty)^2\over f_2(r=\infty)^2$ takes the values of $0$ or $\infty$. For either value, the $\phi$ dependence of the integrand disappears. If (for example) $|l_1|>|l_2|$, then ${f_1(z,r=0)^2\over f_2(z,r=0)^2}$ is 0 and ${f_1(r=\infty)^2\over f_2(r=\infty)^2}$ is $\infty$, so that\footnote{Notice that ${\cos\Big((l_0-l_2)\phi\Big)\over\sqrt{0+\cos\Big((l_0-l_2)\phi\Big)^2}}$ evaluates to 1. Thus, we have $\sqrt{\cos\Big((l_0-l_2)\phi\Big)^2}=\cos\Big((l_0-l_2)\phi\Big)$ and not $\sqrt{\cos\Big((l_0-l_2)\phi\Big)^2}=|\cos\Big((l_0-l_2)\phi\Big)|$. The definition of the square root function $\sqrt{f(z)}=|f(z)|$ is not complex differentiable in the complex variable $z$ i.e. it does not define an analytic function. The $\sqrt{f(z)}=f(z)$ is an analytic function of $z$. This is a crucial element of our topological invariant's definition and is incorporated in the first step of Figure \ref{fig:Maptypes_Supp}(b), labeled 'Smoothen geometry.' }
\bea
N_{124}&=&{(l_0-l_1)\over 4\pi}\int_0^{2\pi} d\phi\left[{\cos\Big((l_0-l_2)\phi\Big)\over\sqrt{\infty+\cos\Big((l_0-l_2)\phi\Big)^2}}-{\cos\Big((l_0-l_2)\phi\Big)\over\sqrt{0+\cos\Big((l_0-l_2)\phi\Big)^2}}\right]\cr\cr
&=&{(l_0-l_1)\over 4\pi}\int_0^{2\pi} d\phi\left[0-1\right]\cr\cr
&=&-{l_0-l_1\over 2}\label{124ans_2}
\eea
This is valued in the half integers, which precludes an interpretation as the wrapping number of an $S^2\to S^2$ mapping. To find a sensible explanation of this result, return to the map (\ref{IntMap}). To define a valid $S^2\to S^2$ mapping, we need to compactify the $(r,\phi)$ plane so that we are mapping from an $S^2$ and we normalize $\vec{S}$ as $\vec{S}\cdot\vec{S}=1$ which is natural since we are mapping to an $S^2$. To compactify the plane to a sphere, a single point (the north pole) is added at infinity. The stereographic projection then maps every point on the extended plane to a corresponding point on the sphere. This is illustrated for the easily visualized compactification of a line to a circle (on the left hand side) and for the compactification of the plane $R^2$ to an $S^2$ (on the right hand side) of Fig.~\ref{fig:Maptypes_Supp} (a). The key point is that we must identify all points obtained as $r\to\infty$ with the north pole. This is where our mapping fails: it is not defined on the one point compactified plane. To make the point as clearly as possible, we start with an example that does define an $S^2\to S^2$ mapping and use it to show how the map (\ref{IntMap}) fails. The $S^2\to S^2$ mapping we consider is\footnote{This example is not contrived - it is essentially the mapping in (\ref{NES}). To see this set $v=r^2$, flip the sign of $n_3$ and set $k=l_0=l_1$.}
\bea
\vec{n}&=&\left(\frac{2\sqrt{r}\cos(k\phi )}{r+1}\,,\,-\frac{2\sqrt{r}\sin(k\phi)}{r+1}\,,\,\frac{r-1}{r+1}\right)
\eea
Focus on the surface traced out by the tip of vector $\vec{n}$, which is a unit sphere. At $r=0$ and any $\phi$, we have $\vec{n}=(0,0,-1)$ which is the south pole. At $r=4$ we have
\bea
\vec{n}&=&\left(\frac{4\sin(k\phi )}{5}\,,\,\frac{4\cos(k\phi)}{5}\,,\,\frac{3}{5}\right)
\eea
which traces $k$ times over the line of latitude defined by $z=\frac{3}{5}$, as $\phi$ runs from 0 to $2\pi$. Now look at $r=\infty$ and any $\phi$. In this case we have $\vec{n}=(0,0,1)$ so that all points with $r=\infty$ and any $\phi$ map to the north pole. That is why $\vec{n}$ defines a sphere - it is a non-trivial fact that $\vec{n}$ becomes independent of $\phi$ as $r\to\infty$ and it is this property that allows us to identify all points at $r=\infty$ with the north pole\footnote{Since $\vec{n}$ takes the same value for all of these distinct ($r=\infty$ and any $\phi)$ points, it sensible to identify them as a single point. This is called the one point compactification of the plane into the sphere.}. This is not the case for most fields $\vec{n}(r,z,\phi)$ that can be defined.

\begin{figure*}[t]
\includegraphics[width=1\linewidth]{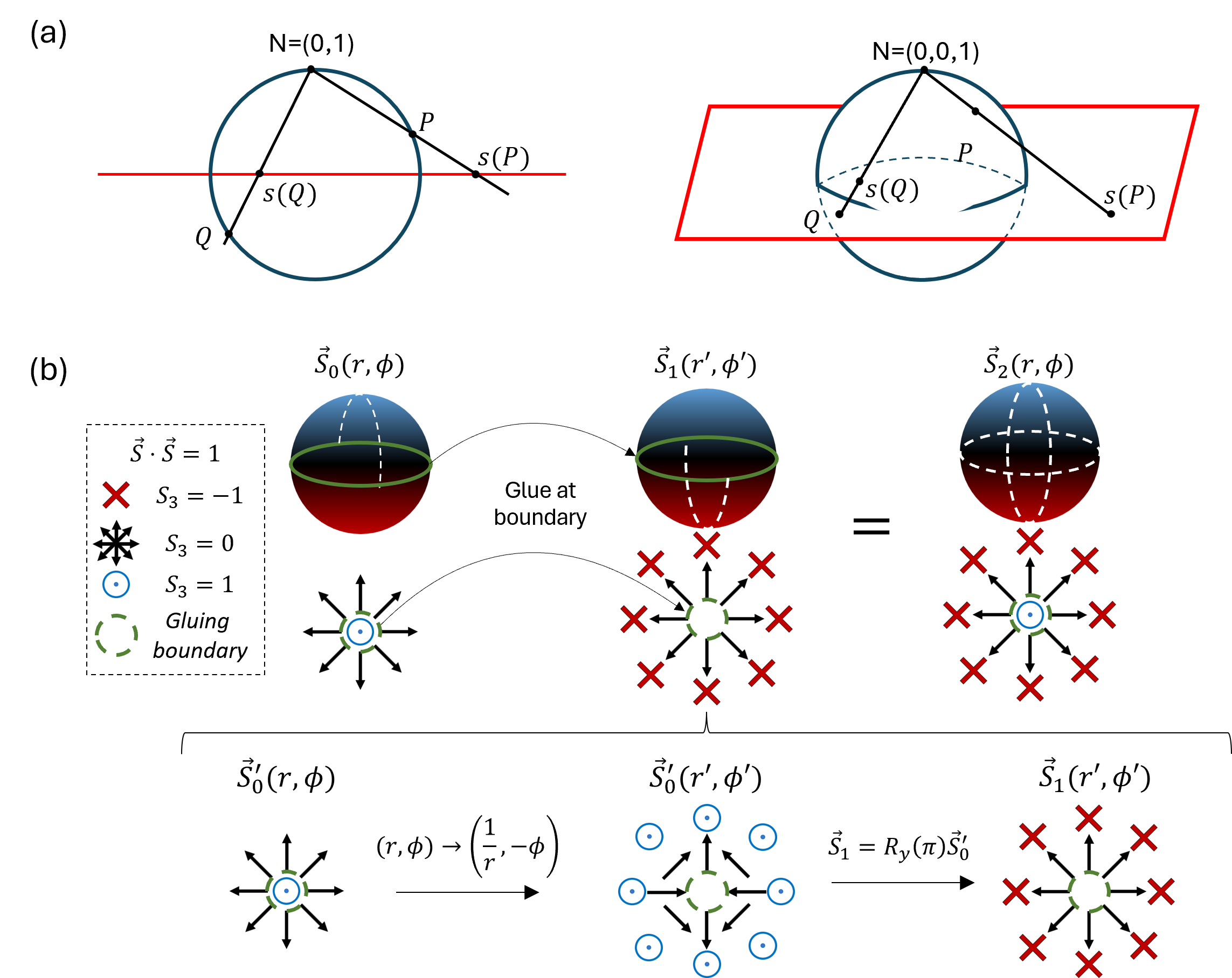}
\caption{\textbf{Construction of exotic mappings.} (a) The LHS figure illustrates the one point compactification of the line (red) to obtain the circle $S^1$ (blue). Points inside the interval $[-1,1]$ (look at $s(Q)$) map to the southern hemisphere (look at $Q$). Points more and more distant (look at $s(P)$) map ever closer (look at $P$) to the north pole. The points at $\pm\infty$ on the line both map to the north pole. Thus, a well defined mapping from the line to the circle maps $\pm\infty$ to the same point. The RHS figure illustrates the one point compactification of the plane $R^2$ to the two dimensional sphere $S^2$. The plane cuts the sphere at the equator. Points on the equator of the sphere are identified with points on the unit circle of the plane. Points inside the unit circle (look at $s(Q)$) map into the southern hemisphere of the $S^2$ (look at $Q$). Points outside the unit sphere (look at $s(P)$) map to points in the northern hemisphere of the $S^2$ (look at $P$). Points with $r=\infty$ and any $\phi$ map to the north pole of the sphere. 
(b) Construction of, $\vec{S}_2(r,\phi)$, which maps $S^2\to S^2$, involves gluing the initial map, $\vec{S}_0(r,\phi)$ which maps $D^2\to D^2$, to its own duplicate denoted by, $\vec{S}_0^{'}(r,\phi)$, along their boundaries. To demonstrate how this results in the desired final mapping, we perform a pair of smooth transformations on  $\vec{S}_0^{'}(r,\phi)$ to transform it to
$\vec{S}_1(r',\phi')$, which covers the missing lower hemisphere that $\vec{S}_0(r,phi)$ does not map to, without changing its topological number. The pair of transformations involve a coordinate transformation $(r,\phi) \to (r',\phi') = (\frac{1}{r}, -\phi)$ followed by a rotation of the vector field with respect to the y-axis.}
\label{fig:Maptypes_Supp}
\end{figure*}

Return to our mapping (\ref{IntMap}). If (as above) we assume that $|l_1|>|l_2|$, we have ${f_1(0)^2\over f_2(0)^2}=0$. In this case\footnote{It is also perfectly consistent to take $S_3=-1$, by choosing the second sheet of the square root function. The net effect is a change in sign of the topological invariant, so that this choice in sign is equivalent to a choice of what we mean by positive or negative wrapping. The choice here must be the same as the choice in the integrand of Eq.~\ref{124ans_2} for consistency.}
\bea
S_1(0,\phi)={{f_1(0)\over f_2(0)}\cos\Big(\phi(l_0-l_1)\Big)\over \sqrt{{f_1(0)^2\over f_2(0)^2}+\cos\Big(\phi(l_0-l_2)\Big)^2}}=0\qquad 
S_2(0,\phi)=-{{f_1(0)\over f_2(0)}\sin \Big(\phi(l_0-l_1)\Big)\over \sqrt{{f_1(0)^2\over f_2(0)^2}+\cos\Big(\phi(l_0-l_2)\Big)^2}}=0\cr\cr
S_3(0,\phi)={\cos\Big(\phi  (l_0-l_2)\Big)\over \sqrt{{f_1(0)^2\over f_2(0)^2}+\cos\Big(\phi  (l_0-l_2)\Big)^2}}=1
\eea
so that $r=0$ maps to the north pole. For $|l_1|>|l_2|$, we also have ${f_1(\infty)^2\over f_2(\infty)^2}=\infty$, so that
\bea
S_1(\infty,\phi)&=&{{f_1(\infty)\over f_2(\infty)}  \cos \Big(\phi  (l_0-l_1)\Big)\over \sqrt{{f_1(\infty)^2\over f_2(\infty)^2}+\cos\Big(\phi  (l_0-l_2)\Big)^2}}\,\,=\,\,\cos \Big(\phi  (l_0-l_1)\Big)\cr\cr 
S_2(\infty,\phi)&=&-{{f_1(\infty)\over f_2(\infty)}  \sin \Big(\phi  (l_0-l_1)\Big)\over \sqrt{{f_1(\infty)^2\over f_2(\infty)^2}+\cos\Big(\phi  (l_0-l_2)\Big)^2}}\,\,=\,\,-\sin \Big(\phi  (l_0-l_1)\Big)\cr\cr
S_3(\infty,\phi)&=&{\cos\Big(\phi  (l_0-l_2)\Big)\over \sqrt{{f_1(\infty)^2\over f_2(\infty)^2}+\cos\Big(\phi  (l_0-l_2)\Big)^2}}=0
\eea
The failure is evident: $r=\infty$ does not map to a single point. Rather, the unit vector revolves $l_0-l_1$ times around the equator (at $z=0$) as $\phi$ is varied from $0$ to $2\pi$. It is no longer possible to perform a one point compactification of the plane to obtain a sphere, so that in this case the plane is (topologically speaking) a disk $D^2$ and not a two sphere $S^2$. The tip of the unit vector is not located at a point as $r\to\infty$: it runs multiple times along a circle, which itself is the boundary of a disk. As $r,\phi$ is varied, the tip of the unit vector does not trace out a sphere, but rather it traces out a disk, with a boundary that is the unit circle. This boundary of the disk is the image of the points at $r=\infty$. We learn that the mapping (\ref{IntMap}) is not a sphere to sphere map ($S^2\to S^2$), but rather, it is a mapping from a disk to a disk ($D^2\to D^2$).

This presents a second challenge: the wrapping number is a topological invariant. Thus, it is invariant under any smooth deformation and it cannot change without tearing, unwrapping, and then repairing the map. However, a map between disks can be continuously deformed (unwrapped) without tearing. Additionally, our invariant takes half-integer values, meaning it is not simply counting something. To construct a true topological invariant, we can glue together two copies of the disk maps (\ref{IntMap}), forming a map from a sphere $S^2$ (obtained by gluing two planes) to another sphere $S^2$ (formed by gluing two disks each defined by the tip of $\vec{S}$, along their boundaries). This process is depicted in Fig.~\ref{fig:Maptypes_Supp}(b). This extension allows us to extend our original $D^2\to D^2$ mapping to an $S^2\to S^2$ mapping. The topological invariant of the original disk map is then half of the wrapping number of the extended sphere map. This follows because the wrapping number of the extended map consists of two identical contributions -- one from each disk -- since both disk maps are identical. For the example we studied above, the wrapping number (denoted with a hat) of the extended $S^2\to S^2$ map is double the invariant we computed for the $D^2\to D^2$ map (denoted without a hat)
\bea
\hat{N}_{124}=2N_{124}=-(l_0-l_1)
\eea
which is an integer. In general, the wrapping number of the completed map is defined by the formula
\bea
\hat{N}_{abc}&=&{1\over 2\pi}\int\vec{S}\cdot\left({\partial\vec{S}\over\partial r}\times{\partial\vec{S}\over\partial\phi}\right)dr d\phi
\eea 
for the unit vector
\bea
\vec{S}=\left({m_a\over\sqrt{m_a^2+m_b^2+m_c^2}},{m_b\over\sqrt{m_a^2+m_b^2+m_c^2}},{m_c\over\sqrt{m_a^2+m_b^2+m_c^2}}\right)\nonumber
\eea
defining a $D^2\to D^2$ mapping. This is a new construction of a topological invariant from the $d=3$ wave function and it is an integer.

We can list 15 new invariants, given by (${\rm sgn}(a-b)\equiv\theta(a-b)-\theta(b-a)$)
\bea
\begin{array}{ccc}
\hat N_{124}=(l_0-l_1){\rm sgn}(|l_2|-|l_1|)&{\hskip 1.0cm} &
\hat N_{125}=-(l_0-l_1){\rm sgn}(|l_2|-|l_1|)\\
\\
\hat N_{126}=(l_0-l_1){\rm sgn}(|l_2|-|l_0|)&{\hskip 1.0cm} &
\hat N_{127}=-(l_0-l_1){\rm sgn}(|l_2|-|l_0|)\\
\\
\hat N_{451}=-(l_0-l_2){\rm sgn}(|l_1|-|l_2|)&{\hskip 1.0cm} &
\hat N_{452}=(l_0-l_2){\rm sgn}(|l_1|-|l_2|)\\
\\
\hat N_{456}=-(l_0-l_2){\rm sgn}(|l_1|-|l_0|)&{\hskip 1.0cm} &
\hat N_{457}=(l_0-l_2){\rm sgn}(|l_1|-|l_0|)\\
\\
\hat N_{671}=(l_1-l_2){\rm sgn}(|l_0|-|l_2|)&{\hskip 1.0cm} &
\hat N_{672}=-(l_1-l_2){\rm sgn}(|l_0|-|l_2|)\\
\\
\hat N_{674}=(l_1-l_2){\rm sgn}(|l_0|-|l_1|)&{\hskip 1.0cm} &
\hat N_{675}=-(l_1-l_2){\rm sgn}(|l_0|-|l_1|)
\end{array}
\eea
\bea
\hat N_{12*}&=&(l_0-l_1)2{\rm sgn}(|l_0|-|l_2))(\theta (|l_0|+|l_1|-2|l_2|)+\theta(|l_0|-|l_1|)+\theta(|l_1|-|l_0|)+\theta(2|l_2|-|l_0|-|l_1|)\cr\cr
&&-\theta(|l_0|-|l_1|)\theta (2|l_2|-|l_0|-|l_1|)-\theta (|l_1|-|l_0|) \theta(|l_0|+|l_1|-2 |l_2|))\cr\cr
&&\quad{\rm where}\quad S_*={1\over 2}(m_3+\sqrt{3}m_8)\cr\cr
\hat N_{453}&=&(l_0-l_2){\rm sgn}(|l_0|-|l_1|)\Big(\theta(|l_0|-|l_2|)+\theta(2|l_1|-|l_0|-|l_2|)-\theta(|l_0|-|l_2|)\theta(2|l_1|-|l_0|-|l_2|)
\cr\cr
&&+\theta(|l_2|-|l_0|)+\theta(|l_0|+|l_2|-2|l_1|)-\theta(|l_2|-|l_0|)\theta(|l_0|+|l_2|-2|l_1|)\Big)\cr\cr
\hat N_{673}&=&(l_2-l_1){\rm sgn}(|l_1|-|l_0|)\Big(\theta(2|l_0|-|l_1|-|l_2|)+\theta(|l_1|-|l_2|)-\theta(2|l_0|-|l_1|-|l_2|)\theta(|l_1|-|l_2|)\cr\cr
&&+\theta(|l_1|+|l_2|-2|l_0|)+\theta(|l_2|-|l_1|)-\theta(|l_1|+|l_2|-2|l_0|)\theta(|l_2|-|l_1|)\Big)
\eea
Here we have listed the wrapping number of the extended $S^2\to S^2$ map. Notice that these invariants depend on all three numbers $|l_0|$, $|l_1|$ and $|l_2|$, clearly demonstrating that the new invariants probe more than just bipartite entanglement. Indeed, $\hat{N}_{124}$ vanishes if either $l_0=l_1$ (when $|0\rangle$ and $|1\rangle$ are not entangled) or $l_1=l_2$ (when $|1\rangle$ and $|2\rangle$ are not entangled). Thus, $\hat{N}_{124}$ is only non-vanishing if $|1\rangle$ is entangled with both $|0\rangle$ and $|2\rangle$. This clearly is a different measure of entanglement as compared to the usual $S^2\to S^2$ mapping which diagnoses entanglement between a pair of states.

\subsubsection{The $d=3$ Qutrit invariants} 

The results of the previous two sections collectively provide a set of 18 topological invariants. In this section, we investigate how many of these invariants are truly independent. To address this question, we evaluate the 18 invariants for various choices of $l_0$, $l_1$ and $l_2$. Analyzing the resulting lists for linear dependence then determines the number of independent invariants.

As a first step, by inspection we immediately see that
\bea
\hat N_{124}&=&\hat N_{125}\qquad 
\hat N_{126}\,\,=\,\,\hat N_{127}\cr\cr
\hat N_{451}&=&\hat N_{452}\qquad
\hat N_{456}\,\,=\,\,\hat N_{457}\cr\cr
\hat N_{671}&=&\hat N_{672}\qquad
\hat N_{674}\,\,=\,\,\hat N_{675}
\label{simpledependences}
\eea
reducing the maximum number of possible independent invariants to 12.

Lists of invariants were generated by allowing each of $\{l_0,l_1,l_2\}$ to run from -30 to 30. Within this set of $(61)^3$ vectors, a total of 9 vectors are linearly independent. Since a maximum of three invariants are related to the invariants that generalize the qubit ($d=2$) invariant, we have genuinely new invariants in $d=3$. The linearly dependent combinations found in this way are
\bea
N_{123}-\hat N_{456}+\hat N_{674}&=&0\cr\cr
\hat N_{671}-N_{45*}-\hat N_{126}&=&0\qquad{\rm where}\qquad \tilde{S}_*={1\over 3}(\sqrt{3}m_8+m_3)\cr\cr
N_{67*}+\hat N_{451}-\hat N_{124}&=&0\qquad{\rm where}\qquad \tilde{S}_*={1\over 3}(\sqrt{3}m_8-m_3)\label{nontrivialdependences}
\eea
It is easy to verify these relations by hand. Notice that one way to select the independent set of invariants, is simply to drop the usual $S^2\to S^2$ invariants from the list.

To complete the discussion an omission in the analysis must be addressed. We have not considered the general choice of triple $\vec{\tilde{S}}=(m_a,m_b,m_c)$ when defining the candidate map. To keep the discussion concrete focus on the example
\bea
\vec{\tilde{S}}&=&(m_1,m_4,m_6)
\eea
for which ($l_{ab}\equiv l_a-l_b$)
\bea
S_1&=&\frac{f_0(r)f_1(r)\cos(l_{01}\phi)}{\sqrt{f_0(r)^2 \left(f_1(r)^2\cos^2(l_{01}\phi)+f_2(r)^2\cos ^2(l_{02}\phi)\right)+f_1(r)^2f_2(r)^2\cos^2(l_{12}\phi)}}\cr\cr
S_2&=&\frac{f_0(r)f_2(r)\cos(l_{02}\phi)}{\sqrt{f_0(r)^2 \left(f_1(r)^2\cos^2(l_{01}\phi)+f_2(r)^2\cos^2(l_{02}\phi)\right)+f_1(r)^2 f_2(r)^2\cos^2(l_{12}\phi)}}\cr\cr
S_3&=&\frac{f_1(r) f_2(r)\cos(l_{12}\phi)}{\sqrt{f_0(r)^2 \left(f_1(r)^2\cos^2(l_{01}\phi)+f_2(r)^2\cos^2(l_{02}\phi)\right)+f_1(r)^2 f_2(r)^2\cos ^2(l_{12}\phi)}}\label{crazymap}
\eea
This triple has not been considered above. We focus on this example as it illustrates the point we will address. Using the above components it is straightforward to compute the topological number density, which is given by the left hand side of (\ref{topnumbdens}). After integrating we obtain a topological invariant, that is, a quantity that is invariant under smooth deformations of the map defined by Equation~(\ref{crazymap}). 

Choosing arbitrary triples to define $\vec{S}$, we generally find that the topological number density is not a total derivative, making it impossible to obtain analytic formulas for the integral\footnote{The only triples that do lead to topological number densities that are a total derivative are constructed using the pairs we have identified above.}. Consequently, we rely on numerical integration, which remains feasible since we have explicit closed-form expressions for the integrand. Moreover, the numerical results provide valuable insights -- for example, integer values for the results of the numerical integration would strongly suggest the existence of new topological invariants worth further investigation. In such cases, a natural question arises: what does the invariant count?

For the mapping in (\ref{crazymap}), assuming for concreteness that $|l_0|<|l_1|<|l_2|$ we find that as $r\to\infty$ we have $\vec{S}=(0,0,1)$ so it is sensible to perform a one point compactification of the plane. Thus (\ref{crazymap}) defines an $S^2\to S^2$ mapping and the topological invariant must be a wrapping number. There is an important difference between the map (\ref{crazymap}) we consider here and the map (\ref{NES}) considered above. In the map (\ref{NES}), the components $S_1$ (proportional to $\cos(\phi(l_0-l_1))$) and $S_2$ (proportional to $\sin(\phi(l_0-l_1))$ both change sign, but they do so in a coordinated way: the phase of $S_1+iS_2$ is linear in $\phi$. Consequently, following the tip of $\vec{v}=(S_1,S_2)$ as $\phi$ increases from 0 to $2\pi$, the tip never retraces its path, but rather completes an integer number of rotations which traces out a closed circle. In contrast to this, the three vector components appearing in (\ref{crazymap}) change signs in an uncoordinated way. When the three components of the $\vec{S}$ change in this uncorrelated way, the tip of the unit vector retraces its path without closing, producing in the end a wrapping number of zero. Direct numerical integration of the topological number density confirms that the wrapping number is indeed zero. 

In life accidents often happen. Topology is not exempt, and it is possible for some triples to become a non-zero invariant, by an accident of how $l_0$, $l_1$ and $l_2$ are chosen. Two types of accidental invariants are possible. An example of the first type of accidental invariant is provided by $\tilde{S}=(m_1,m_3,m_5)$. Recall from (\ref{explicitms}) that we have
\bea
m_1=2 f_0(r) f_1(r) \cos \Big(\phi  (l_0-l_1)\Big) \qquad\quad m_3=f_0(r)^2-f_1(r)^2 \qquad\qquad m_5=-2f_0(r)f_2(r)\sin\Big(\phi  (l_0-l_2)\Big)
\eea
We have always focused on the case that $l_0\ne l_1$, $l_0\ne l_2$ and $l_1\ne l_2$ since this is needed to ensure that all three states $|0\rangle_B$, $|1\rangle_B$ and $|2\rangle_B$ are entangled. Setting any two (or more) angular momenta equal, drastically reduces the entanglement in the state. To illustrate our point, we must drop this assumption. If we choose $l_2=l_1$ the above triple becomes
\bea
m_1=2 f_0(r) f_1(r) \cos \Big(\phi  (l_0-l_1)\Big) \qquad\quad m_3=f_0(r)^2-f_1(r)^2 \qquad\qquad m_5=-2f_0(r)f_1(r)\sin\Big(\phi  (l_0-l_1)\Big)
\eea
which defines a usual $S^2\to S^2$ mapping with a non-zero wrapping number determined by $l_0-l_1$. In the process, $(S_1,S_2)=(m_1,m_5)$ becomes a nice pair. Only a single invariant -- the one given above -- can be defined with this nice pair as a consequence of the fact that the wave function has bipartite entanglement between $|0\rangle_B$ and the sum $|1\rangle_B+|2\rangle_B$. When $l_1=l_2$ $(S_1,S_2)=(m_1,m_5)$ and $(S_1,S_2)=(m_2,m_4)$ becomes nice pairs, when $l_0=l_1$ $(S_1,S_2)=(m_4,m_7)$ and $(S_1,S_2)=(m_5,m_6)$ become nice pairs and when $l_0=l_2$ $(S_1,S_2)=(m_1,m_7)$ and $(S_1,S_2)=(m_2,m_6)$ become nice pairs.

This first type of accidental invariant will not affect our analysis at all, since our focus is on states that have all three angular momenta distinct. The second accidental invariant can occur, even when the three angular momenta are distinct. It only affects triples with a $\sin(\cdot)$ appearing in one component and a $\cos(\cdot)$ in another. We can again illustrate it with the example $\tilde{S}=(m_1,m_3,m_5)$. For this triple, choosing $l_0=1$, $l_1=2=l_3$ gives $N_{135}=-1$, choosing $l_0=1$, $l_2=4$, $l_3=2$ also gives a wrapping number $N_{135}=-1$ and finally choosing $l_0=1$, $l_2=4$, $l_3=6$ also gives a wrapping number $N_{135}=-1$. For these three cases we have
\bea
(l_0=1,l_1=2,l_2=2)\qquad
m_1=2 f_0(r) f_1(r) \cos \Big(\phi\Big) \qquad\quad  m_5=-2f_0(r)f_1(r)\sin\Big(\phi\Big)\cr\cr
(l_0=1,l_1=2,l_2=2)\qquad
m_1=2 f_0(r) f_1(r) \cos \Big(\phi\Big) \qquad\quad  m_5=-2f_0(r)f_1(r)\sin\Big(3\phi\Big)\cr\cr
(l_0=1,l_1=2,l_2=2)\qquad
m_1=2 f_0(r) f_1(r) \cos \Big(5\phi\Big) \qquad\quad  m_5=-2f_0(r)f_1(r)\sin\Big(3\phi\Big)
\eea
so that we find the same wrapping number for all three choices. This is a generic features of these accidental invariants: they have only a weak dependence on the angular momenta. To understand these numbers, consider Figure \ref{fig:accident}. There we explain why these maps all have a wrapping number of 1. This happens whenever the two frequencies $(l_0-l_1)$ and $(l_0-l_2)$ are both odd, or when both are a given common integer $p$ times odd integers. The wrapping number is then $\pm p$. 

As mentioned above, these accidental invariants only affect triples that have a term with with a $\sin(\cdot)$ and a $\cos(\cdot)$ dependence. Consequently, there are a total of at most 18 triples that are affected.

\begin{figure*}[t]
\includegraphics[width=1\linewidth]{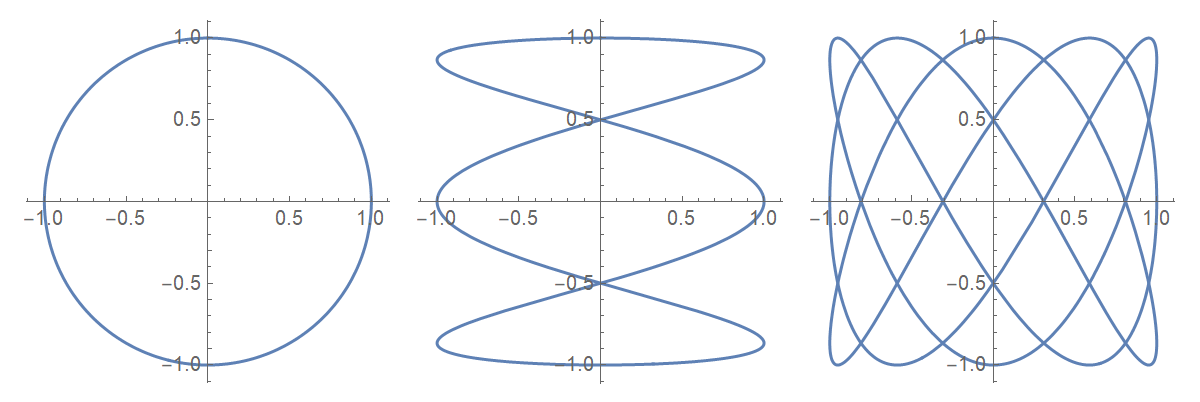}
\caption{\textbf{Accidental wrapping.} Here we plot the curves $x=\cos(\phi)$ $y=\sin(\phi)$ (left), $x=\cos(\phi)$ $y=\sin(3\phi)$ (middle) and $x=\cos(5\phi)$, $y=\sin(3\phi)$. All three curves wrap the origin exactly once.}
\label{fig:accident}
\end{figure*}

For $d=3$ we have systematically scanned over all possible triples $(m_a,m_b,m_c)$ that we have not yet considered\footnote{The triples we have considered are all constructed using a nice pair.} and over the mutually distinct values of $(l_0,l_1,l_2)$ and values for which the second accident can't happen, each from $-5$ to $5$\footnote{This ensures that no triple becomes non-zero by accident, along the lines of the above discussion.}. We find no additional invariants among the triples that were omitted from the discussion above.

\subsection{Topological invariants for Qudit (any $d$) states}\label{generald}

In this section the results obtained above for the qudit ($d=2$) and qutrit ($d=3$) invariants are used to explain the structure of the topological invariants for any $d$. This is possible because the results obtained above establish an interesting algebraic structure, most easily exhibited using the Cartan structure of semisimple Lie algebras. The ``Cartan structure" of a semisimple Lie algebra refers to the decomposition of the algebra into a direct sum of its Cartan subalgebra (a maximal commutative subalgebra where all elements are simultaneously diagonalizable) and the root spaces associated with the linear functionals on the Cartan subalgebra, essentially describing the structure of the algebra through the action of the Cartan subalgebra on the root spaces. This allows for a detailed analysis of the algebra's representation theory and classification based on its root system. Here we will simply review the bare minimum of the theory we need and refer the reader to \cite{Hall2015} for a very readable account with all the details.

The ``density matrix'' defined by the qudit wave function Equation~(\ref{Eq: SpatialDoFSimpEntState}) is a $d\times d$ hermitian matrix with unit trace\footnote{This density matrix describes photon B after we partially collapse the total wave function by performing a position measurement of photon A.}. Thus, it can uniquely be written as
\bea
\rho = {1\over d}{\bf 1}_{d} + \hat{\rho}
\eea
where ${\bf 1}_{d}$ is the $d\times d$ unit matrix and $\hat{\rho}$ is an arbitrary traceless and Hermitian matrix. The set of $d\times d$ traceless and Hermitian matrices forms a vector space\footnote{Addition of vectors is matrix addition, multiplication of a vector by a scalar is multiplication of a matrix by a scalar and the inner product between two vectors $A$ and $B$ is given by ${\rm Tr}(A^\dagger B)$ where $\dagger$ complex conjugates each element and transposes the matrix as usual.} and a basis for this vector space is provided by the Lie algebra $su(d)$ of the Lie group $SU(d)$\footnote{In what follows it will be helpful to keep in mind that as a vector space $su(d)$ has dimension $d^2-1$.}. This is why semisimple Lie algebras are relevant for our problem.

The Lie algebra is a vector space $V$ with an additional product structure $V\times V\to V$, called the Lie bracket. For our application we study a matrix Lie algebra and the Lie bracket between $A$ and $B$ is simply the matrix commutator $[A,B]$. If the Lie algebra has a basis $T_a$ we can write
\bea
[T_a,T_b]=f_{ab}{}^c\,T_c\label{LAbasis}
\eea
where $f_{ab}{}^c$ are the structure constants of the Lie algebra. The algebra of the Pauli matrices is a familiar example of a Lie algebra, which is $su(2)$. The structure constants completely determine the Lie brackets of all elements of the Lie algebra, so they define which algebra we study. The Cartan structure of the Lie algebra arises by choosing a particular basis for the algebra. We decompose the Lie algebra into a maximal set of generators which commute with themselves
\bea
[H_a,H_b]=0
\eea
called the Cartan subalgebra. For $su(2)$ the Cartan subalgebra only contains $\sigma_3$, while for $su(3)$ it contains $\lambda_3$ and $\lambda_8$. This illustrates two facts we will need: (1) for $su(d)$ the Cartan subalgebra $H_i$ contains $d-1$ elements and (2) these elements can all be taken diagonal\footnote{Of course, we could choose a different basis for the vector space $W$ on which the matrices of $su(d)$ act. By saying ``these elements can all be taken diagonal'' we are stating what basis we are taking for this vector space $W$. Our wave function is naturally an element of $W$.}. The remaining elements of the basis of the Lie algebra can be grouped into ${d(d-1)\over 2}$ pairs of matrices $E_a$ and $E_{-a}$ with $a=1,2,\cdots{d(d-1)\over 2}$. $E_a$ and $E_{-a}$ are called roots\footnote{The number of roots plus the number of elements in the Cartan subalgebra adds up to the dimension of the Lie algebra as it must: $2{d(d-1)\over 2}+d-1=d^2-1$. Note also that the roots themselves are not hermitian. The reader searching for a bit of excitement in life might try to prove that $E_{-a}=E_a^\dagger$.}. The defining feature of the roots is that the Lie bracket $[E_a,E_{-a}]$ belongs to the Cartan subalgebra for all $a$, and the Lie bracket between a root and any element of the Cartan subalgebra is proportional to the root
\bea
[E_{\pm a},H_i]=\pm\lambda_{a,i}E_{\pm a}
\eea
where $\lambda_{a,i}$ is a number. The roots for $su(2)$ are the raising and lowering operators $\sigma_\pm={1\over 2}(\sigma_1\pm i\sigma_2)$ of $\sigma_3$. In this new language, the Lie algebra provided by the three Pauli matrices is replaced by a pair of roots $\sigma_\pm$ and a single element in the Cartan subalgebra $\sigma_3$, while the Lie algebra provided by the Gell-Mann matrices is replaced by 3 pairs of roots $E_{\pm 1},E_{\pm 2},E_{\pm 3}$ and two elements in the Cartan subalgebra $H_1,H_2$. The roots $E_{\pm 1},E_{\pm 2},E_{\pm 3}$ would raise and lower the eigenvalues of $H_1,H_2$ in exactly the same way that $\sigma_\pm$ raise and lower the eigenvalue of $\sigma_3$. At the risk of laboring the point, all we have done is change basis for the Lie algebra.

With this background we can now return to the topological invariants we wish to describe. In our discussion above we were influenced by the fact that certain choices for the vector components lead to a topological number density that can be written as a total derivative. As remarked in Equation~(\ref{remarkablepairs}) this can be traced back to a property of the pair $(S_1,S_2)$. In the language of the above discussion, these ``nice pairs'' are obtained from a pairs of roots $E_{\pm a}$ as
\bea
\tilde{S}_1=\langle\psi(\vec{r})|{E_a+E_{-a}\over 2}|\psi(\vec{r})\rangle
\qquad 
\tilde{S}_2=\langle\psi(\vec{r})|{E_a-E_{-a}\over 2i}|\psi(\vec{r})\rangle
\label{RootsandS}
\eea
For $su(d)$ there are ${d(d-1)\over 2}$ pairs of roots and so we learn that there are ${d(d-1)\over 2}$ ``nice pairs''. These nice pairs are used to construct either (1) the ``usual $S^2\to S^2$ maps'' or (2) the ``more exotic invariants'' related to $D^2\to D^2$ maps. We know that the Lie bracket of the two roots $E_{\pm a}$ is in the Cartan subalgebra. Taking
\bea
\tilde{S}_3=\langle\psi(\vec{r})|[E_a,E_{-a}]|\psi(\vec{r})\rangle\label{idCSA}
\eea
gives a triple of vectors that define one of the ``usual $S^2\to S^2$ maps''. For the three $S^2\to S^2$ maps that can be defined for $d=3$ we have 
\bea
&&E_1={1\over 2}(\lambda_1+i\lambda_2)\qquad E_{-1}={1\over 2}(\lambda_1-i\lambda_2)\qquad [E_1,E_{-1}]=\lambda_3\cr\cr
&&E_2={1\over 2}(\lambda_4+i\lambda_5)\qquad E_{-2}={1\over 2}(\lambda_4-i\lambda_5)\qquad [E_2,E_{-2}]={1\over 2}(\lambda_3+\sqrt{3}\lambda_8)\cr\cr
&&E_3={1\over 2}(\lambda_6+i\lambda_7)\qquad E_{-3}={1\over 2}(\lambda_6-i\lambda_7)\qquad [E_3,E_{-3}]={1\over 2}(-\lambda_3+\sqrt{3}\lambda_8)
\eea
It is instructive use Equation~(\ref{idCSA}) to compare the last entry on each line above to the value of $S_3$ used to obtain the invariants in Equations~(\ref{frsts2s2}), (\ref{scnds2s2}) and (\ref{thrds2s2}).

To understand the origin of the relation between the ``nice pairs'' and the roots, we note that each root $E_{a}$ ($a>0$) is a matrix of zeroes with a single off diagonal entry equal to 1. This off diagonal entry appears above the diagonal, in row $i$ and column $j$, with $j>i$. For the specific form (\ref{Eq: SpatialDoFSimpEntState}) for the wave function, using (\ref{RootsandS}) we find
\bea
\tilde{S}_1=2f_{i-1}(r)f_{j-1}(r)\cos \big((l_{i-1}-l_{j-1})\phi\big)\qquad \tilde{S}_2=2f_{i-1}(r)f_{j-1}(r)\sin \big((l_{i-1}-l_{j-1})\phi\big)
\eea
where $f(r)$ is a function of $r$, independent of $\phi$. This recovers precisely the structure (\ref{defnicepairs}) of nice pairs, so that we have proved the connection (defined by Equation (\ref{RootsandS})) between the roots of $su(d)$ and the ``nice'' pairs. 

To understand why the choice (\ref{idCSA}) leads to one of the nice $S^2\to S^2$ maps, note that $[E_a,E_{-a}]$ is a traceless diagonal matrix with 1 in position $i$ on the diagonal and $-1$ in position $j$. Again using the specific form (\ref{Eq: SpatialDoFSimpEntState}) for the wave function and using (\ref{idCSA}) we find
\bea
   \tilde{S}_3=\big(f_{i-1}(r)\big)^2-\big(f_{j-1}(r)\big)^2
\eea
After normalizing we recover precisely the form given in (\ref{NES}). This gives a complete derivation of the connection (defined by Equations (\ref{idCSA}) and (\ref{RootsandS})) between the Cartan structure of $su(d)$ and the usual $S^2\to S^2$ maps. Thus, for the $d$ dimensional case there are ${d(d-1)\over 2}$ independent $S^2\to S^2$ maps that can be defined.

To describe the more exotic invariants, as above we choose $(S_1,S_2)$ to be a nice pair, but depart from the above analysis by changing our choice for $S_3$. We allow $S_3$ to range over all roots except for $E_{a}$ and $E_{-a}$ used to produce the nice pair, and over all elements of the Cartan subalgebra, except for $[E_{a},E_{-a}]$. When choosing $S_3$ to be a root we take the real forms 
\bea
S_3={1\over 2}(E_{b}+E_{-b}) \qquad {\rm or}\qquad S_3={1\over 2i}(E_{b}-E_{-b})\label{twochoiceS3}
\eea
as opposed to $S_3=E_{\pm b}$ which is complex. This defines a total of ${1\over 2}d(d-1)(d(d-1)-2+d-2)$ maps. For $d=3$ this counting gives 15 and the maps obtained are those appearing in Section \ref{exoticmaps}. The total number of maps (usual $S^2\to S^2$ plus the exotic maps) we obtain is ${1\over 2}d(d-1)(d^2-3)$. This counting gives 18 for $d=3$ which counts the maps appearing in Section \ref{usualmaps} and Section \ref{exoticmaps}.

The relations (\ref{simpledependences}) obtained by inspection, showed that not all of the wrapping numbers computed from these maps are independent. Translated into $d$ dimensions, the analog of (\ref{simpledependences}) shows that the maps obtained from the two choices in (\ref{twochoiceS3}) are not independent and only one choice should be listed. This leads to a total of ${1\over 2}d(d-1)\left({1\over 2}d(d-1)-1+d-2\right)$ maps. The analog of (\ref{nontrivialdependences}) for the $d$ dimensional case again implies that the simplest way to construct a list of independent invariants is simply to drop the usual $S^2\to S^2$ invariants. Thus, we find a total of ${1\over 4}d(d-1)(d-2)(d+3)$ independent topological invariants are captured by our wave functions for $d>2$, while the wave function captures a single topological invariant for $d=2$.

There are a number of comments that should be made to put the analysis of this section into perspective. Our analysis for general $d>2$ is possible thanks both to (1) the connection between the wave function and the Lie algebra $su(d)$ that we have exhibited and (2) the specific form of the wave function spelled out in Equation~(\ref{Eq: SpatialDoFSimpEntState}). It is for this specific state that we have a total of ${1\over 4}d(d-1)(d-2)(d+3)$ independent topological invariants. For more general states the connection between the wave function and the Lie algebra $su(d)$ will continue to hold, but other elements of the argument including the `nice pairs' and their connection to the roots, will have to be modified. The total number of ways in which a triple defining $\vec{S}$ can be selected is given by $d^2-1\choose 3$, and these all define invariants that might encode topological information about the wave function. A more careful choice of wave function will almost certainly lead to a different and perhaps far richer topological spectrum. This is well worth investigating further.

\subsection{Non-Abelian Higgs Potential}\label{NAHP}

In this section, we provide an alternative perspective on the topological invariants introduced above. This viewpoint offers deeper insight into these invariants, naturally linking them to well-established quantities in non-Abelian gauge theories. In particular, it clarifies why a spectrum of topological invariants appears.

Consider a coupled Yang-Mills-Higgs theory. Our first observation is that both the Higgs potential $\phi$ and the gauge potential $A_\mu$ are matrix-valued fields that take values in the Lie algebra of the gauge group. Expanding these fields in the basis introduced in (\ref{LAbasis}), we obtain  
\begin{equation}
\phi^a = {\rm Tr}(T^a \phi), \qquad A_\mu^a = {\rm Tr}(T^a A_\mu).
\end{equation}
This formulation immediately suggests a connection between the density matrix and the Higgs potential. Notably, the gauge field carries both a Lie algebra index $a$ and a spacetime vector index $\mu$, while the Higgs potential carries only a Lie algebra index. This structure mirrors that of the density matrix, which itself carries a single Lie algebra index. To develop a precise connection, note that the dynamics of the Yang-Mills-Higgs theory is defined by the Lagrangian density
\begin{eqnarray}
{\cal L}&=&-{1\over 4}F^a_{\mu\nu}F^{a\mu\nu}+{1\over 2}D^\mu\phi^aD_\mu\phi^a-{\lambda\over 4}(\phi^a\phi^a-1)^2\label{Lgngn}
\end{eqnarray}
where the non-Abelian field strength tensor $F^a_{\mu\nu}$ and the covariant derivative of the Higgs potential are given by
\begin{eqnarray}
F^a_{\mu\nu}&=&\partial_\mu A^a_\nu-\partial_\nu A^a_\mu+g_{YM}^2\epsilon^{abc}A^b_\mu A^c_\nu\qquad
D_\mu\phi^a\,\,=\,\,\partial\phi^a+g_{YM}^2\epsilon^{abc}A_\mu^b\phi^c\label{dervs}
\end{eqnarray}
We make use of Section 1.4 of \cite{Harvey:1996ur} and, from this point onward, specialize to the gauge group $SU(2)$. This choice restricts the Lie algebra index $a$ to take values in $\{1,2,3\}$, implying that both the gauge potential and the Higgs potential can be represented as vectors, denoted by $\vec{A}_\mu$ and $\vec{\phi}$, respectively. The formulas (\ref{dervs}) has been specialized to $SU(2)$.

The perturbative spectrum of the theory consists of a massless spin-1 gauge field, as well as massive spin-1 gauge bosons with mass $g_{YM}$. Additionally, the Higgs field acquires a mass given by $m_H = \sqrt{2\lambda}$. The vacuum of the Higgs sector consists of all field configurations that minimize the potential, forming a target space given by the sphere  
\begin{equation}
\vec{\phi} \cdot \vec{\phi} = 1.
\end{equation}

Our density matrix takes values in $su(d)$. We identify an $su(2)$ subalgebra within $su(d)$ that corresponds to the $su(2)$ structure appearing in the Yang-Mills-Higgs theory. Each such $su(2)$ subalgebra is defined by a pair of root generators, $E_{\pm a}$, which serve as the analogs of the usual $su(2)$ raising and lowering operators, along with an element of the Cartan subalgebra, given by the commutator $[E_a, E_{-a}]$ which plays the role of $J_3$. These three generators isolate three specific components of the density matrix, collectively forming the vector $\vec{S}$ associated with the usual $S^2 \to S^2$ mappings. We naturally identify this vector $\vec{S}$ with the Higgs potential $\vec{\phi}$.

In general, such an identification is not always possible, making its consistency in our case highly non-trivial. This consistency relies on two key observations: First, both vectors are unit-normalized, satisfying 
\begin{equation}
\vec{\phi} \cdot \vec{\phi} = 1 = \vec{S} \cdot \vec{S}.
\end{equation}
As a result, the vector $\vec{S}$ naturally belongs to the Higgs vacuum. Second, the topological invariant that quantifies the magnetic charge of the 't Hooft-Polyakov monopole depends solely on the Higgs potential. Remarkably, its precise formulation exactly coincides with the wrapping number of the usual $S^2 \to S^2$ maps we defined earlier. To see the equality, we begin by noting that the topological invariant which defines the magnetic charge is given by
\begin{equation}
N_{\rm 't\, Hooft-Polyakov}={1\over 8\pi}\int d\Omega^i \epsilon^{ijk}\epsilon^{abc}\phi^a{\partial\phi^b\over\partial x^j}{\partial\phi^c\over\partial x^k}\label{tHPM}
\end{equation}
where $d\Omega^i$ is an area element. Now consider the topological invariant we have used above, given by
\begin{eqnarray}
    N&=&{1\over 4\pi}\int \vec{S}\cdot\left({\partial\vec{S}\over\partial x}\times{\partial\vec{S}\over\partial y}\right)\,dx\, dy\cr\cr
    &=&{1\over 4\pi}\int \epsilon^{abc}S^a {\partial S^b\over\partial x}{\partial S^c\over\partial y}\,dx\, dy
\end{eqnarray}
Collect the space coordinates into a vector $(x^1,x^2,x^3)=(x,y,z)$. Further we can define three natural area elements
\begin{equation}
d\Omega^1=dy\, dz=dx^2\, dx^3\qquad d\Omega^2=dx\, dz=dx^1\,dx^3\qquad d\Omega^3=dx\, dy=dx^1\, dx^2
\end{equation}
The vector index on $d\Omega^i$ is assigned so that areas on a plane are perpendicular to the plane as usual. With these area  elements defined and recalling that our vectors $\vec{S}$ depend only on $x^1=x$ and $x^2=y$, we immediately find
\begin{eqnarray}
    N&=&{1\over 8\pi}\int d\Omega^i \epsilon^{ijk}\epsilon^{abc}S^a 
    {\partial S^b\over\partial x^j}{\partial S^c\over\partial x^k}\label{rosn}
\end{eqnarray}
The equality between (\ref{tHPM}) and (\ref{rosn}) under the identification $\vec{\phi}=\vec{S}$ is now manifest.

Notice that to make this connection we had to choose an $su(2)$ sub-Lie algebra of $su(d)$. This choice is not unique. The analysis of the Yang-Mills-Higgs theory with gauge group $SU(d)$ is very similar to what we have reviewed above\footnote{The only changes are that the Lie algebra index $a$ now runs from 1 to $d^2-1$ and in the formula (\ref{dervs}) $\epsilon^{abc}$ should be replaced by the structure constant $f^{abc}$ introduced in (\ref{LAbasis}).}. In particular, the monopole charge is associated with an $SU(2)$ subgroup of the gauge group. The choice of which $SU(2)$ subgroup to consider is not unique and different embeddings yield distinct topological invariants and so distinct monopole charges \cite{Irwin:1997ew}. Thus, the Higgs potential is now characterized by a spectrum of magnetic charges. Using the correspondence we developed above, we immediately conclude that the density matrix itself is characterized by a spectrum of topological invariants. Our choice of ``nice pairs'' corresponds to choosing an $SU(2)$ subgroup of $SU(d)$ exhibiting a clear parallel with the discussion in Section \ref{generald}.

\subsection{Final Comments}

With a vast set of invariants, each dependent on numerous parameters ${l_0,l_1,\cdots,l_d}$, subtle properties may yet be uncovered. Our analysis serves as a first step, and further exploration is needed to refine and expand the theoretical framework.

A plethora of topological invariants have appeared in the literature. While often referred to as a Skyrmion number, the invariant studied here differs from Skyrme’s original topological charge, which is defined via the current\cite{skyrme1962unified}
\begin{equation}
J^\mu = \epsilon^{\mu\nu\alpha\beta} \, {\rm Tr}(L_\nu L_\alpha L_\beta),\qquad{\rm with}\qquad
L_\mu = U^\dagger \partial_\mu U,
\end{equation}
where $U$ is an element of $SU(2)$. Unlike Skyrme’s invariant, which applies to fields defined over four spacetime dimensions, our invariant arises in a different setting. It corresponds exactly to the topological charge of the 't Hooft-Polyakov monopole in non-Abelian Yang-Mills theory\cite{Harvey:1996ur} as demonstrated above in Section \ref{NAHP}. 

Another related but distinct topological invariant is the Hopf charge, which has been explored in the Faddeev model\cite{faddeev1997knots} and realized experimentally in solitonic structures known as Hopfions\cite{ackerman2017static}. These three invariants are fundamentally different, each arising from a distinct class of topological maps:
\begin{itemize}
    \item {\bf Faddeev model}: The Hopf invariant, associated with the homotopy group $\pi_3(S^2)$, classifies maps $S^3 \to S^2$.
    \item {\bf Skyrme model}: The Skyrmion number is derived from maps $\mathbb{R}^3 \to S^3$, corresponding to $\pi_3(S^3)$. This follows from compactifying $\mathbb{R}^3$ to $S^3$.
    \item {\bf Non-Abelian Yang-Mills monopoles}: The relevant topological charge corresponds to maps $S^2 \to S^2$, classified by $\pi_2(S^2)$. After compactifying $\mathbb{R}^2$ to $S^2$, this is precisely the invariant relevant to our work.
\end{itemize}

\section*{Supplementary: Experiment and data analysis}

\subsection{Experimental setup}
A schematic of the experimental setup is shown in Fig.~\ref{fig:ExpFig}. An ultraviolet $\lambda = 404.2$ nm wavelength pump beam (produced by a Coherent OBIS continuous wavelength laser operating at $100$ mW)  with a Gaussian profile ($\ell = 0$) is imaged and demagnified with two lenses to a beam radius of 0.3 mm onto a type 1, Periodically Poled KTP (PPKTP) non-linear crystal (NLC), of length $2$ mm, to produce a pair of entangled photons of wavelength $\lambda = 808.4$ nm via spontaneous parametric downconversion (SPDC). The two photons, signal (photon A) and idler (photon B) were each imaged and magnified ($\times10$) with two lenses, L1 (f=$50$ mm) and L2 (f=$500$ mm), from the crystal plane to the plane of the spatial light modulators (SLM A and SLM B respectively- Holoeye Pluto 2.1), with the unconverted pump beam being filtered using a bandpass filter (BPF, transmitting a band centered at $808$ nm with a FWHM of $10$ nm). The magnified beam radius of the SPDC is at 3 mm on the SLM screens. Each of the SLM screens consists of $960 \times 960$ pixels with each pixel having dimensions $8 \times 8$ $\mu$m$^2$.
Spatial projective measurements can be simultaneously performed on both photons using the coupled detection system consisting of the SLMs and a pair of single mode fibres (SMFs). In this process the photons were demagnified by a factor of $500$ using lenses, L3 (f=$1000$ mm) and L4 (f=$2$ mm), before being coupling into the SMFs, with fibre core widths of 4.4 $\mu$m. The photons were then measured in coincidence using a pair of independent Avalanche-Photo Diodes (APDs, Excelitas SPCM-AQRH Single-Photon Counting Module) connected to a coincidence counter (CC, Swabian: Time Tagger).
\begin{figure*}[t]
\includegraphics[width=1\linewidth]{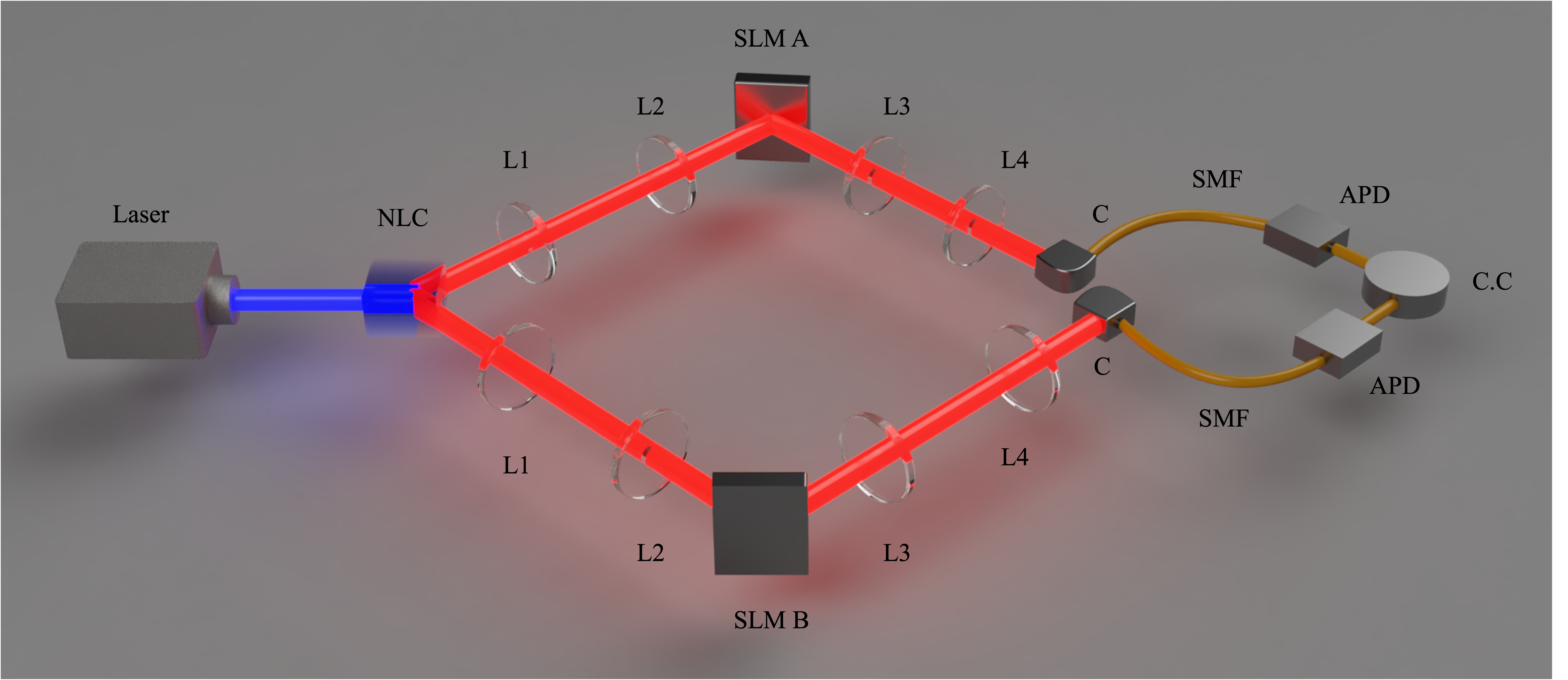}
\caption{\textbf{Experimental setup for the creation and detection of d-dimensional OAM-OAM entangled biphoton states.} A laser was used to produce a pump photon which was incident on the non-linear crystal (NLC) whereby SPDC took place which resulted in a pair of degenerate photons being produced which were entangled in the OAM degree of freedom. Photon A and photon B were imaged onto SLM A and SLM B respectively with the use of lenses L1 (f=$50$ mm) and L2 (f=$500$ mm). The photons were then imaged onto their respective single mode fibres (SMFs) which are connected to couplers (C). Lenses L3 (f=$1000$ mm) and L4 (f=$2$ mm) were used for this imaging. Each SMF is connected to an Avalanche Photo-Diode (APD) which were in turn sent signals to a coincidence counter (C.C).}
\label{fig:ExpFig}
\end{figure*}

\subsection{Experimental considerations}
\begin{itemize}
    \item From the SPDC process the quantum state that is generated can be given by:
    \begin{equation}
\ket{\psi}_{AB}=\sum\limits_{\ell=-n}^{n} \sum\limits_{\ell'=-n}^{n} c_{\ell,\ell'}\ket{\ell}_A\ket{\ell'}_B, \label{Eq:OAMOAM_EntState}
    \end{equation}
    whereby $c_{\ell,\ell'}$ follows a Gaussian distribution across an OAM domain centered about $\ell=0$. Note that the topological charge of the pump photons is $\ell=0$. This causes the fidelity of a state studied if the qubit cases to decrease as the difference in OAM values increased.
    \item The projective measurements onto the OAM basis were performed through phase only projective measurements. The effects of unbalanced coincidence count measurements can be seen in the coincidence matrices.
    \item A phase only technique is used for the projective measurements along with a single mode fiber, this causes challenges when performing projections onto superposition states. The beam waist of the superposition mode on the SLM is required to be of an appropriate size such that the weighting coefficients of the modes match the corresponding weighting coefficients measured when the OAM bandwidth measurement was performed.
    The scaling of the weighting coefficients follows a relationship
    \begin{equation}
        \omega_{SLM}\propto
        \frac{\omega}{\sqrt{|\ell|+1}},
    \end{equation}
    where the $\omega_{SLM}$ corresponds to the beam waist of the mode encoded onto the SLM, while $\omega$ corresponds to the beam waist of the SPDC profile on the SLM.
\end{itemize}

\subsection{High-dimensional quantum state tomography}

\noindent In high dimensional composite Hilbert Spaces the operators that form a complete basis can be given by the Kronecker products of the GGMs \cite{bertlmann2008bloch}:
    
\noindent these matrices are operators which are able to perform rotations on a state living in a $d \times d$ dimensional Hilbert Space \cite{agnew2011tomography}. Therefore a d-dimensional bi-photon state, described by a density matrix $\rho$ can be expressed 

\begin{equation}
\rho=\ket{\psi}_{AB\,AB }\bra{\psi}=A\sum_{i=0}^{d-1}{\sum_{j=0}^{d-1}{\rho_{ij}[\lambda_i \otimes \lambda_j]}},  \end{equation}

\noindent where $A$ represents a normalisation factor, and $\rho_{ij}$ represents the expectation value of $\hat{\tau}_i \otimes \hat{\tau}_j$

\begin{equation}
    \rho_{ij}=tr(\rho[\lambda_i \otimes \lambda_j])=_{AB}\bra{\psi}\lambda_i\otimes \lambda_j\ket{\psi}_{AB}=\sum_{k=0}^{d-1}{\sum_{l=0}^{d-1}\gamma_i^k\gamma_j^l|_{AB}\braket{\psi|\tau_i^k}\ket{\tau_j^l}|^2}, 
\label{Eq.Densitymatrixdecomposition}
\end{equation}
where $\gamma_{i(j)}^{k(l)}$ represents the $k(l)^{th}$ eigenvalue of the corresponding $k(l)^{th}$ eigenvector $\tau_{i(j)}^{k(l)}$ of the GGMs ($\lambda_{i(j)}$). 
In a similar manner the density matrix for a high dimensional bi-photon state can be reconstructed using the procedure followed for the bi-photon qu$d$it case. The number of measurements performed scales as: $N=(4 {d\choose2} +d)^2$.

\subsection{Procedure used to obtain a density matrix from coincidence measurements}
\noindent Experimental projective measurements are performed in order to obtain amplitude and phase information related to the basis states in a pure quantum state. For an individual photon, the projections are performed with the use of a SLM and a SMF. In order to project onto an OAM state ($\ket{\ell}$) the conjugate of the OAM state ($\bra{\ell}$) is encoded onto the SLM causing the desired OAM basis states to be transformed into Gaussian ($\ell=0$) modes which are accepted and propagate through the SMF. Independent OAM projections are performed on photon A (which is incident on SLM A Fig. \ref{fig:ExpFig}) and photon B (which is incident on SLM B Fig. \ref{fig:ExpFig}) thus resulting in a projection being performed on the bi-photon state. A matrix is compiled with the measurement results called the coincidence/projection matrix ($C^M$). A density matrix is guessed (assuming the decomposition in Eq. \ref{Eq.Densitymatrixdecomposition}) initially and using this density matrix a predicted coincidence matrix can be inferred. An optimisation procedure is performed in order to reduced the square of the difference between the coincidence matrix which was measured experimentally ($C^M$) and the predicted coincidence matrix ($C^P(\rho)$). An optimal solution for the density matrix is found ($\hat{\rho}_{opt}$) such that the predicted coincidence matrix obtained minimises $\chi^2=\sum_i^N{\frac{(C^M_i-C^P_i(\rho))^2}{C_i^P(\rho)}}$. Modal cross-talk is observed in the experimental data that was obtained. In order to reduce the affect that that these modal contributions have on the density matrices and the topological spectra, a thresholding procedure is implemented to the optimisation procedure. The thresholding procedure sets a limit to the smallest magnitude of the density matrix components by a value $\epsilon$;  (abs$(\rho_i)=0|$abs$(\rho_i)\leq \, \epsilon $). The value $\epsilon$ is informed by the OAM bandwidth measurement. Let us consider a 3D OAM subspace ($\ket{\ell_1}_A\ket{-\ell_1}_B \, , \, \ket{\ell_2}_A\ket{-\ell_2}_B \, , \, \ket{\ell_3}_A\ket{-\ell_3}_B$) over which the density matrix is constructed, the modal cross talk would be contributed from the following projections made on the bi-photon state: $\ket{\ell_2}_A\ket{-\ell_1}_B \, , \, \ket{\ell_3}_A\ket{-\ell_1}_B \,
, \, \ket{\ell_1}_A\ket{-\ell_2}_B  \, , \ket{\ell_3}_A\ket{-\ell_2}_B
, \, \ket{\ell_1}_A\ket{-\ell_3}_B  \, , \ket{\ell_2}_A\ket{-\ell_3}_B$ . The quantities for these projections are obtained from the OAM bandwidth which is normalised to the maximum. The maximum of these coincidence measurements are assigned as $\epsilon$. This method can be generalised to d-dimensions.

Explicitly, the procedure follows:
\begin{enumerate}
    \item A density matrix is guessed and a predicted coincidence matrix which corresponds to the density matrix is obtained.
    \item The predicted coincidence matrix is then compared to the experimentally measured coincidence matrix. The discrepancy between these two matrices is calculated using a least squares method.
    \item A new prediction for the density matrix is obtained in an effort to reduce $\chi^2$ between the corresponding predicted coincidence matrix and the experimentally measured coincidence matrix.
    \item The procedure is iterative and therefore improving the prediction of the density matrix which corresponds to the experimentally measured coincidence matrix.
\end{enumerate}

\noindent Fig.~\ref{fig:labelingconvention} illustrates the order in which the projective measurements are performed on the quantum states for the projection matrices, looking at the projections for each photon, the projections are performed in descending order starting with the basis OAM state $(\ket{\ell}=[1,0]^T$ and $\ket{-\ell}=[0,1]^T)$ projections first which can be seen as the $d\times d$ submatrix in the top left of the projection matrix. The superpositions are indicated under the marked "\textbf{MUBs}" list.  ($\frac{1}{\sqrt{2}}\left(\ket{\ell}+e^{i\theta}\ket{-\ell}\right)$ with $\theta \in [0,\pi/2,\pi, 3\pi/2] $)

\begin{figure*}[t]
\includegraphics[width=1\linewidth]{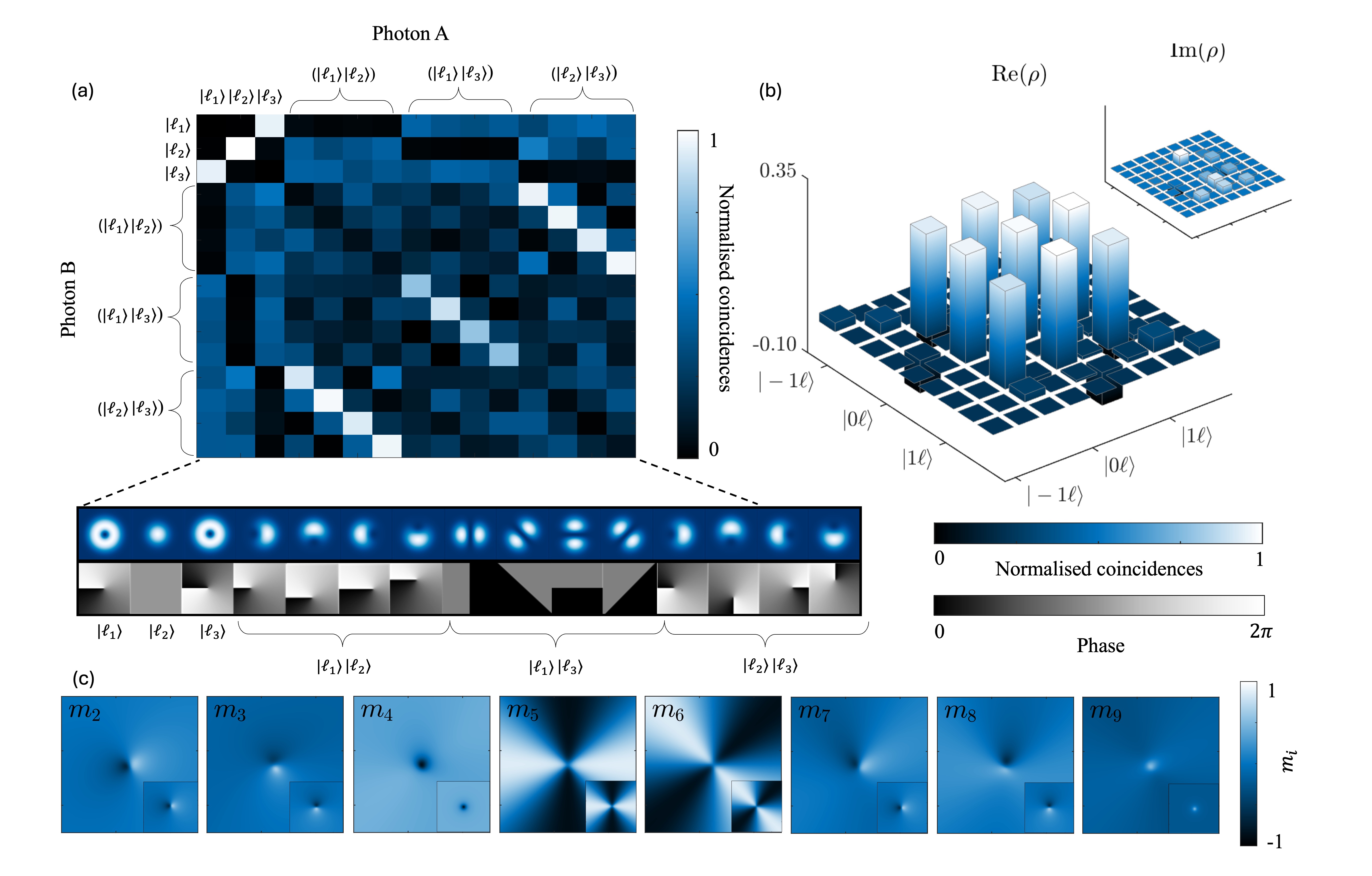}
\caption{\textbf{Quantum state tomography and associated density matrix.} (a) Represents a coincidence matrix which is constructed in the following way whereby projections are performed on OAM basis states (e.g $\ell_1$, $\ell_2$ and $\ell_3$) which represent the orthogonal states. Projections are also performed on the mutually unbiased bases states (MUBs) for a given pair of OAM basis states ($\ell_n$ and $\ell_m$) the following MUBs can be constructed: $\ket{+}=\frac{1}{\sqrt{2}}(\ket{\ell_n}+\ket{\ell_m})$,
$\ket{i}=\frac{1}{\sqrt{2}}(\ket{\ell_n}+i\ket{\ell_m})$
$\ket{-}=\frac{1}{\sqrt{2}}(\ket{\ell_n}-\ket{\ell_m})$,  and $\ket{-i}=\frac{1}{\sqrt{2}}(\ket{\ell_n}-i\ket{\ell_m})$ (b) Represents the density matrix (real and imaginary components) reconstructed from the coincidence matrix represented in (a). The labelling for the density matrix components goes as follows: the density matrix is composed of $d^2\times d^2$ elements, looking along the x and y axes , each the $d^2$ elements are grouped into $d$ elements. In the example density matrix shown $\ket{\ell_1\ell}$ (where the three elements of the axis are $\ket{\ell_1}\ket{\ell_1}, \ket{\ell_1}\ket{\ell_2}, \ket{\ell_1}\ket{\ell_3}$ and so the rest of the axis ticks are labelled in the same way. In this example $\ket{\ell_1}=\ket{\ell=-1}, \ket{\ell_2}=\ket{\ell=0}, \ket{\ell_3}=\ket{\ell=1}$ (c) Represents the spatially varying Gellmann matrices ($m_i$) obtained from the density matrix in (b). }
\label{fig:labelingconvention}
\end{figure*}

\subsection{Concurrence, Fidelity and purity}
\noindent To quantify the quality and degree of entanglement of our produced states, we used the concurrence (for 2D states), fidelity and purity as our figures of merit.\\

\noindent The concurrence was used to calculate the degree of entanglement between the 2D entangled states. It was calculated using 
\begin{equation}
    C(\rho) = \text{max} \{ 0, \lambda_1 -\lambda_2- \lambda_3 - \lambda_4 \},
\end{equation}
where $\lambda_i$ are eigenvalues of the operator $R = \text{Tr} \left( \sqrt{\sqrt{\rho} \tilde{\rho} \sqrt{\rho}}\right)$ in descending order and $\tilde{\rho} = \sigma_{y} \otimes \sigma_{y} \rho^* \sigma_{y} \otimes \sigma_{y}$. The concurrence ranges from 0 for separable states to 1 for entangled states.

The fidelity was calculated using
\begin{equation}
    F =\left( \text{Tr}  \left( \sqrt{  \sqrt{\rho_T}\rho_M \sqrt{\rho_T}  }  \right) \right)^2 ,
\end{equation}
where $\rho_T$ is the target density matrix while $\rho_M$ is the measured density matrix. The fidelity is 0 if the states are not identical or 1 when they are identical up to a global phase.

The state purity was calculated using
\begin{equation}
     P = \text{Tr} \left(\rho^2\right).
\end{equation}
The purity is $\frac{1}{d^2}$ for a maximally mixed state and 1 for a pure state.



\subsection{Experimental Data}





\section*{Supplementary: Numerical evaluation of Topological Spectra}

\noindent Here we discuss the extraction of topological information from the reconstructed biphoton density matrices. For 2D states, there is only one topological signal that is calculated, however for $d>2$ we calculate several topological signals forming a "topological spectrum". To calculate these signals we follow the following procedure: 
\begin{itemize}
    \item[(1)] Perform a change of basis on one of the photons from the OAM basis to position basis. Since we have used OAM-OAM entangled photons in this work, the choice of photon which must undergo a basis change is arbitrary, with the main consequence being a change in the sign over the entire topological spectrum (assuming the produced state is that given in Eq.~\ref{Eq:OAMOAM_EntState}). This operation can be seen as follows $\rho_{AB} \to \rho(\vec{r}) = {}_A\langle r |\rho_{AB} | r \rangle_A$, where $$ \rho_{AB} = \sum_{ijmn}c_{ijmn}\ket{\ell_i}_A\ket{\ell_j}_B{}_A\bra{\ell_m}{}_B\bra{\ell_n}. $$ Then using $\langle r|\ell_i\rangle = \text{LG}_{\ell_i}(\vec{r})$, the spatially varying density matrix becomes $$ \rho(\vec{r}_A) = \sum_{im}\left[\text{LG}_{\ell_i}(\vec{r}) \, \text{LG}^*_{\ell_m}(\vec{r}) \right] \sum_{jn} c_{ijmn} \ket{\ell_j}_B{}_B\bra{\ell_n}, $$ where the superpositions of complex scalar fields, $\text{LG}_{\ell_i}(\vec{r}) \, \text{LG}^*_{\ell_m}$, become the field coefficient components of the density matrix whose elements are labeled by $\ket{\ell_j}_B{}_B\bra{\ell_n}$.
    \item[(2)] Calculate the vector field components using $m_i(\vec{r}) = \text{Tr}\left(\lambda_i\rho(\vec{r}) \right)$, where $\lambda_i$ is a given observable for $\text{SU}(d)$.
    \item[(3)] The topological spectrum is then calculated by selecting and normalizing every possible combination of vector field triplets, $\vec{\tilde{S}}=(m_i,m_j,m_k)$, and performing the integral given in Eq.\ref{Eq: SkyNumSupp}. Every topological signal is calculated from a chosen triplet that forms a candidate map and therefore can form part of the final topological spectrum representation. In the main text, we have only shown the part of the spectrum which is "generically" non-zero as listed in the Table~\ref{tab:GVecDefs} below.
\end{itemize}

\begin{table}[h!]
\centering
\renewcommand{\arraystretch}{1.5} 
\begin{tabular}{|>{\centering\arraybackslash}m{3cm}|>{\centering\arraybackslash}m{4cm}|}
\hline
\textbf{$n^\text{th}$ topological signal} & \textbf{Gellman Vector triplet \newline $\{m_i, m_j, m_k\}$} \\ \hline
1                                         & $\{m_1, m_2, m_3\}$                                         \\ \hline
2                                         & $\{m_4, m_5, \frac{m_3 + \sqrt{3}m_8}{2}\}$                 \\ \hline
3                                         & $\{m_6, m_7, \frac{-m_3 + \sqrt{3}m_8}{2}\}$                \\ \hline
4                                         & $\{m_1, m_2, m_4\}$                                         \\ \hline
5                                         & $\{m_1, m_2, m_5\}$                                         \\ \hline
6                                         & $\{m_1, m_2, m_6\}$                                         \\ \hline
7                                         & $\{m_1, m_2, m_7\}$                                         \\ \hline
8                                         & $\{m_1, m_2, m_8\}$                                         \\ \hline
9                                         & $\{m_4, m_5, m_1\}$                                         \\ \hline
10                                        & $\{m_4, m_5, m_2\}$                                         \\ \hline
11                                        & $\{m_4, m_5, m_3\}$                                         \\ \hline
12                                        & $\{m_4, m_5, m_6\}$                                         \\ \hline
13                                        & $\{m_4, m_5, m_7\}$                                         \\ \hline
14                                        & $\{m_6, m_7, m_1\}$                                         \\ \hline
15                                        & $\{m_6, m_7, m_2\}$                                         \\ \hline
16                                        & $\{m_6, m_7, m_3\}$                                         \\ \hline
17                                        & $\{m_6, m_7, m_4\}$                                         \\ \hline
18                                        & $\{m_6, m_7, m_5\}$                                         \\ \hline
\end{tabular}
\caption{Gellman Vector Triplet definitions for Topological Signals for topological spectra associated with 3D states}
\label{tab:GVecDefs}
\end{table}

\noindent There are three caveats that should be considered when calculating the topological signals: (1) first we must choose an appropriate definition for the behaviour of the map at $r = 0$, (2) secondly, when reporting the topological signals of the $D^2\to D^2$ a choice should be made as to whether the topological number of the initial or extended map is to be reported, a choice which simply amounts to whether the initial map's topological signal is multiplied by a factor of 2 or not and (3) lastly, numerical integration may be unstable due to singularities present in the integrand of Eq.\ref{Eq: SkyNumSupp}. The latter is addressed in the proceeding section. To illustrate results of the choices made with regards to (1) and (2), Figure~\ref{fig: NumFixes} shows the mapping that emerges when choosing the vector triplet $\vec{\tilde{S}}=(\tilde{S}_1,\tilde{S}_2,\tilde{S}_3)=(m_1,m_2,m_4)$ for the state with $\ell_2=1,\ell_1=2, \ell_0=3$. At $r\to0$ initial behaviour of the map  is to flip between positive and negative values for $\tilde{S}_3$ as $\phi$ changes. This can be seen in  Figure~\ref{fig: NumFixes}(a) where the tip of the vector near the point $r=0$ oscillates between pointing towards the north pole and south pole with changing $\phi$, thus resulting in the "wedges" appearing above and below the equator. This culminates in a $\phi$ dependence of whether the vector points towards the north pole or south pole at $r=0$, and further results in a false calculation of the Skyrmion number. Thus a choice for the definition of $\vec{\tilde{S}}$ at $r=0$ must be made to correct for the behaviour of the map. One choice is to take $\tilde{S}_3\to |\tilde{S}_3|$ which has the effect of gluing the ``wedges" of map together by their common boundaries (``red", $r\to0$ and ``purple", $r\to\infty$) in the plane, as shown in Figure~\ref{fig: NumFixes}(b). Secondly, we define a second copy of the map which is identical to the first, and glue the original and the copy at their equatorial boundary, as shown in Figure~\ref{fig: NumFixes}(c). It is easy to verify that the result of this action is to simply multiply the initial invariant by a factor of 2. Throughout the main text, the choice was made to report the integer invariant whilst showing the features of the disk mapping.

\begin{figure*}[t]
\includegraphics[width=\linewidth]{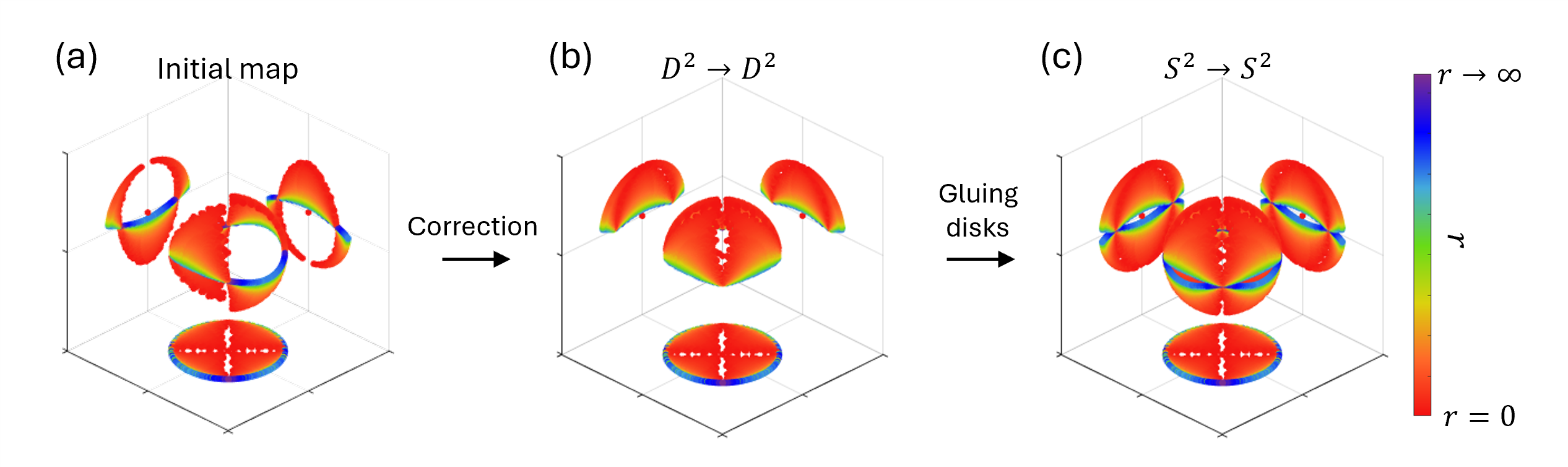}
\caption{Steps taken to (a-b) correct the initial map and (b-c) extend it from a $D^2\to D^2$ mapping into an $S^2 \to S^2$ mapping for states with $\ell_2=1,\ell_1=2, \ell_0=3$ and a chosen map constructed from the triplet $(\tilde{S}_1,\tilde{S}_2,\tilde{S}_3)=(m_1,m_2,m_4)$. The unit vector has been plotted as a point cloud with the colourmap (``red", $r\to0$ and ``purple", $r\to\infty$) indicating the radial coordinate at which the corresponding unit vector is found on the plane. 
}
\label{fig: NumFixes}
\end{figure*}

\subsection{Singular considerations}

Our topological invariants are computed by numerical integration of an integrand that is written in terms of a measured vector $\vec{S}$. In this section we explore the issue of how well behaved this numerical integration is. Our key finding is a demonstration that some of the integrands can develop singularities. The fact that we can determine the presence of these singularities apriori, is a useful input to extracting invariants from the experimental data, as it suggests where the analysis may be subtle.

For convenience we focus on the $d=3$ wave function, but it is straight forward to reach analogous conclusions for any $d$. In this case our topological invariants are constructed from components of the 8 dimensional vector
\bea
m_a=\langle\psi(\vec{r})|\lambda_a|\psi(\vec{r})\rangle\qquad\qquad a=1,2,\cdots,8
\eea
defined using the Gell-Mann matrices $\lambda_a$. The explicit expressions for the $m_a$ are given in (\ref{explicitms}). The topological invariant is given by choosing a triple of gellman observables
\bea
\vec{S}(m_{a_1},m_{a_2},m_{a_3})\equiv\left({m_{a_1}\over\sqrt{m_{a_1}^2+m_{a_2}^2+m_{a_3}^2}},{m_{a_2}\over\sqrt{m_{a_1}^2+m_{a_2}^2+m_{a_3}^2}},{m_{a_3}\over\sqrt{m_{a_1}^2+m_{a_2}^2+m_{a_3}^2}}\right)
\eea
and then evaluating the integral
\bea
\int \vec{S}\cdot\left({\partial\vec{S}\over\partial r}\times{\partial\vec{S}\over\partial\phi}\right) dr\, d\phi\label{IntToDo}
\eea

We now examine the sensitivity of the numerical integration needed to evaluate the topological invariant. A crucial question is whether the integral (\ref{IntToDo}) develops singularities. If such singularities arise, the numerical integration based on measured data, may be delicate. To approach this question, return to Cartesian coordinates
\bea
\int \vec{S}\cdot\left({\partial\vec{S}\over\partial r}\times{\partial\vec{S}\over\partial\phi}\right) dr\, d\phi&=&\int {1\over r}\vec{S}\cdot\left({\partial\vec{S}\over\partial r}\times{\partial\vec{S}\over\partial\phi}\right) dx\, dy
\eea
where, as usual $x=r\cos\phi$, $y=r\sin\phi$ and $dx dy=r dr d\phi$. Does the integrand
\bea
i_{a_1a_2a_3}&=&{1\over r}\vec{S}\cdot \left({\partial\vec{S}\over\partial\phi} \times{\partial\vec{S}\over\partial r}\right)
\eea
develop singularities? The point $r=0$ is clearly a potential location of a singularity. A singularity is only avoided if $\vec{S}\cdot\left( {\partial\vec{S}\over\partial r}\times{\partial\vec{S}\over\partial r}\right)$ vanishes at $r=0$, as $r^q$ for any real $q\ge 1$. We will demonstrate that although many of the integrands are free of singularities, it is possible for singularities to develop at $r=0$. These are the only possible singularities. Our approach is straightforward: we directly evaluate the relevant integrands and systematically examine them for singular behavior. We start with two examples, one which is non-singular and one which is singular, to illustrate how we performed our analysis. We then list some results.

Consider the invariant $N_{128}$, determined by selecting the triple $\vec{S}(m_1,m_2,m_8)$. Direct evaluation gives (we set $l_{ab}=l_a-l_b$, use (\ref{explicitms}) and assume all $l_a>0$)
\bea
\vec{S}\cdot\left({\partial\vec{S}\over\partial\phi}\times{\partial\vec{S}\over\partial r}\right)&=&\frac{12 (l_0-l_1) r ^{2 l_0+2 l_1-1} \left(2 r ^{2 l_2} (l_0+l_1-2 l_2)+(l_0-l_1) \left(r ^{2 l_0}-r ^{2 l_1}\right)\right)}{\left(14 r ^{2 (l_0+l_1)}-4 r ^{2 (l_0+l_2)}+r ^{4 l_0}-4 r ^{2 (l_1+l_2)}+r ^{4 l_1}+4 r ^{4 l_2}\right)^{3/2}}
\eea
For simplicity, specialize to $l_0=1$, $l_1=2$ and $l_2=3$ and take the limit $r\to 0$. This expression behaves as
\bea
\vec{S}\cdot\left({\partial\vec{S}\over\partial\phi}\times{\partial\vec{S}\over\partial r}\right)&=&12r +O(r^2)
\eea
so that the integrand
\bea
i_{128}\sim {1\over r}\vec{S}\cdot\left({\partial\vec{S}\over\partial\phi}
\times{\partial\vec{S}\over\partial r}\right)&=&12+O(r)
\eea
is not singular at $r=0$. This regular behaviour of the integrand suggests that the numerical evaluation of the invariant will be robust.

To confirm that singularities can and do develop, consider the invariant $N_{124}$, determined by selecting the triple $\vec{S}(m_1,m_2,m_4)$ of Gell-man observables. Direct evaluation gives (we define $l_{ab}=l_a-l_b$, use (\ref{explicitms}) and assume all $l_a>0$)
\bea
\vec{S}\cdot\left({\partial\vec{S}\over\partial\phi}\times{\partial\vec{S}\over\partial r}\right)&=&-\frac{(l_0-l_1) (l_1-l_2) r ^{2 l_1+l_2-1} \cos (\phi  (l_0-l_2))}{\left(r ^{2 l_2} \cos ^2(\phi  (l_0-l_2))+r ^{2 l_1}\right)^{3/2}}
\eea
This expression does not vanish as $r\to 0$. For simplicity, again set $l_0=1$, $l_1=2$ and $l_2=3$ and take the limit $r\to 0$. We find that the integrand behaves as
\bea
i_{124}\sim{1\over r}\vec{S}\cdot\left({\partial\vec{S}\over\partial\phi}\times{\partial\vec{S}\over\partial r}\right)&=&-{\cos (2\phi)\over r}+O(1)
\eea
which clearly is singular. This potentially makes the numerical integration needed to obtain the invariant delicate.

To get an indication about how frequent these singularities here, we provide a list of the behavior of the integrand (denoted $i_{12x}$) in the $r\to 0$ limit, after setting $l_0=1$, $l_1=2$ and $l_2=3$, relevant for the $N_{12x}$ invariants:
\bea
i_{123}\sim{1\over r}\vec{S}\cdot\left({\partial\vec{S}\over\partial\phi}\times{\partial\vec{S}\over\partial r}\right)&=&4+O(r)\cr\cr
i_{124}\sim{1\over r}\vec{S}\cdot\left({\partial\vec{S}\over\partial\phi}\times{\partial\vec{S}\over\partial r}\right)&=&-{\cos (2\phi)\over r}+O(1)\cr\cr
i_{125}\sim{1\over r}\vec{S}\cdot\left({\partial\vec{S}\over\partial\phi}\times{\partial\vec{S}\over\partial r}\right)&=&-{\sin (2\phi)\over r}+O(1)\cr\cr
i_{126}\sim{1\over r}\vec{S}\cdot\left({\partial\vec{S}\over\partial\phi}\times{\partial\vec{S}\over\partial r}\right)&=&-2\cos (\phi)+O(r)\cr\cr
i_{127}\sim{1\over r}\vec{S}\cdot\left({\partial\vec{S}\over\partial\phi}\times{\partial\vec{S}\over\partial r}\right)&=&-2\sin (\phi)+O(r)\cr\cr
i_{128}\sim{1\over r}\vec{S}\cdot\left({\partial\vec{S}\over\partial\phi}\times{\partial\vec{S}\over\partial r}\right)&=&12+O(r)
\eea
Of the six integrands, two develop singularities. The presence of the singularity is dependent upon the value of the orbital angular momenta. To demonstrate this, flip the values of $l_1$ and $l_2$ and repeat the exercise. The result is
\bea
i_{123}\sim{1\over r}\vec{S}\cdot\left( {\partial\vec{S}\over\partial\phi}\times{\partial\vec{S}\over\partial r}\right)&=&16r^2+O(r^3)\cr\cr
i_{124}\sim{1\over r}\vec{S}\cdot\left( {\partial\vec{S}\over\partial\phi}\times{\partial\vec{S}\over\partial r}\right)&=&2 \sec ^2(\phi )+O(r)\cr\cr
i_{125}\sim{1\over r}\vec{S}\cdot\left( {\partial\vec{S}\over\partial\phi}\times{\partial\vec{S}\over\partial r}\right)&=&2\csc ^2(\phi )+O(r)\cr\cr
i_{126}\sim{1\over r}\vec{S}\cdot\left( {\partial\vec{S}\over\partial\phi}\times{\partial\vec{S}\over\partial r}\right)&=&-\frac{2 \cos (\phi )}{r }+O(1)\cr\cr
i_{127}\sim{1\over r}\vec{S}\cdot\left( {\partial\vec{S}\over\partial\phi}\times{\partial\vec{S}\over\partial r}\right)&=&\frac{2 \sin (\phi )}{r }+O(1)\cr\cr
i_{128}\sim{1\over r}\vec{S}\cdot\left( {\partial\vec{S}\over\partial\phi}\times{\partial\vec{S}\over\partial r}\right)&=&48r^2+O(r)
\eea
Again, only two integrands develop singularities, but the singularities now appear in different invariants.

For $d=3$ the next subsection provides an exhaustive analysis of the singularities.

\subsubsection{Regular and singular invariants}

In this section we explore the integrands of the complete list of topological invariants.

\begin{itemize}
\item[1.] The integrand is
\bea
i_{123}&=&\frac{4 (l_0-l_1)(|l_0|-|l_1|) r ^{2 (|l_0|+|l_1|-1)}}{\left(r ^{2 |l_0|}+r ^{2 |l_1|}\right)^2}
\eea
demonstrating that there are no singularities.

\item[2.] The integrands
\bea
i_{124}&=&-\frac{(l_0-l_1) (|l_1|-|l_2|) r ^{2 |l_1|+|l_2|-2} \cos (\phi  (l_0-l_2))}{\left(r ^{2 |l_2|} \cos ^2(\phi  (l_0-l_2))+r ^{2|l_1|}\right)^{3/2}}
\eea
\bea
i_{125}&=&\frac{(l_0-l_1) (|l_1|-|l_2|) r ^{2|l_1|+|l_2|-2} \sin (\phi  (l_0-l_2))}{\left(r ^{2|l_2|} \sin ^2(\phi  (l_0-l_2))+r ^{2|l_1|}\right)^{3/2}}
\eea
develop singularities if and only if $|l_2|=|l_1|+1$.

\item[3.] The integrands
\bea
i_{126}&=&-\frac{(l_0-l_1) (|l_0|-|l_2|) r ^{2|l_0|+|l_2|-2} \cos (\phi  (l_1-l_2))}{\left(r ^{2|l_0|}+r ^{2|l_2|} \cos ^2(\phi  (l_1-l_2))\right)^{3/2}}
\eea
\bea
i_{127}&=&\frac{(l_0-l_1) (|l_0|-|l_2|) r ^{2|l_0|+|l_2|-2} \sin (\phi  (l_1-l_2))}{\left(r ^{2|l_0|}+r ^{2|l_2|} \sin ^2(\phi  (l_1-l_2))\right)^{3/2}}
\eea
develop singularities if and only if $|l_2|=|l_0|+1$.

\item[4.] The integrand
\bea
i_{128}&=&\frac{12 (l_0-l_1) r ^{2 (l_0+l_1-1)} \left(2 r ^{2 l_2} (l_0+l_1-2 l_2)+(l_0-l_1) \left(r ^{2 l_0}-r ^{2 l_1}\right)\right)}{\left(14 r ^{2 (l_0+l_1)}-4 r ^{2 (l_0+l_2)}+r ^{4 l_0}-4 r ^{2 (l_1+l_2)}+r ^{4 l_1}+4 r ^{4 l_2}\right)^{3/2}}
\eea
does not develop singularities.

\item[5.] The integrands
\bea
i_{451}&=&\frac{(l_0-l_2) (| l_1| -| l_2| ) r ^{| l_1| +2 | l_2| -2} \cos (\phi  (l_0-l_1))}{\left(r ^{2 | l_1| } \cos ^2(\phi  (l_0-l_1))+r ^{2 | l_2| }\right)^{3/2}}
\eea
\bea
i_{452}&=&-\frac{(l_0-l_2) (| l_1| -| l_2| ) r ^{| l_1| +2 | l_2| -2} \sin (\phi  (l_0-l_1))}{\left(r ^{2 | l_1| } \sin ^2(\phi  (l_0-l_1))+r ^{2 | l_2| }\right)^{3/2}}
\eea
are singular if and only if $|l_1|=|l_2|+1$.

\item[6.] The integrand
\bea
i_{453}&=&\frac{4 (l_0-l_2) r ^{2 (| l_0| +| l_2| -1)} \left(r ^{2 | l_1| } (| l_0| -2 | l_1| +| l_2| )+(| l_0| -| l_2| ) r ^{2 | l_0| }\right)}{\left(\left(r ^{2 | l_0| }-r ^{2 | l_1| }\right)^2+4 r ^{2 (| l_0| +| l_2| )}\right)^{3/2}}
\eea
develops singularities if and only if $|l_2|<|l_1|<|l_0|$, $2|l_1|=|l_0|+|l_2|+1$ or $|l_2|<|l_0|<|l_1|$, $|l_0|=|l_2|+1$.

\item[7.] The integrands
\bea
i_{456}&=&-\frac{(l_0-l_2) (| l_0| -| l_1| ) r ^{2 | l_0| +| l_1| -2} \cos (\phi  (l_1-l_2))}{\left(r ^{2 | l_0| }+r ^{2 | l_1| } \cos ^2(\phi  (l_1-l_2))\right)^{3/2}}
\eea
\bea
i_{457}&=&\frac{(l_0-l_2) (| l_0| -| l_1| ) r^{2 |l_0| +|l_1| -2} \sin (\phi  (l_1-l_2))}{\left(r ^{2 | l_0| }+r ^{2 | l_1| } \sin^2(\phi  (l_1-l_2))\right)^{3/2}}
\eea
are singular if and only if $|l_1|=|l_0|+1$.

\item[8.] The integrand
\bea
i_{458}&=&-12 (l_0-l_2) r ^{2 (| l_0| +| l_2| -1)}\times \cr\cr
&&\qquad\frac{\left(| l_0|  \left(-r ^{2 | l_0| }+r ^{2 | l_1| }-2 r ^{2 | l_2| }\right)+| l_2|  \left(r ^{2 | l_0| }+r ^{2 | l_1| }+2 r ^{2 | l_2| }\right)-2 | l_1|  r ^{2 | l_1| }\right)}{\left(4 r ^{2 | l_2| } \left(2 r ^{2 | l_0| }-r ^{2 | l_1| }\right)+\left(r ^{2 | l_0| }+r ^{2 | l_1| }\right)^2+4 r ^{4 | l_2| }\right)^{3/2}}
\eea
is free from singularities.

\item[9.] The integrands
\bea
i_{671}&=&\frac{(l_1-l_2) (| l_0| -| l_2| ) r ^{| l_0| +2 | l_2| -2} \cos (\phi  (l_0-l_1))}{\left(r ^{2 | l_0| } \cos ^2(\phi  (l_0-l_1))+r ^{2 | l_2| }\right)^{3/2}}
\eea
\bea
i_{672}&=&-\frac{(l_1-l_2) (| l_0| -| l_2| ) r ^{| l_0| +2 | l_2| -2} \sin (\phi  (l_0-l_1))}{\left(r ^{2 | l_0| } \sin ^2(\phi  (l_0-l_1))+r ^{2 | l_2| }\right)^{3/2}}
\eea
develop singularities if and only if $|l_0|=|l_2|+1$.

\item[10.] The integrand
\bea
i_{673}&=&\frac{4 (l_1-l_2) r ^{2 (| l_1| +| l_2| -1)} \left(r ^{2 | l_0| } (2 | l_0| -| l_1| -| l_2| )+(| l_2| -| l_1| ) r ^{2 | l_1| }\right)}{\left(\left(r ^{2 | l_0| }-r ^{2 | l_1| }\right)^2+4 r ^{2 (| l_1| +| l_2| )}\right)^{3/2}}
\eea
is singular if and only if $|l_2|<|l_0|<|l_1|$, $2|l_0|=|l_1|+|l_2|+1$ or $|l_2|<|l_1|<|l_0|$, $|l_1|=|l_2|+1$.

\item[11.] The integrands
\bea
i_{674}&=&\frac{(l_1-l_2) (| l_0| -| l_1| ) r ^{| l_0| +2 | l_1| -2} \cos (\phi  (l_0-l_2))}{\left(r ^{2 | l_0| } \cos ^2(\phi  (l_0-l_2))+r ^{2 | l_1| }\right)^{3/2}}
\eea
\bea
i_{675}&=&-\frac{(l_1-l_2) (| l_0| -| l_1| ) r ^{| l_0| +2 | l_1| -2} \sin (\phi  (l_0-l_2))}{\left(r ^{2 | l_0| } \sin ^2(\phi  (l_0-l_2))+r ^{2 | l_1| }\right)^{3/2}}
\eea
develop singularities if and only if $|l_0|=|l_1|+1$.

\item[12.] The integrand
\bea
i_{678}&=&-12 (l_1-l_2) r ^{2 (| l_1| +| l_2| -1)}\times \cr\cr
&&\qquad\frac{\left(| l_1|  \left(r ^{2 | l_0| }-r ^{2 | l_1| }-2 r ^{2 | l_2| }\right)+| l_2|  \left(r ^{2 | l_0| }+r ^{2 | l_1| }+2 r ^{2 | l_2| }\right)-2 | l_0|  r ^{2 | l_0| }\right)}{\left(-4 r ^{2 | l_2| } \left(r ^{2 | l_0| }-2 r ^{2 | l_1| }\right)+\left(r ^{2 | l_0| }+r ^{2 | l_1| }\right)^2+4 r ^{4 | l_2| }\right)^{3/2}}
\eea
is always regular.

\subsection{Similarity scores}

To evaluate the fidelity of the topological spectrum we computed the residual similarity and the cosine similarity,  $\text{Residual Similarity}=1-\frac{\sum{(|A_i|-|E_i|)^2}}{\sum{|A_i|}}$ $\text{Cosine Similarity}=\frac{\sum{(\textbf{A}\cdot \textbf{E})}}{||\vec{A}||_2\cdot ||\vec{E}||_2} $ where $\textbf{A}=\frac{\vec{A}}{||\vec{A}||_1}$ $\textbf{E}=\frac{\vec{E}}{||\vec{E}||_1}$. $\vec{A}$ represents the analytical topological spectrum for a maximally entangled state while $\vec{E}$ represents the the experimentally obtained topological spectrum. The residual similarity presents a normalised squared difference between the topological signals which is subtracted from $1$, this similarity score is sensitive to the scaling of topological spectra and is an ideal measure to use when comparing the similarities of topological spectra which have the same topological signals however these signals are scaled by a constant factor. 
The cosine similarity compares the similarity of the vectors $\vec{A}$ and $\vec{E}$ in a u-dimensional space, where u represents the number of topological signals in the topological spectrum.  

\begin{table}[h!]
    \centering
    \begin{tabular}{| c | c | c |} 
     \hline
       Topological charges in state &  Residual Similarity &  Cosine Similarity \\ [0.5ex] 
     \hline\hline
      -1,0,1 & 0.92  & 0.98 \\ 
     \hline
      -3,0,3 & 0.93 & 0.99 \\
    \hline
    -5,4,5 & 0.78 & 0.99 \\
    [1ex] 
     \hline
    \end{tabular}
    \caption{Represents the similarity scores for the topological spectra for the corresponding states   }
    \label{tab:Similarity_scores_3D}
\end{table}

\end{itemize}

\section*{Supplementary: Emergence of topology under entanglement perturbation}

\noindent In real-world scenarios we are likely to encounter quantum channels which severely alter the entanglement distribution of a given quantum state. In our previous work \cite{ornelas2024non} we observed that, for 2D entangled states, as long as some entanglement persists, the initial topological signal would remain intact. However, the topological spectrum for states with $d>2$ is derived from different partitions of the state, thus potentially providing new avenues for the emergence of topology under entanglement distribution alterations between subspaces that live within the entire space of the biphoton state. Here we observe how the topology is affected by the alteration of the entanglement distribution of our state through a few simple proof of principle simulations, whose results are shown in Figure~\ref{fig:Noise_simulations}. In Figure~\ref{fig:Noise_simulations}(a) a simulation is shown for the density matrix, extended topological spectrum and Gellman vector field components derived from the maximally entangled state of the form 
\begin{equation}
    |\tilde{\psi} (\vec{r})\rangle_{AB} = \sum_{i=0}^{d-1} F_i(\vec{r}_A)|i\rangle_B
    \label{Eq: Ideal}
\end{equation}
with $i=\ell_i\in\{-3,0,3\}$.

\begin{figure*}[t]
\includegraphics[width=1\linewidth]{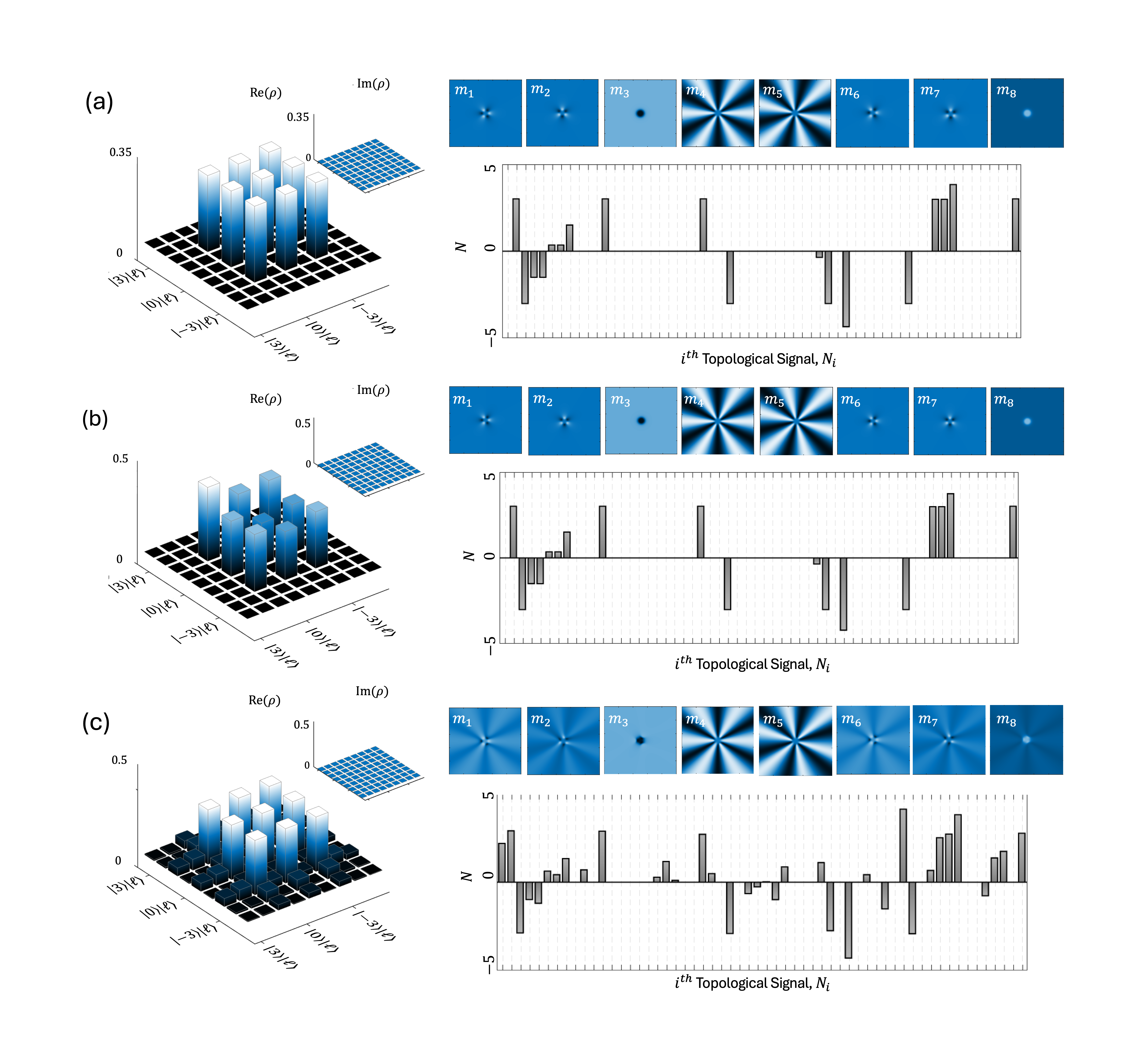}
\caption{\textbf{Emergence of topology under entanglement perturbation}(a) Represents the density matrix with the corresponding spatially-varying Gellmann matrices and the topological spectrum for a simulated maximally entangled state of $\ket{\Psi}_{AB} = \frac{1}{\sqrt{3}}\left(\ket{-3}_A\ket{3}_B + \ket{0}_A\ket{0}_B + \ket{3}_A\ket{-3}_B\right)$ (b) Represents the density matrix with the corresponding spatially-varying Gellmann matrices and the topological spectrum for a state $\ket{\Psi}_{AB} = 0.3333\ket{-3}_A\ket{3}_B + 0.2857\ket{0}_A\ket{0}_B + 0.3810\ket{3}_A\ket{-3}_B$ (c) Represents the density matrix with the corresponding spatially-varying Gellmann matrices and the topological spectrum for a state of the form $|\tilde{\psi} (\vec{r})\rangle_{AB} = |\psi (\vec{r}_A)\rangle_{AB} + |\Omega(\vec{r}_A)\rangle_{AB}$, where $|\Omega(\vec{r}_A)\rangle_{AB} = \sum_{j,k=0,\;j\neq k}^{d-1}  \delta_j F_j(\vec{r}_A)|k\rangle_B$ and $0.025\leq\delta_j\leq0.051$}
\label{fig:Noise_simulations}
\end{figure*}

\subsection{Non-maximally entangled states}
\noindent As a first step we consider partially separable states of the form
\begin{equation}
    |\tilde{\psi} (\vec{r})\rangle_{AB} = \sum_{i=0}^{d-1} c_i F_i(\vec{r}_A)|\ell_i\rangle_B,
    \label{Eq: NonMaxEnt}
\end{equation}
where the scalar coefficients, $c_i$, control the degree of non-separability of the state. Such states naturally arise from typical photon entanglement sources such as SPDC \cite{nape2023quantum}, where the probability of measuring different correlated states changes depending on the process used to produce the state. In Figure~\ref{fig:Noise_simulations}(b) a simulation is shown for a state with coefficients, $c_0\in\{0.3333,0.2857,0.3810\}$. The affect of altering the coefficients of the state is evident in the change in the density matrix however one can verify empirically that there was no change in the Gellman vector components nor the topological spectrum. This is further verified by computing a Cosine similarity for the spectrum of $C_s=1.00$, confirming that the spectra for the state in Figure~\ref{fig:Noise_simulations}(a) and (b) match. Therefore, altering the entangled state through changing its coefficients does not alter the topological information of the state.

\subsection{Entanglement distribution within biphoton subspaces}

\noindent Now we consider distributing entanglement across different subspaces within the full biphoton hilbert space. To capture this we consider the following model
\begin{equation}
    |\tilde{\psi} (\vec{r})\rangle_{AB} = (1-\delta)|\psi (\vec{r}_A)\rangle_{AB} + \delta|\Omega(\vec{r}_A)\rangle_{AB},
    \label{Eq: EntDistribution}
\end{equation}
where $|\psi_i (\vec{r}_A)\rangle_{AB} = \sum_{i=0}^{d-1} F_i(\vec{r}_A)|i\rangle_B$ is the initial maximally entangled pure state, $|\Omega(\vec{r}_A)\rangle_{AB} = \sum_{j,k=0,\;j\neq k}^{d-1}  F_j(\vec{r}_A)|k\rangle_B$ are the remaining states required to describe a state in the full biphoton hilbert space and $\delta\in[0,1]$ controls the contribution of these states. When $\delta=0$ we have a maximally entangled state in the subspace labeled by the basis states $\{|i\rangle\}$. However, when $0<\delta<1$ our state is partially non-separable due to the emergence of entanglement in other subspaces that live within the full biphoton hilbert space. In Figure~\ref{fig:Noise_simulations}(c) a simulation is shown for a state with randomly chosen $\delta$ values for each state in the full Hilbert space. The affect of this is evident in the density matrix where previously zero-valued components within the density matrix now have non-zero values. Furthermore, the Gellman vector components have been altered significantly which resulted in a significant change in the topological spectrum of the state when compared to the spectrum given in Figure~\ref{fig:Noise_simulations}(a). Besides the change in the magnitude over several topological signals, some initially non-zero signals were "turned on" by the redistribution of entanglement across the subspaces within the full Hilbert space. This deviation was captured by a drop in $C_s$ to $0.87$ when compared against the initial maximally entangled state. Therefore, by introducing entanglement into subspaces of the full biphoton Hilbert space, we have observed the emergence of topological information within the high-dimensional entangled state.\\


\end{widetext}


\newpage
\begin{thebibliography}{10}

\bibitem{ashbridge2022knotting}
Z.~Ashbridge, S.~D. Fielden, D.~A. Leigh, L.~Pirvu, F.~Schaufelberger, and L.~Zhang, ``Knotting matters: orderly molecular entanglements,'' {\em Chemical Society Reviews}, vol.~51, no.~18, pp.~7779--7809, 2022.

\bibitem{cruz2007cosmic}
M.~Cruz, N.~Turok, P.~Vielva, E.~Martinez-Gonzalez, and M.~Hobson, ``A cosmic microwave background feature consistent with a cosmic texture,'' {\em Science}, vol.~318, no.~5856, pp.~1612--1614, 2007.

\bibitem{ozawa2019topological}
T.~Ozawa and H.~M. Price, ``Topological quantum matter in synthetic dimensions,'' {\em Nature Reviews Physics}, vol.~1, no.~5, pp.~349--357, 2019.

\bibitem{eto2024tying}
M.~Eto, Y.~Hamada, and M.~Nitta, ``Tying knots in particle physics,'' {\em arXiv preprint arXiv:2407.11731}, 2024.

\bibitem{faddeev1997knots}
L.~Faddeev and A.~J. Niemi, ``Knots and particles,'' {\em Nature}, vol.~387, no.~arXiv: hep-th/9610193, p.~58, 1997.

\bibitem{hall2016tying}
D.~S. Hall, M.~W. Ray, K.~Tiurev, E.~Ruokokoski, A.~H. Gheorghe, and M.~M{\"o}tt{\"o}nen, ``Tying quantum knots,'' {\em Nature physics}, vol.~12, no.~5, pp.~478--483, 2016.

\bibitem{ge2021observation}
H.~Ge, X.-Y. Xu, L.~Liu, R.~Xu, Z.-K. Lin, S.-Y. Yu, M.~Bao, J.-H. Jiang, M.-H. Lu, and Y.-F. Chen, ``Observation of acoustic skyrmions,'' {\em Physical Review Letters}, vol.~127, no.~14, p.~144502, 2021.

\bibitem{xue2022topological}
H.~Xue, Y.~Yang, and B.~Zhang, ``Topological acoustics,'' {\em Nature Reviews Materials}, vol.~7, no.~12, pp.~974--990, 2022.

\bibitem{wang2025topological}
B.~Wang, Z.~Che, C.~Cheng, C.~Tong, L.~Shi, Y.~Shen, K.~Y. Bliokh, and J.~Zi, ``Topological water-wave structures manipulating particles,'' {\em Nature}, pp.~1--7, 2025.

\bibitem{shen2024optical}
Y.~Shen, Q.~Zhang, P.~Shi, L.~Du, X.~Yuan, and A.~V. Zayats, ``Optical skyrmions and other topological quasiparticles of light,'' {\em Nature Photonics}, vol.~18, no.~1, pp.~15--25, 2024.

\bibitem{tsesses2018optical}
S.~Tsesses, E.~Ostrovsky, K.~Cohen, B.~Gjonaj, N.~Lindner, and G.~Bartal, ``Optical skyrmion lattice in evanescent electromagnetic fields,'' {\em Science}, vol.~361, no.~6406, pp.~993--996, 2018.

\bibitem{gutierrez2021optical}
R.~Guti{\'e}rrez-Cuevas and E.~Pisanty, ``Optical polarization skyrmionic fields in free space,'' {\em Journal of Optics}, vol.~23, no.~2, p.~024004, 2021.

\bibitem{du2019deep}
L.~Du, A.~Yang, A.~V. Zayats, and X.~Yuan, ``Deep-subwavelength features of photonic skyrmions in a confined electromagnetic field with orbital angular momentum,'' {\em Nature Physics}, vol.~15, no.~7, pp.~650--654, 2019.

\bibitem{ornelas2024non}
P.~Ornelas, I.~Nape, R.~de~Mello~Koch, and A.~Forbes, ``Non-local skyrmions as topologically resilient quantum entangled states of light,'' {\em Nature Photonics}, vol.~18, no.~3, pp.~258--266, 2024.

\bibitem{wang2024topological}
A.~A. Wang, Z.~Zhao, Y.~Ma, Y.~Cai, R.~Zhang, X.~Shang, Y.~Zhang, J.~Qin, Z.-K. Pong, T.~Marozs{\'a}k, {\em et~al.}, ``Topological protection of optical skyrmions through complex media,'' {\em Light: Science \& Applications}, vol.~13, no.~1, p.~314, 2024.

\bibitem{yu2017room}
G.~Yu, P.~Upadhyaya, Q.~Shao, H.~Wu, G.~Yin, X.~Li, C.~He, W.~Jiang, X.~Han, P.~K. Amiri, {\em et~al.}, ``Room-temperature skyrmion shift device for memory application,'' {\em Nano letters}, vol.~17, no.~1, pp.~261--268, 2017.

\bibitem{blanco2018topological}
A.~Blanco-Redondo, B.~Bell, D.~Oren, B.~J. Eggleton, and M.~Segev, ``Topological protection of biphoton states,'' {\em Science}, vol.~362, no.~6414, pp.~568--571, 2018.

\bibitem{zhao2019non}
H.~Zhao, X.~Qiao, T.~Wu, B.~Midya, S.~Longhi, and L.~Feng, ``Non-hermitian topological light steering,'' {\em Science}, vol.~365, no.~6458, pp.~1163--1166, 2019.

\bibitem{song2020skyrmion}
K.~M. Song, J.-S. Jeong, B.~Pan, X.~Zhang, J.~Xia, S.~Cha, T.-E. Park, K.~Kim, S.~Finizio, J.~Raabe, {\em et~al.}, ``Skyrmion-based artificial synapses for neuromorphic computing,'' {\em Nature Electronics}, vol.~3, no.~3, pp.~148--155, 2020.

\bibitem{gobel2021beyond}
B.~G{\"o}bel, I.~Mertig, and O.~A. Tretiakov, ``Beyond skyrmions: Review and perspectives of alternative magnetic quasiparticles,'' {\em Physics Reports}, vol.~895, pp.~1--28, 2021.

\bibitem{benalcazar2017quantized}
W.~A. Benalcazar, B.~A. Bernevig, and T.~L. Hughes, ``Quantized electric multipole insulators,'' {\em Science}, vol.~357, no.~6346, pp.~61--66, 2017.

\bibitem{lustig2021topological}
E.~Lustig and M.~Segev, ``Topological photonics in synthetic dimensions,'' {\em Advances in Optics and Photonics}, vol.~13, no.~2, pp.~426--461, 2021.

\bibitem{tai2019three}
J.-S.~B. Tai and I.~I. Smalyukh, ``Three-dimensional crystals of adaptive knots,'' {\em Science}, vol.~365, no.~6460, pp.~1449--1453, 2019.

\bibitem{tsesses2025four}
S.~Tsesses, P.~Dreher, D.~Janoschka, A.~Neuhaus, K.~Cohen, T.~C. Meiler, T.~Bucher, S.~Sapir, B.~Frank, T.~J. Davis, {\em et~al.}, ``Four-dimensional conserved topological charge vectors in plasmonic quasicrystals,'' {\em Science}, vol.~387, no.~6734, pp.~644--648, 2025.

\bibitem{sugic2021particle}
D.~Sugic, R.~Droop, E.~Otte, D.~Ehrmanntraut, F.~Nori, J.~Ruostekoski, C.~Denz, and M.~R. Dennis, ``Particle-like topologies in light,'' {\em Nature communications}, vol.~12, no.~1, pp.~1--10, 2021.

\bibitem{shen2023topological}
Y.~Shen, B.~Yu, H.~Wu, C.~Li, Z.~Zhu, and A.~V. Zayats, ``Topological transformation and free-space transport of photonic hopfions,'' {\em Advanced Photonics}, vol.~5, no.~1, p.~015001, 2023.

\bibitem{ehrmanntraut2023optical}
D.~Ehrmanntraut, R.~Droop, D.~Sugic, E.~Otte, M.~R. Dennis, and C.~Denz, ``Optical second-order skyrmionic hopfion,'' {\em Optica}, vol.~10, no.~6, pp.~725--731, 2023.

\bibitem{marco2022optical}
D.~Marco and M.~A. Alonso, ``Optical fields spanning the 4d space of nonparaxial polarization,'' {\em arXiv preprint arXiv:2212.01366}, 2022.

\bibitem{lin2024space}
W.~Lin, N.~Mata-Cervera, Y.~Ota, Y.~Shen, and S.~Iwamoto, ``Space-time hopfion crystals,'' {\em arXiv preprint arXiv:2406.06096}, 2024.

\bibitem{Harvey:1996ur}
J.~A. Harvey, ``{Magnetic monopoles, duality and supersymmetry},'' in {\em {ICTP Summer School in High-energy Physics and Cosmology}}, 3 1996.

\bibitem{Irwin:1997ew}
P.~Irwin, ``{SU(3) monopoles and their fields},'' {\em Phys. Rev. D}, vol.~56, pp.~5200--5208, 1997.

\bibitem{lei2021photonic}
X.~Lei, A.~Yang, P.~Shi, Z.~Xie, L.~Du, A.~V. Zayats, and X.~Yuan, ``Photonic spin lattices: symmetry constraints for skyrmion and meron topologies,'' {\em Physical Review Letters}, vol.~127, no.~23, p.~237403, 2021.

\bibitem{forbes2024orbital}
A.~Forbes, L.~Mkhumbuza, and L.~Feng, ``Orbital angular momentum lasers,'' {\em Nature Reviews Physics}, pp.~1--13, 2024.

\bibitem{yao2024multi}
J.~Yao, Y.~Shen, J.~Hu, and Y.~Yang, ``Multi-degree-of-freedom hybrid optical skyrmions,'' {\em arXiv preprint arXiv:2409.05689}, 2024.

\bibitem{agnew2011tomography}
M.~Agnew, J.~Leach, M.~McLaren, F.~S. Roux, and R.~W. Boyd, ``Tomography of the quantum state of photons entangled in high dimensions,'' {\em Physical Review A—Atomic, Molecular, and Optical Physics}, vol.~84, no.~6, p.~062101, 2011.

\bibitem{zhou2020solids}
H.~Zhou, H.~Polshyn, T.~Taniguchi, K.~Watanabe, and A.~Young, ``Solids of quantum hall skyrmions in graphene,'' {\em Nature Physics}, vol.~16, no.~2, pp.~154--158, 2020.

\bibitem{psaroudaki2021skyrmion}
C.~Psaroudaki and C.~Panagopoulos, ``Skyrmion qubits: A new class of quantum logic elements based on nanoscale magnetization,'' {\em Physical Review Letters}, vol.~127, no.~6, p.~067201, 2021.

\bibitem{nape2023quantum}
I.~Nape, B.~Sephton, P.~Ornelas, C.~Moodley, and A.~Forbes, ``Quantum structured light in high dimensions,'' {\em APL Photonics}, vol.~8, no.~5, 2023.

\bibitem{erhard2018twisted}
M.~Erhard, R.~Fickler, M.~Krenn, and A.~Zeilinger, ``Twisted photons: new quantum perspectives in high dimensions,'' {\em Light: Science \& Applications}, vol.~7, no.~3, pp.~17146--17146, 2018.

\bibitem{bornman2021optimal}
N.~Bornman, W.~Tavares~Buono, M.~Lovemore, and A.~Forbes, ``Optimal pump shaping for entanglement control in any countable basis,'' {\em Advanced Quantum Technologies}, vol.~4, no.~10, p.~2100066, 2021.

\bibitem{bredon}
G.~E. Bredon, ``Topology and geometry,'' {\em Graduate Texts in Mathematics}, 1993.

\bibitem{lee2010introduction}
J.~Lee, {\em Introduction to Topological Manifolds}.
\newblock Graduate Texts in Mathematics, Springer New York, 2010.

\bibitem{tao2006perelman}
T.~Tao, ``Perelman's proof of the poincar$\backslash$'e conjecture: a nonlinear pde perspective,'' {\em arXiv preprint math/0610903}, 2006.

\bibitem{thurston1997three}
W.~P.~H. Thurston, {\em Three-Dimensional Geometry and Topology, Volume 1: Volume 1}.
\newblock Princeton university press, 1997.

\bibitem{donaldson1983application}
S.~K. Donaldson, ``An application of gauge theory to four-dimensional topology,'' {\em Journal of Differential Geometry}, vol.~18, no.~2, pp.~279--315, 1983.

\bibitem{witten1994monopoles}
E.~Witten, ``Monopoles and four-manifolds,'' {\em arXiv preprint hep-th/9411102}, 1994.

\bibitem{vonk2005mini}
M.~Vonk, ``A mini-course on topological strings,'' {\em arXiv preprint hep-th/0504147}, 2005.

\bibitem{gao2020paraxial}
S.~Gao, F.~C. Speirits, F.~Castellucci, S.~Franke-Arnold, S.~M. Barnett, and J.~B. G{\"o}tte, ``Paraxial skyrmionic beams,'' {\em Physical Review A}, vol.~102, no.~5, p.~053513, 2020.

\bibitem{Hall2015}
B.~C. Hall, ``Lie groups, lie algebras, and representations: An elementary introduction,,'' {\em Graduate Texts in Mathematics}, vol.~222, 2015.

\bibitem{skyrme1962unified}
T.~H.~R. Skyrme, ``A unified field theory of mesons and baryons,'' {\em Nuclear Physics}, vol.~31, pp.~556--569, 1962.

\bibitem{ackerman2017static}
P.~J. Ackerman and I.~I. Smalyukh, ``Static three-dimensional topological solitons in fluid chiral ferromagnets and colloids,'' {\em Nature materials}, vol.~16, no.~4, pp.~426--432, 2017.

\bibitem{bertlmann2008bloch}
R.~A. Bertlmann and P.~Krammer, ``Bloch vectors for qudits,'' {\em Journal of Physics A: Mathematical and Theoretical}, vol.~41, no.~23, p.~235303, 2008.

\end{thebibliography}

\makeatletter
\newenvironment{footnotebib}{%
  \begin{thebibliography}{99}%
\@ifnextchar\bibitem{\@gobble\@bibitem}{}}%
  {\end{thebibliography}}
\makeatother

\begin{footnotebib}
\bibitem{notes}
\end{footnotebib}

\end{document}